\pdfoutput=1
\documentclass[]{aa}

\usepackage{newtxtext}
\usepackage[varg]{newtxmath}
\usepackage{subcaption}
\usepackage{upgreek} % Provide the upright mu used by A&A
\usepackage{makecell}
\usepackage{siunitx}
\usepackage{booktabs}
\usepackage{graphicx}
\usepackage{orcidlink}
\usepackage{tikz}
\usetikzlibrary{shapes.geometric, arrows, calc, arrows.meta}
\usepackage{natbib}
\bibpunct{(}{)}{;}{a}{}{,} % natbib format for A&A and ApJ
\usepackage{hyperref}
\hypersetup{
    colorlinks,
    linkcolor={red!50!black},
    citecolor={blue!50!black},
    urlcolor={blue!80!black},
    pdftitle={Using debris disk observations to infer substellar companions
     orbiting within or outside a parent planetesimal belt},
    pdfauthor={T. A. Stuber, T. Löhne, S. Wolf},
    pdfkeywords={debris disk, exoplanet, planetesimals, exo-kuiper belts,
     dust dynamics, N-body simulations, collisional simulations,
     observational simulations, ELT, MICADO, METIS, JWST, NIRCam, MIRI,
     Herschel, PACS, ALMA, VLT, SPHERE, planet-disk interactions,
     circumstellar matter, interplanetary medium, infrared: planetary systems,
     submillimeter: planetary systems, methods: numerical}
}
\usepackage{subfiles}

\newcommand{\sbs}[1]{\ensuremath{_\text{#1}}}
\newcommand{\ACE}{\texttt{ACE}}
\newcommand{\DMS}{\texttt{DMS}}
\newcommand{\run}[1]{\texttt{#1}}
\newcommand{\SNR}{$S\textrm{/}N$}

\begin{document}

%%%%%%%%%%%%%%%%%%%%%%%%%%%%%%%%%%%%%%%%%%%%%%%%%%%%%%%%%%%%%%%%%%%%%%%%%%%%%%%%
%%%%%%%%%%%%%%%%%%%%%%%%%%%%%%%%%%%%%%%%%%%%%%%%%%%%%%%%%%%%%%%%%%%%%%%%%%%%%%%%

\title{Using debris disk observations to infer substellar companions orbiting
       within or outside a parent planetesimal belt}
\titlerunning{Infer substellar companions from debris disk observations}
\author{T. A. Stuber\inst{\ref{inst_kiel}} \orcidlink{0000-0003-2185-0525}
        \and T. Löhne\inst{\ref{inst_jena}} \orcidlink{0000-0003-2584-5280}
        \and S. Wolf\inst{\ref{inst_kiel}} \orcidlink{0000-0001-7841-3452}
}
\authorrunning{T. A. Stuber et al.}
\institute{Institut für Theoretische Physik und Astrophysik,
           Christian-Albrechts-Universität zu Kiel,
           Leibnizstr. 15, 24118 Kiel, Germany\\
           \email{tstuber@astrophysik.uni-kiel.de}
           \label{inst_kiel}
           \and Astrophysikalisches Institut und Universitätssternwarte,
           Friedrich-Schiller-Universität Jena, Schillergässchen 2–3,
           07745 Jena, Germany\label{inst_jena}
          }

\date{Received 1 February 2022 /
Accepted 2 October 2022}

\abstract
{Alongside a debris disk, substellar companions often exist in the same
system. The companions influence the dust dynamics via their
gravitational potential.}
{We analyze whether the effects of secular perturbations,
originating from a substellar companion, on the dust dynamics can be
investigated with spatially resolved observations.}
{We numerically simulated the collisional evolution of narrow and eccentric
cold planetesimal belts around a star of spectral type \mbox{A3$\,$V} that are
secularly perturbed by a substellar companion that orbits either closer to or
farther from the star than the belt. Our model requires a perturber on an
eccentric orbit \mbox{($e \gtrsim \num{0.3}$)} that is both far from and more
massive than the collisionally dominated belt around a luminous central star.
Based on the resulting spatial dust distributions, we simulated spatially
resolved maps of their surface brightness in the $K$, $N$, and $Q$~bands and at
wavelengths of \mbox{\SI{70}{\upmu\m}} and \mbox{\SI{1300}{\upmu\m}}.}
{Assuming a nearby debris disk seen face-on, we find that the surface
brightness distribution varies significantly with observing wavelength, for
example between the $N$ and $Q$~band. This can be explained by the varying
relative contribution of the emission of the smallest
grains near the blowout limit. The orbits of both the small grains that form
the halo and the large grains close to the parent belt precess due to the
secular perturbations induced by a substellar companion orbiting inward of the
belt. The halo, being composed of older grains, trails the belt. The magnitude
of the trailing decreases with increasing perturber mass and hence with
increasing strength of the perturbations. We recovered this trend in synthetic
maps of surface brightness by fitting ellipses to lines of constant
brightness. Systems with an outer perturber do not show a uniform halo
precession since the orbits of small grains are strongly altered.
We identified features of the brightness distributions suitable for
distinguishing between systems with a potentially detectable inner or outer
perturber, especially with a combined observation with JWST/MIRI in the
$Q$~band tracing small grain emission and with ALMA at millimeter wavelengths
tracing the position of the parent planetesimal belt.}
{}


\maketitle

%%%%%%%%%%%%%%%%%%%%%%%%%%%%%%%%%%%%%%%%%%%%%%%%%%%%%%%%%%%%%%%%%%%%%%%%%%%%%%%%
%%%%%%%%%%%%%%%%%%%%%%%%%%%%%%%%%%%%%%%%%%%%%%%%%%%%%%%%%%%%%%%%%%%%%%%%%%%%%%%%

\section{Introduction}\label{introduction}

 Various close-by main-sequence stars with debris disks have been found to host
 exoplanets, for example, $\beta$
 Pictoris \citep{10.1051/0004-6361:200811325, 10.1126/science.1187187,
 10.1038/s41550-019-0857-1, 10.1051/0004-6361/202039039}, $\epsilon$ Eridani
 \citep{10.1086/317319}, or AU Microscopii \citep{10.1038/s41586-020-2400-z,
 10.1051/0004-6361/202040235}.
 Several techniques, direct and indirect, were used to detect the
 aforementioned exoplanets: the exoplanets $\beta$~Pic~c and $\epsilon$~Eri~b
 were inferred by measuring the radial velocity of the host star
 \citep{1952Obs....72..199S}; AU~Mic~b and c were detected by measuring the
 light curve of the host star while the planet transits the line of sight
 \citep[e.g.,][]{1952Obs....72..199S, 10.1007/978-3-319-55333-7_117}; and
 $\beta$~Pic~b and c were detected by direct imaging
 \citep{10.1088/1538-3873/128/968/102001}. The first two
 techniques are sensitive to planets orbiting relatively close to the host
 star: for the radial velocity method, planets with long orbital periods are
 difficult to detect because the amplitude of the radial velocity signal
 decreases with increasing distance between the planet and the host star
 and due to the sheer time span required to observationally cover an orbit
 \citep[e.g.,][]{2010exop.book...27L}; for the
 transit method, the larger the orbit of a planet, the smaller the
 angular cross section the planet is blocking in front of the star and the less
 likely a sufficient alignment of the host star, the orbiting planet, and the
 observer is. The technique of direct imaging is capable of finding substellar
 companions on larger orbits, $\gtrsim \SI{100}{au}$, but requires the planets
 to  still be intrinsically bright and to not have cooled down since formation.
 This, as such, favors young systems as targets, that is, T~Tauri and Herbig
 Ae/Be stars as well as young moving group members
 \citep{10.1088/1538-3873/128/968/102001}; older, already cooled planets are
 difficult to detect. Astrometry that uses the data obtained by \textit{Gaia}
 \citep{10.1051/0004-6361/201629272} is expected to add thousands
 of exoplanet detections, but the orbital period of the planets discovered
 is limited by the mission lifetime of approximately ten years
 \citep{10.1051/0004-6361:20078997, 10.1088/0004-637X/797/1/14,
 10.1051/0004-6361/201730921}. An exoplanet hunting method without these
 biases is gravitational microlensing \citep[e.g.,][]{10.1086/186066,
 10.1086/171700, 10.1088/1674-4527/12/8/005, 10.3390/geosciences8100365}, but
 with this method systems at distances on the order of kiloparsecs are
 probed, too distant to be spatially resolved.
 In summary, we lack a planet hunting method to find old, and hence
 intrinsically dark, far-out planets in close-by stellar systems.

 In addition to exoplanets, stars have often been found to host
 debris disks \citep[e.g.,][]{10.1051/0004-6361/201218800,
 10.1051/0004-6361/201525764, 10.1051/0004-6361/201323058}, a common component
 in stellar systems beyond the proto-planetary phase \citep[e.g.,][]{
 10.1086/508649, 10.1051/0004-6361/201321050, 10.1051/0004-6361/201628329,
 10.1093/mnras/stx3188}. They are produced and continuously replenished by
 mutually colliding planetesimals that grind themselves down to dust in a
 collisional cascade and are characterized as being optically thin
 \citep[for recent reviews, see][]{
 10.2458/azu_uapress_9780816531240-ch023, 10.1146/annurev-astro-081817-052035,
 10.1016/B978-0-12-816490-7.00016-3}. The disks are usually observed in the
 near-infrared via the stellar light scattered off the dust and in the
 mid-infrared, far-infrared, and (sub)millimeter wavelength range via the
 thermal emission of the dust itself.

 Planets orbiting in a debris disk system have an impact on the planetesimal
 and dust grain dynamics via their gravitational potential. Therefore, by
 observing the dust emission one can potentially draw conclusions regarding the
 possibly unseen planets orbiting the central star
 \citep[e.g.,][]{10.1086/308093, 10.1016/j.pss.2006.04.035,
 10.1088/1674-4527/10/5/001, 10.3847/0004-637X/827/2/125}.
 The strength of the perturbing effect that a substellar companion has on
 the orbits of planetesimals and dust primarily depends on the distance of
 the perturber to the perturbed objects. Therefore, the spatial dust
 distribution produced by planetesimal collisions can be a signpost of old and
 far-out planets as well.
 Hence, analyses of spatially resolved observations of debris disks potentially
 serve as a planet hunting method that is complementary to the well-established
 methods that make use of stellar radial velocity, transits, and direct imaging
 to find exoplanets in close-by stellar systems
 \citep[e.g.,][]{2022A&A...659A.135P}.

 Narrow, eccentric debris rings are particularly promising tracers of long-term
 planetary perturbations. The deviation from circularity suggests that
 perturbations have happened, while the narrowness excludes violent short-term
 causes such as stellar flybys
 \citep[e.g.,][]{10.1046/j.1365-8711.2001.04212.x, 10.1006/icar.2001.6700}
 or (repeated) close encounters with planets
 \citep[e.g.,][]{10.1038/nature03676}. For long-term perturbations, where
 timescales are much longer than individual orbital periods, the orbits of belt
 objects are affected more coherently, with little spread in orbital elements.
 In contrast, instantaneous positions along the orbits are important in
 short-term perturbation events, resulting in a wider spread in orbital
 elements and wider disks. A narrow yet eccentric disk can only be compatible
 with a disk-crossing planet if the thus-excited wide disk component is
 subsequently removed \citep{10.1093/mnras/stab760}.
 The belts around Fomalhaut \citep[e.g.,][]{10.1038/nature03601,
 10.1088/0004-637X/775/1/56, 10.1088/2041-8205/750/1/L21,
 10.3847/1538-4357/aa71ae}, HD~202628 \citep{10.1088/0004-6256/144/2/45,
 10.3847/0004-6256/152/3/64, 10.3847/1538-3881/ab3ec1}, HD~53143
 \citep{2022ApJ...933L...1M}, and the younger system HR~4796\,A
 \citep[e.g.,][]{10.1051/0004-6361/201014881, 10.1093/mnras/sty135} are
 well-resolved examples of narrow, eccentric disks.

 To accomplish the task of using spatially resolved observations of dust
 emission to infer exoplanets and their properties, two
 key ingredients are necessary: first, the
 planet-disk interaction, that is, how perturbing planets shape the spatial
 dust distributions, and second, how these dust distributions appear in
 observations. With such a framework, in addition to searching for hints of
 exoplanets, we can use the known exoplanet--debris disk systems as test beds
 to better constrain debris disk properties such as collisional timescales,
 planetesimal stirring \citep[e.g.,][]{10.1093/mnras/stab771}, or
 self-gravitation \citep{10.3847/1538-4357/abda46} as well as planetesimal
 and dust material properties such as the critical energy for fragmentation,
 $Q\sbs{D}^\star$ \citep[e.g.,][]{10.1051/0004-6361/201833061}.

 This paper is organized as follows: In Sect.~\ref{sect_ace} we present the
 numerical methods applied to collisionally evolve planetesimal belts secularly
 perturbed by a substellar companion and discuss the resulting spatial grain
 distributions for different perturber--belt combinations. Based on these
 results, we show in Sect.~\ref{sect_obs_appearance} how we simulated flux
 density maps and explore the relative contribution of different grain sizes to
 the total radiation as well as how the halo of small grains on very eccentric
 orbits can be investigated observationally. In Sect.~\ref{sect_inner_vs_outer}
 we search for observable features to distinguish between systems with a
 substellar companion orbiting inside or outside a parent planetesimal belt and
 present our results in a simple decision tree. Lastly, in
 Sect.~\ref{sect_discussion} we discuss the results and in
 Sect.~\ref{sect_summary} briefly summarize our findings.

%%%%%%%%%%%%%%%%%%%%%%%%%%%%%%%%%%%%%%%%%%%%%%%%%%%%%%%%%%%%%%%%%%%%%%%%%%%%%%%
%%%%%%%%%%%%%%%%%%%%%%%%%%%%%%%%%%%%%%%%%%%%%%%%%%%%%%%%%%%%%%%%%%%%%%%%%%%%%%%

\section{ACE simulations}\label{sect_ace}

 We used the code \textbf{A}nalysis of \textbf{C}ollisional \textbf{E}volution
 \citep[\ACE,][]{10.1016/j.icarus.2004.10.003,
 10.1051/0004-6361:20064907, 10.1051/0004-6361/201014208,
 10.1051/0004-6361/201630297,10.1051/0004-6361/201935199} to evolve the radial,
 azimuthal, and size distribution of the material in debris disks. \ACE\
 follows a statistical approach, grouping particles in category bins according
 to their masses, $m$, and three orbital elements: pericenter distances, $q$,
 eccentricities, $e$, and longitudes of periapse,
 $\varpi = \Omega + \omega$.

 Mutual collisions can lead to four different outcomes in \ACE, depending on
 the masses and relative velocities of the colliders. At the highest impact
 energies, both colliders are shattered and the fragments dispersed such that
 the largest remaining fragment has less than half the original mass. Below the
 energy threshold for disruption and dispersal, a larger fragment remains,
 either as a direct remnant of the bigger object or as a consequence of
 gravitational re-accumulation. A cloud of smaller fragments is produced. If
 neither of the two colliders is shattered by the impact, both were assumed to
 rebound, resulting in two large remnants and a fragment cloud. In the
 unreached case of an impact velocity below $\sim 1$~m/s, the colliders would
 stick. These four regimes assumed in \ACE\ represent a simplification of the
 zoo of outcomes mapped by \citet{10.1051/0004-6361/200912852}.
 In addition to the collisional cascade, we took into account stellar radiation
 and wind pressure, the accompanying drag forces, and secular gravitational
 perturbation by a substellar companion.

 In the following subsections we describe the recent improvements made to the
 \ACE\ code, motivate and detail the simulation parameters, and present
 the resulting distributions.

%%%%%%%%%%%%%%%%%%%%%%%%%%%%%%%%%%%%%%%%%%%%%%%%%%%%%%%%%%%%%%%%%%%%%%%%%%%%%%%

 \subsection{Improved advection scheme}\label{subsect_ace_new}
 In \citet{10.1051/0004-6361/201014208} and subsequent work
 \citep[e.g.,][]{10.1051/0004-6361/201015328, 10.1051/0004-6361/201423523,
 10.1051/0004-6361/201630297} we used the upwind advection scheme to propagate
 material through the $q$--$e$ grid of orbital elements under the influence of
 Poynting--Robertson (PR) and stellar wind drag.
 \citet{10.1051/0004-6361/201935199} applied that scheme to the modeling of
 secular perturbations, where $q$, $e$, and $\varpi$  are affected while the
 semimajor axes are constant. In the following we refer to both PR drag and
 secular perturbations as transport.

 The upwind scheme moves material across the borders from one bin to its
 neighbors based on the coordinate velocities and amount of material in that
 bin. The scheme is linear in time and has the advantage that the transport
 gains and losses can be added simply to the collisional ones. For the PR
 effect, where drag leads to smooth and monotonous inward spread and
 circularization from the parent belt and halo, this scheme is sufficient.
 However, a narrow, eccentric parent belt under the influence of (periodic)
 secular perturbations requires the translation of sharp features in $q$, $e$,
 and $\varpi$ across the grid.
 The upwind scheme smears the sharp features out too quickly, inducing an
 unwanted widening and dynamical excitation of the disk
 \citep{10.1051/0004-6361/201935199}. To reduce the effect of this numerical
 dispersion, we introduced an operator splitting to the \ACE\ code, where
 collisions and transport (caused by drag and secular perturbation) are
 integrated one after the other for every time step. The transport part is
 integrated using a total variance diminishing (TVD) scheme
 \citep{10.1016/0021-9991(83)90136-5} with the \texttt{superbee} flux limiter
 \citep{10.1146/annurev.fl.18.010186.002005}. The contributions from
 PR drag and secular perturbation to the change rates
 $\dot q$, $\dot e$, and $\dot\varpi$ are summed up. For each time step
 $\Delta t$, the flow in three dimensions is again subdivided into five stages:
 $\Delta t/2$ in $q$, $\Delta t/2$ in $e$, $\Delta t$ in $\varpi$, $\Delta t/2$
 in $e$, and $\Delta t/2$ in $q$. A comparison of the resulting amounts of
 numerical dispersion in the TVD and the upwind schemes is shown in
 Appendix~\ref{app_ace_dispersion}.

%%%%%%%%%%%%%%%%%%%%%%%%%%%%%%%%%%%%%%%%%%%%%%%%%%%%%%%%%%%%%%%%%%%%%%%%%%%%%%%

 \subsection{Common parameters}\label{subsect_ace_params}
 The distribution of observable dust grains is determined by a range of
 parameters, including not only parameters of the dust, the disk, and the
 perturber, but also of the host star. The dust material was not varied in our
 study as we deem the discussed tracers of planetary perturbations unaffected.
 We chose a material blend of equal volume fractions of water ice
 \citet{1998A&A...331..291L} and astronomical silicate \citep{10.1086/379118},
 assuming a bulk density of \SI{2.35}{\g\per\cubic\cm} for the blend. The
 refractive indices are combined with the Maxwell--Garnett mixing rule
 \citep{10.1098/rsta.1904.0024}. Radiation pressure efficiency
 as well as absorption and scattering cross sections were calculated assuming
 compact spheres, using the Mie theory \citep{10.1002/andp.19083300302}
 algorithm \texttt{miex} \citep{10.1016/j.cpc.2004.06.070}.
 Below a limiting grain radius $s\sbs{bo}$, which depends on the
 dust material and the stellar luminosity, radiation pressure overcomes the
 gravitational pull and removes the grains on short timescales
 \citep[e.g.,][]{10.1016/0019-1035(79)90050-2}.
 We assumed the same critical specific energy for disruption and dispersal,
 $Q\sbs{D}^*$, as in \citet[their Eq.~12]{10.1051/0004-6361/201630297}: a sum
 of three power laws for strength- and gravity-dominated shocks and a pure
 gravitational bond, respectively.

 The grid of object radii extended from \SI{0.36}{\upmu\m} to
 \SI{481}{\metre}. At the upper end, a factor of \num{2.3} separated
 neighboring size bins. This factor reduced to \num{1.23} near the blowout
 limit. The according mass grid follows Eq.~(26) of
 \citet{10.1051/0004-6361/201935199}. Material in the initial dust belt covered
 only radii that exceeded \SI{100}{\upmu\m}, with a power-law index of
 \num{-3.66}, close to the steady-state slope expected in the strength regime
 \citep{10.1016/S0019-1035(03)00145-3}. The lower size bins were filled in the
 subsequent evolution. The logarithmic grid of pericenter distances had 40 bins
 between \SI{40}{au} and \SI{160}{au}. The eccentricity grid was linear for
 $0.4 \lesssim e \lesssim 1$ and logarithmic outside of this range, following
 Eq.~(29) of \citet{10.1051/0004-6361/201630297}. The 36 bins in the grid of
 orbit orientations, $\varpi$, were each \ang{10} wide.

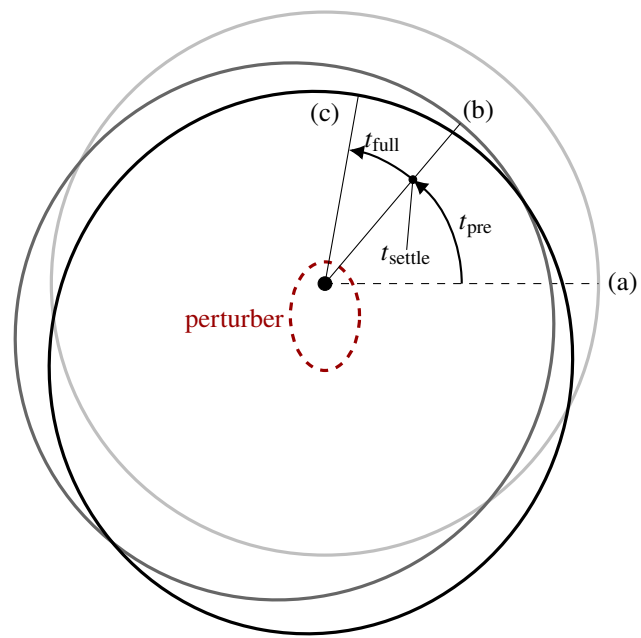
\begin{figure}\centering%
  \begin{tikzpicture}[baseline,scale=3.6]
    \draw[fill=black] (0,0) circle (0.25mm);
    \draw[color=red!60!black, very thick, dashed] (270:0.6*0.2cm) ellipse (0.632*0.2cm and 0.2cm);
    \draw[very thick,color=black!25] (0,0) circle (1cm);
    \draw[dashed] (0,0) -- (0:1);
    \begin{scope}[rotate=230]
     \draw[very thick,color=black!60] (0:{sin(5.0*180.0/18.0)*0.3}) ellipse (1.0cm and 0.973cm);
     \draw (0,0) -- (180:{1-sin(5.0*180.0/18.0)*0.3});
    \end{scope}
    \begin{scope}[rotate=260]
     \draw[very thick] (0:{sin(8.0*180.0/18.0)*0.3}) ellipse (1.0cm and 0.955cm);
     \draw (0,0) -- (180:{1-sin(8.0*180.0/18.0)*0.3});
    \end{scope}
    \draw[thick,-{Latex[length=2mm, width=2mm]}] (0:0.5) arc (0:50:0.5cm) node [midway,anchor=west] {$t\sbs{pre}$};
    \draw[thick,-{Latex[length=2mm, width=2mm]}] (50:0.5) arc (50:80:0.5cm) node [midway,anchor=south] {$t\sbs{full}$};
    \draw[fill=black] (50:0.5) circle (0.14mm) -- +(-95:0.26) +(-95:0.28) node {$t\sbs{settle}$};
    \draw (0:1) node [right] {(a)};
    \draw (50:0.73) node [anchor=south west] {(b)};
    \draw (90:0.7) node [below] {(c)};
    \draw[color=red!60!black] (-130:0.18) node [left] {perturber};
  \end{tikzpicture}%
  \caption{Schematic representation of the belt orbits (solid gray and black)
           and the planet orbit    (red)  at the different stages of the ACE
           simulations:
           (a) the initial, circular, unperturbed belt;
           (a--b) the belt perturbed by the planet, but not modified by
           collisions;
           (b) the belt modified by collisions, but not by the perturber;
           (b--c) the belt modified by both the perturber and collisions.}
  \label{fig_ace_sim_times}
 \end{figure}

 The belts were assumed to start unperturbed but pre-stirred; initially
 circular with a maximum free eccentricity $e_\text{max} = 0.1$. The
 distributions in $q$, $e$, and $\varpi$ were jointly initialized from a random
 sample of uniformly distributed semimajor axes, eccentricities, and longitudes
 of ascending nodes. This random sample was then propagated for a time
 $t_\text{pre}$ under the sole influence of transport, that is, PR drag and
 secular perturbation. After this propagation the resulting sample was injected
 into the grid. From this time on, the distribution was allowed to settle into
 collisional equilibrium for $t_\text{settle} = 20$~Myr. Only after
 $t_\text{pre} + t_\text{settle}$ passed would the combined simulation of
 transport and collisions begin, lasting for a time $t\sbs{full}$. The procedure
 ensures that (a) the numerical dispersion is further reduced by solving the
 initial perturbation before the discretization of the grid is imposed and (b)
 the size distribution of dust grains has time to reach a quasi-steady state.
 Figure~\ref{fig_ace_sim_times} illustrates the constant orbit of the planet and
 the mean belt orbit at the different stages of the simulation runs.
 Stages $t\sbs{pre}$ and $t\sbs{full}$ were tuned such that the mean belt
 orientation (and eccentricity) was the same for all simulation runs. This is
 to mimic the normal case of an observed disk of given eccentricity and
 orientation and an unseen perturber with unknown mass and orbital parameters.

 Our simulations resulted in snapshots of narrow belts that have not reached
 a dynamical equilibrium with the perturber yet, undergoing further evolution.
 If not prevented by the self-gravity of the belt,
 differential secular precession between belt inner and outer edges would
 widen the eccentricity distribution to a point where it is only compatible
 with the broad disks that are in the majority among those observed
 \citep[e.g.,][]{10.3847/1538-4357/aabcc4, 10.1146/annurev-astro-081817-052035}.
 The combined collisional and dynamical steady state of broad disks was
 simulated by \citet{10.1051/0004-6361/201219962}. \citet{10.1098/rsos.200063}
 discussed the potential dynamical history of both broad and narrow disks.

 We assumed the perturber to be a point mass on a constant,
 eccentric orbit closer to or further away from the star than the belt.
 The disk was assumed not to exert secular perturbation itself,
 neglecting the effects of secular resonances internal to the disk
 or between disk and planet
 \citep[cf.][]{10.1093/mnras/stv1847,10.3847/1538-4357/abda46}.
 The model is only applicable to systems where the perturber is more
 massive than the disk. Because the estimated total masses of the best-studied,
 brightest debris disks can exceed tens or hundreds of Earth masses
 \citep{10.1093/mnras/stx2932,10.1093/mnras/staa2385},
 we limited our study to perturber masses on the order of a Jupiter mass or
 above.

 The stellar mass and luminosity determine orbital velocities and the blowout
 limit. For stars with a higher
 luminosity-to-mass ratio, the blowout limit is at larger grain radii. Barely
 bound grains slightly above that limit populate extended orbits to form the
 disk halo. In addition, the lower size cutoff induces a characteristic wave in
 the grain size distribution \citep{10.1016/0032-0633(94)90008-6,
 10.1051/0004-6361:20031017, 10.1051/0004-6361:20064907} that can translate to
 observable spectro-photometric features \citep{10.1051/0004-6361:20077709}.
 Both the spectral ranges at which the halo and the wave in the size
 distribution are most prominent are determined by the stellar spectral type.
 However, that influence is well understood and mostly qualitative. The
 differences from one host star to another at a constant wavelength (of light)
 are similar to the differences from one wavelength (of light) to another for a
 single host star. We therefore modeled only one
 specific host star with a mass of $\SI{1.92}{M_\odot}$ and a luminosity of
 $\SI{16.6}{L_\odot}$, roughly matching the A3$\,$V star Fomalhaut. We assumed
 the spectrum of a modeled stellar atmosphere with effective temperature
 $T\sbs{eff} = \SI{8600}{\K}$, surface gravity
 $\log\sbs{10} (g [\SI{}{\cm\per\s\squared}]) = 4.0$, and metallicity
 $\left[ \textrm{Fe/H} \right] = 0.0$ \citep{10.1086/306745}. The results are
 insensitive to the last two parameters. The corresponding blowout limit is
 $s\sbs{bo} \approx \SI{4}{\upmu\m}$. For main-sequence stars of mass lower
 than the Sun, radiation pressure is too weak to produce blowout grains
 \citep[e.g.,][]{%
 10.1007/s00159-006-0028-0,%
 10.1051/0004-6361/201220486,%
 10.3847/1538-3881/ab095e}. The lack of a natural lower size cut-off for late-type stars can
 lead to transport processes \citep[e.g.,][]{10.1051/0004-6361/201015328}
 and nondisruptive collisions \citep{10.1051/0004-6361/201423862, 10.1051/0004-6361/201527626}
 becoming more important, potentially resulting in observable features that are qualitatively
 different from the ones presented here. We do not cover this regime here.

%%%%%%%%%%%%%%%%%%%%%%%%%%%%%%%%%%%%%%%%%%%%%%%%%%%%%%%%%%%%%%%%%%%%%%%%%%%%%%%

 \subsection{Varied parameters}\label{subsect_ace_runs}

 We varied a total of five physically motivated parameters in our study:
 the disk mass $M\sbs{b}$, the belt widths, $\Delta a\sbs{b}$, as well as
 the perturber mass, $M\sbs{p}$, semimajor axis, $a\sbs{p}$, and eccentricity,
 $e\sbs{p}$. The parameter combinations assumed for belts and perturbers are
 summarized in Tables~\ref{table_belt} and \ref{table_perturber}, respectively,
 together with the collisional settling time, $t\sbs{settle}$,
 the initial perturbation time, $t\sbs{pre}$, and the
 period of full simulation of perturbations and collisions, $t\sbs{full}$.
 We use the abbreviations given in the first columns of these tables to
 refer to individual model runs. For example, the run that combined the wider
 parent belt with an inner high-mass perturber is denoted \run{w-i-M3}, while the
 combination of a narrower belt with the low-eccentricity, high-mass inner perturber
 is denoted \run{n-i-M3-le}.

 \begin{table}\centering%
 \caption{Assumed parameters for the parent belt.\label{table_belt}}
 \begin{tabular}{rcccccl}
    \toprule
    Id.      & $M\sbs{b}$   & $\Delta a\sbs{b}$ & $t\sbs{settle}$ & Description \\
             & $[M_\oplus]$ & [\SI{}{au}]       & [Myr]           & \\
    \midrule
    \run{n}  & $0.09$       & 10                & 20              & reference \\
    \run{w}  & $0.09$       & 20                & 20              & wide \\
    \run{m2} & $0.28$       & 10                & 6.4             & high mass \\
    \run{m3} & $1.4$        & 10                & 1.3             & very high mass \\
    \bottomrule
 \end{tabular}
 \tablefoot{$M\sbs{b}$ is the total belt mass in objects with radii
 $s < \SI{500}{\m}$, $\Delta a\sbs{b}$ the spread in orbital semimajor axes,
 and $t\sbs{settle}$ the time during which the size distribution is allowed to settle
 to a collisional equilibrium before collisions and perturbations were modeled jointly.
 See text for details.}
 \end{table}

 \begin{table*}\centering%
 \caption{Parameter combinations for the perturbers.\label{table_perturber}}
 \begin{tabular}{rcccccl}
    \toprule
     Id.        & $M\sbs{p}$     & $a\sbs{p}$  & $e\sbs{p}$ & $t\sbs{pre}$ & $t\sbs{full}$ & Description \\
                & $[M\sbs{Jup}]$ & [\SI{}{au}] &            & [Myr]        & [Myr]         & \\
     \midrule
     \run{i-M1} & $0.5$          & $20$        & $0.6$      & 25           & 15            & inner, low mass \\
     \run{i-M2} & $2.5$          & $20$        & $0.6$      & 5            & 3             & inner, medium mass, \textsl{reference} \\
     \run{i-M3} & $12.5$         & $20$        & $0.6$      & 1            & 0.6           & inner, high mass \\
     \run{i-M4} & $62.5$         & $20$        & $0.6$      & 0.2          & 0.12          & inner, very high mass\\
     \midrule
     \run{o-M2} & $2.5$          & $500$       & $0.6$      & 25           & 15            & outer, medium mass \\
     \run{o-M3} & $12.5$         & $500$       & $0.6$      & 5            & 3             & outer, high mass \\
     \run{o-M4} & $62.5$         & $500$       & $0.6$      & 1            & 0.6           & outer, very high mass\\
     \midrule
     \run{-le}  & ---            & ---         & $0.3$      & ---          & ---           & low eccentricity\\
     \run{di}   & $0.49$         & $40$        & $0.31$     & ---          & ---           & degenerate, inner\\
     \run{p0}   & ---            & ---         & ---        & ---          & ---           & no precession \\
     \run{p1}   & ---            & ---         & ---        & ---          & ---           & no ongoing prec. \\
     \bottomrule
 \end{tabular}
 \tablefoot{Where no values are given, the corresponding \textsl{reference} values apply.}
 \end{table*}

 The effects of secular perturbation by a single
 perturber can be reduced to two main quantities: the timescale and the
 amplitude. To leading orders in orbital eccentricities and semimajor axes,
 the time required for a full precession cycle of grains launched from parent
 bodies on near-circular orbits at $a\sbs{b}$ is given by
 \citep{10.1051/0004-6361/201935199}
 \begin{equation}\label{eq:Tprec}
   T\sbs{prec} \approx \frac{4}{3} \frac{M_*}{M\sbs{p}} P\sbs{b}
   \left\{
   \begin{array}{l}
     \left(\frac{a\sbs{b}}{a\sbs{p}}\right)^2
     \frac{(1 - \beta)^4}{(1 - 2\beta)^{7/2}} \propto a\sbs{b}^{7/2}
     a\sbs{p}^{-2} M\sbs{p}^{-1} \quad \text{(inner)}\\
     \left(\frac{a\sbs{p}}{a\sbs{b}}\right)^3
     \frac{(1 - 2\beta)^{3/2}}{1 - \beta} \propto a\sbs{b}^{-3/2}
     a\sbs{p}^3 M\sbs{p}^{-1} \quad \text{(outer)}
   \end{array}\right.
 \end{equation}
 for perturbers distant from the belt, where $M_* = 1.92\,M_\odot$ is the mass
 of the host star, $P\sbs{b}$ the orbital period of the parent bodies, and $\beta$ the
 radiation-pressure-to-gravity ratio of the launched grains. Hence, the
 perturbation timescale is determined by a combination of perturber semimajor axis,
 perturber mass, and belt radius.

 The amplitude of the perturbations is controlled by the forced orbital
 eccentricity that is induced by the perturber,
 \begin{equation}\label{eq:ef}
   e\sbs{f} \approx \frac{5}{4} e\sbs{p} \left\{
   \begin{array}{l}
     \frac{a\sbs{p}}{a\sbs{b}} \frac{1 - 2\beta}{1 - \beta} \propto a\sbs{b}^{-1}
     a\sbs{p} e\sbs{p} \quad \text{(inner)}\\
     \frac{a\sbs{b}}{a\sbs{p}} \frac{1 - \beta}{1 - 2\beta} \propto a\sbs{b}
     a\sbs{p}^{-1} e\sbs{p} \quad \text{(outer)}\, ,
   \end{array}\right.
 \end{equation}
 around which the actual belt eccentricity evolves. This amplitude is determined
 by belt radius, perturber semimajor axis, and perturber eccentricity.

 With the perturbation problem being only two-dimensional, we reduced the set
 of varied parameters by fixing the mean radius of the belt at \SI{100}{au},
 a typical value for cold debris disks observed in the far-infrared and at
 (sub)millimeter wavelengths
 \citep[e.g.,][]{%
 10.1051/0004-6361/201321050,%
 10.1088/0004-637X/792/1/65,%
 10.1093/mnras/stx1378,%
 10.1093/mnras/stx3188,%
 10.3847/1538-4357/aabcc4,%
 10.1146/annurev-astro-081817-052035,%
 10.1093/mnras/staa3917,%
 10.1051/0004-6361/202140740}.
 For $a\sbs{b} = \text{const}$, which is a given parent belt, the
 timescale is constant for $M\sbs{p} \propto a\sbs{p}^{-2}$ and an inner
 perturber, or $M\sbs{p} \propto a\sbs{p}^3$ and an outer perturber.
 The amplitude is constant for $e\sbs{p} \propto a\sbs{p}^{-1}$ and
 an inner perturber, or $e\sbs{p} \propto a\sbs{p}$ and an outer perturber.
 We expect degenerate behavior for some parameter
 combinations even in our reduced set.

 The runs \run{di}, with an inner perturber closer to the belt, and \run{o-M3},
 with an outer perturber, were constructed as degeneracy checks. Their outcomes
 should be as close to the reference run, \run{n-i-M2}, as possible.
 The perturbation timescales and amplitudes in the middles of the respective
 belts, given in Eqs.~(\ref{eq:Tprec}) and (\ref{eq:ef}), are the same for all
 three parameter sets. For \run{n-di} the parameters listed in
 Tables~\ref{table_belt} and \ref{table_perturber} imply that differential
 precession acts on exactly the same timescale and with the same amplitude as in
 run \run{n-i-M2} throughout the whole disk. The outcomes of the \run{di} runs, which
 had an inner perturber closer to the belt, should therefore be fully
 degenerate with the equivalent \run{M2} runs. For the runs with an outer
 perturber, \run{o-M3}, the degeneracy applies only to the belt center because
 the sense of the differential perturbation is inverted as the exponent to
 $a\sbs{b}$ changes from $+7/2$ to $-3/2$. The dependence on $\beta$ is inverted
 too: the $\beta$-dependent term in Eq.~(\ref{eq:Tprec}) increases with
 increasing $\beta$ for an inner perturber and decreases with increasing $\beta$
 for an outer perturber.

 While $M\sbs{p}$ affects only the perturbation timescale, $a\sbs{p}$ and
 $e\sbs{p}$ affect the overall amplitude of the perturbations. In runs \run{le} we
 therefore lowered the perturber eccentricities to the more moderate value of
 $e\sbs{p} = 0.3$ (from the reference value, $e\sbs{p} = 0.6$).
 In an initially circular belt at \SI{100}{au}, a perturber with $e\sbs{p} = 0.3$
 at \SI{20}{au} induces a maximum belt eccentricity that amounts to approximately
 twice the forced eccentricity given by Eq.~(\ref{eq:ef}),
 that is, $\approx 2\times 5/4 \times 0.3 \times 20/100 = 0.15$
 (compared to $0.30$ for $e\sbs{p} = 0.6$).
 While planetary orbital eccentricities around 0.3 are common among long-period
 exoplanets, eccentricities around 0.6 are rare \citep{10.3847/0004-637X/821/2/89}.
 Likewise, belt eccentricities around 0.15 are closer to the maximum of
 what is derived for observed disks, as exemplified by the aforementioned
 disks around Fomalhaut, HD 202628, and HR~4796\,A. Therefore,
 our reference case provides clearer insights into the expected
 qualitative behavior, while runs~\run{le} are closer to observed disks.

 The perturber determines the rate at which the secular perturbation occurs,
 while the collisional evolution in the belt determines the rate at which the
 small-grain halo is replenished and realigned.
 Collisions occur on a timescale given by the dynamical excitation,
 spatial density, and strength of the material. Instead of varying all these
 quantities, we varied only the disk mass (in runs \run{m2} and \run{m3}) as a
 proxy for the spatial density. An increased disk mass and a reduced perturber
 mass are expected to yield similar results, as in both cases, the ratio between
 the timescales for collisional evolution and for secular perturbation is reduced.
 The total disk masses given in Table~\ref{table_belt} may seem low because
 the simulation runs are limited to object radii $\lesssim 0.5$~km.
 However, when extrapolating to planetesimal radii $\sim 100$~km
 with a typical power-law index of -3.0\ldots-2.8,
 as observed in the Solar System asteroid and Kuiper belts,
 the total masses increase by factors of $(100/0.5)^{1.0\ldots1.2}$ (i.e., by
 2 to 2.5 orders of magnitude).

 The belt width was varied explicitly in runs \run{w}.
 Not only do the belts in runs \run{w} have lower collision rates but also
 increased differential precession and potentially a clearer spatial separation
 of observable features.

 Finally, we set up two runs that allow us to differentiate between the effects
 of the mean belt eccentricity, the differential precession of the belt, and the
 ongoing differential precession of the halo with respect to the belt. Only the
 last would be a sign of a currently present perturber. The first two could,
 for example, be the result of an earlier, ceased perturbation. In runs
 \run{p0}, we assumed belts with the same mean eccentricity and orientation as \run{M2}, but
 without any differential secular perturbation, that is, no twisted belt or halo.
 Such a configuration could result if the perturbations have ceased some time
 ago or the eccentricity was caused by another mechanism, such as a single giant
 breakup. In runs \run{p1}, the belts were initially twisted to the same degree as for
 \run{n-i-M1} to \run{n-i-M4}, but no ongoing precession was assumed to drive
 the twisting of the halo. Ceased perturbation could be a physical motivation
 for that scenario as well. However, the main purpose of \run{p0} and \run{p1}
 was to help interpret the causes of and act as a baseline for features in the
 other simulation runs.

%%%%%%%%%%%%%%%%%%%%%%%%%%%%%%%%%%%%%%%%%%%%%%%%%%%%%%%%%%%%%%%%%%%%%%%%%%%%%%%

 \subsection{Grain distributions for an inner perturber}\label{subsect_ace_inner}
\begin{figure}\centering%
   \includegraphics[width=\linewidth]{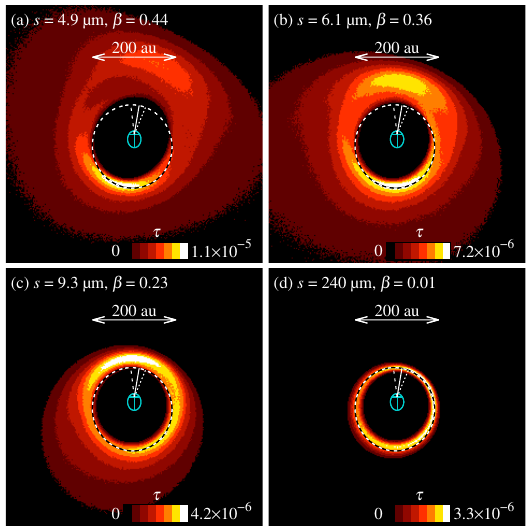}
   \caption{Maps of contributions to face-on optical depth, $\tau$, in run~\run{n-i-M2}
   from grains of four different size bins with mean grain radii $s$ and
   corresponding $\beta$ ratios, as labeled in the top-left corners of the panels.
   The orbits of the perturbers are shown with blue-green ellipses, and the host star
   is at the crossings of the major axes and latera recta.
   Dashed black-and-white ellipses trace the center of the parent belt.
   White lines show the orientations and lengths of
   the pericenters at the belt center (solid)  and the belt edges (dashed).
   Arrows indicate the scale.}
   \label{fig_ace_sizes}
 \end{figure}

 Figure~\ref{fig_ace_sizes} shows distributions of normal geometrical optical
 thickness for grains of different sizes in run \run{n-i-M2}, our reference for
 subsequent comparisons. These maps resulted from a Monte-Carlo sampling
 of the $q$--$e$--$\varpi$ phase space as well as mean anomaly. The maps show
 the contribution per size bin. The total optical thickness peaks at $\approx \num{3e-5}$
 in run \run{n-i-M2}, the total fractional luminosity at a similar value,
 which is a moderate value among the minority of mass-rich disks with brightness above current
 far-infrared detection limits \citep[e.g.,][]{10.1051/0004-6361/201321050, 10.1093/mnras/stx3188}
 and a low value among those above millimeter detection limits \citep[e.g.,][]{10.3847/1538-4357/aabcc4}.

 The big grains in Fig.~\ref{fig_ace_sizes}d represent the parent belt,
 which started out circular and then completed almost half a counterclockwise precession cycle.
 The higher precession rate of the inner belt edge caused the left side to be diluted and
 the right side to be compressed, resulting in an azimuthal brightness asymmetry.
 This geometric effect of differential precession is notable only when the
 width of the belt is resolved. In wider belts, differential precession can create
 spiral density variations \citep{10.1086/377195,10.1086/428954,10.1051/0004-6361:20053391}.
 The effect becomes increasingly prominent over time or reaches a limit set by the
 belt's self-gravity.

 For smaller grains, the effects described in \citet{10.1051/0004-6361/201630297}
 come into play. Figure~\ref{fig_ace_sizes}c shows the distribution of grains with
 radii $s \approx \SI{9}{\upmu\m}$. These grains are produced in or near the
 parent belt, but populate eccentric orbits as a result of their radiation
 pressure-to-gravity ratio $\beta \approx 0.23$. Those grains born near the
 parent belt's pericenter inherit more orbital energy and form a more extended
 halo on the opposite side, beyond the belt's apocenter, in the lower part of the
 panel. At the same time, the alignment of their orbits creates an overdensity of
 mid-sized grains near the belt pericenter. The yet smaller grains in panels (a)
 and (b) tend to become unbound when launched from the parent belt's pericenter,
 while those launched from the belt's apocenter have their apocenters on the
 opposite side, forming a halo beyond the belt's pericenter, in the upper parts
 of the panels. These effects are purely caused by the parent belt being
 eccentric.

 The ongoing differential precession causes a misalignment of the halo with
 respect to the parent belt \citep{10.1051/0004-6361/201935199}. This
 misalignment is more clearly seen in panels (b) and (c), which show a clockwise
 offset of the outer halo with respect to the belt, resulting from the wider
 halo orbits precessing at a lower rate. The population of halo grains is in an
 equilibrium between steady erosion and replenishment. Erosion is caused by
 collisions and PR drag, while replenishment is caused by
 collisions of somewhat bigger grains closer to the belt. The population of
 freshly produced small grains forms a halo that is aligned with the belt, while
 the older grains in the halo trail behind. The ratio of grain lifetime and
 differential precession timescale determines the strength of the misalignment. If
 the secular perturbation is strong or collisions are rare, the misalignment will
 be more pronounced, and vice versa. The smaller the grains, the higher are their
 $\beta$ ratios, the wider their orbits, and the more they trail behind the belt.

\begin{figure}\centering%
  \includegraphics[width=\linewidth]{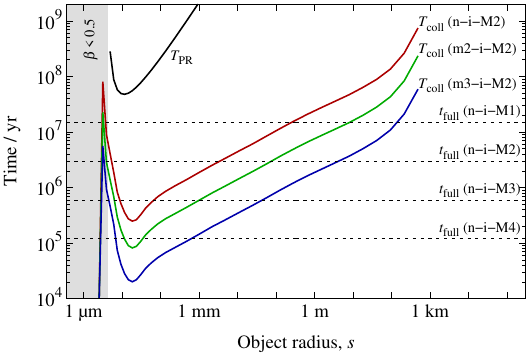}
  \caption{Size dependence of timescales for collisions (solid blue, green, and
           red), PR drag (solid black), and a full secular precession cycle at
           \SI{100}{au} (dashed black) for perturbers of different masses, as
           labeled. The timescales are calculated as described in
           \citet{10.1051/0004-6361/201935199}.}
  \label{fig_ace_timescales}
 \end{figure}

 A comparison of the most important timescales is given in Fig.~\ref{fig_ace_timescales}.
 The collision timescale shown was obtained directly from \ACE\ run~\run{n-i-M2},
 although it was very similar for all runs with the reference belt \run{n}.
 In runs with the wide belt \run{w}, the collision timescales were longer by a factor of two.
 In the more mass-rich belts \run{m2} and \run{m3} the collision timescales were
 correspondingly shorter.
 As a proxy to a PR timescale, Fig.~\ref{fig_ace_timescales} shows
 the $e$-folding time of the pericenter distances of grains with a size-dependent
 $\beta$ ratio, launched from a circular belt of radius $a\sbs{b} = \SI{100}{au}$
 \citep{10.1051/0004-6361/201935199}:
 \begin{equation}
   T\sbs{PR} \equiv \frac{q}{\dot q}(\beta) =
   \num{32}\,\text{Myr} \times \frac{1 - \beta}{\beta (4 - 5\beta) \left[1 - 2\beta \right]^{1.5}}
   \left(\frac{a\sbs{b}}{\SI{100}{au}}\right)^2 \frac{M_\odot}{M_*}\, .
 \end{equation}
 The PR timescale has a lower limit of 50~Myr, obtained for grains with $\beta \approx 0.2$
 (corresponding to radii $s\approx\SI{10}{\upmu\m}$),
 resulting in PR drag being insignificant in the presented model runs.
 A stronger contribution from PR drag would be expected for normal optical
 thickness $\lesssim\num{1e-6}$ \citep{10.1051/0004-6361:20042073,10.1093/mnras/stv453},
 much less than the peak value of $\num{3e-5}$ reached in our disks. With collision timescale
 being proportional to disk mass squared and PR timescale being independent from disk mass,
 lowering the optical thickness (and hence the disk mass) by a factor of 30 would increase the
 collision timescale by a factor of $\approx \num{1000}$, bringing the green
 curve in Fig.~\ref{fig_ace_timescales} close to the black one for small grains.
 After simulations longer than the PR timescale, dust from the belt and halo could travel
 closer to the star, showing up in surface brightness maps at up to mid-infrared wavelengths,
 even for the optical thickness considered here
 \citep{10.1051/0004-6361/201117731,10.1093/mnras/stv453,10.1051/0004-6361/201630297}.

 The time $t\sbs{full}$ is the time during which planetary perturbations
 and collisions were modeled simultaneously in our \ACE\ runs.
 Over this period the complex belt eccentricity covered one sixth of a full
 precession cycle around the eccentricity that was forced by the perturber.
 Depending on the perturber-to-belt mass ratio, the collision timescale shown
 in Fig.~\ref{fig_ace_timescales} can be shorter or longer than the precession
 timescale for the grains that are most abundant and best observable,
 from the blowout limit at $s\approx\SI{4}{\upmu\m}$ to $s\approx\SI{1}{\milli\m}$.
 Where the collision timescale is shorter, the distribution can be considered
 equilibrated. Where the precession timescale is shorter, an equilibrium may only be reached
 after several full precession cycles, when the complex eccentricities are randomized.
 This long-term equilibrium after randomization is studied by \citet{10.1051/0004-6361/201219962},
 who find the resulting disks to be azimuthally symmetric. The numerical
 dispersion prevented us from following the evolution over such long timescales,
 which is why we limited this study to nonequilibrium, ongoing perturbation for
 the cases where precession timescales are shorter, that is, for the higher
 perturber-to-disk mass ratios. In runs \run{n}, grains with radii
 $\lesssim \SI{20}{\upmu \m}$ have $\beta$
 ratios distributing them largely outside of the parent belt, forming
 the halo. Grains around that critical size have the shortest collisional
 lifetimes (see Fig.~\ref{fig_ace_timescales}). For these grains differential
 precession and collisions do not reach an equilibrium if perturber-to-disk mass
 ratios exceed a factor of \num{100}\ldots\num{300}, taking into account the
 extrapolation to largest planetesimal radii of $\sim \SI{100}{\km}$, as
 discussed in Sect.~\ref{subsect_ace_runs}. When considering only the dust
 content, that is, grains up to roughly \SI{1}{\mm} in radius and $M\sbs{b}$ as
 given in Table~\ref{table_belt}, this mass ratio increases to
 \num{7000}\ldots\num{20000}.
 For grains with radii of \SI{8}{\upmu \m}, where the halo extent is
 significant, the equilibrium is not reached when the perturber-to-disk mass
 ratio exceeds \num{10}\ldots\num{30} (or \num{700}\ldots\num{2000} for the dust
 content).

\begin{figure}\centering%
   \includegraphics[width=\linewidth]{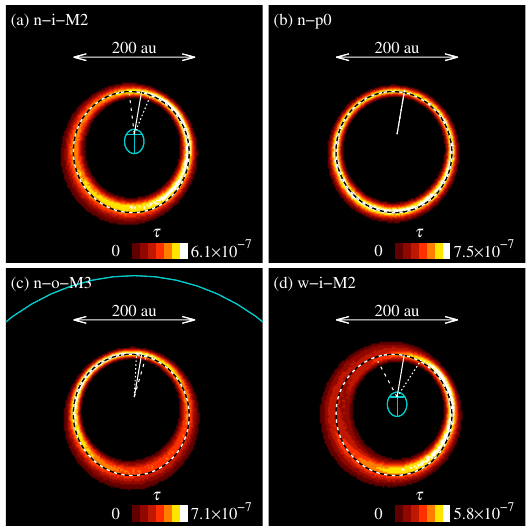}
   \caption{Maps of contributions to face-on optical depth for grains of radius
            $s = \SI{5}{\milli\metre}$ in four different simulation runs:
            (a) \run{n-i-M2}, (b) \run{n-p0}, (d) \run{n-o-M3}, and (e) \run{w-i-M2}.
            See Fig.~\ref{fig_ace_sizes} for a detailed description.}
   \label{fig_ace_degen_big}
 \end{figure}

\begin{figure}\centering%
   \includegraphics[width=\linewidth]{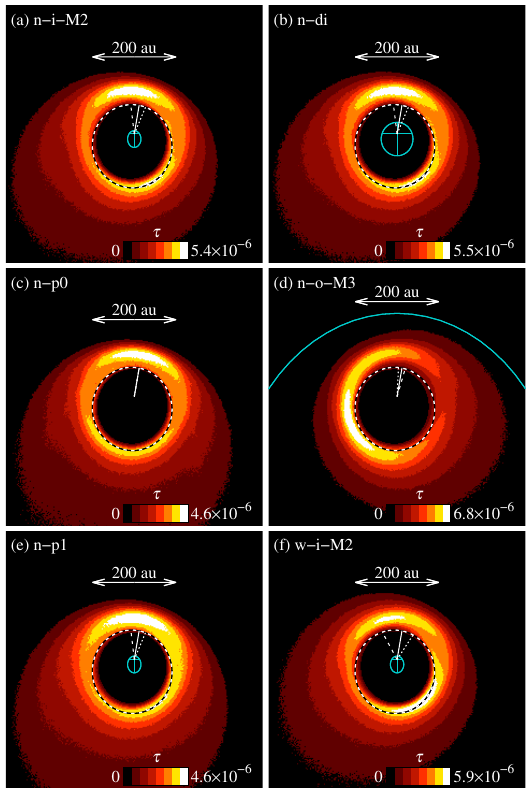}
   \caption{Maps of contributions to face-on optical depth for grains of radius
            $s = \SI{7.5}{\upmu\m}$ in six different simulation runs:
            (a) \run{n-i-M2}, (b) \run{n-di}, (c) \run{n-p0}, (d) \run{n-o-M3},
            (e) \run{n-p1}, and (f) \run{w-i-M2}. See Fig.~\ref{fig_ace_sizes} for a
            detailed description.}
   \label{fig_ace_degen_small}
 \end{figure}

 This leads to the question of how much of the halo asymmetry is actually
 caused by the already asymmetric belt. Figure~\ref{fig_ace_degen_big} shows the
 distributions of bigger grains for runs (a) \run{n-i-M2} and (b) \run{n-p0}.
 While \run{n-i-M2} shows the features already discussed above, the belt in run
 \run{n-p0} is purely eccentric, without the characteristic left-right asymmetry
 caused by differential precession. Run \run{n-p1} is not shown, as it would be
 indistinguishable from \run{n-i-M2}. The corresponding distributions of smaller
 grains in runs \run{n-i-M2}, \run{n-p0}, and \run{n-p1} are shown in panels
 (a), (c), and (e) of Fig.~\ref{fig_ace_degen_small}, respectively. As expected,
 the small-grain halo in run \run{n-p0} shows no additional asymmetry. Run
 \run{n-p1} has a belt that shows the same degree of asymmetry as that in run
 \run{n-i-M2}, but no ongoing precession that could further twist the halo. The
 slight misalignment of the small-grain halo in panel (e) is purely caused by
 the left--right asymmetry in density, and hence, collision rates in the parent
 belt. The \run{n-p1} halo is already similar to the case of a low-mass
 perturber modeled in \run{n-i-M1}, shown in Fig.~\ref{fig_ace_masses_small}a,
 indicating that the effect of differential perturbations is weak compared to
 collisions in run \run{n-i-M1}. For the given combination of belt parameters
 and perturber orbit, the mass of a perturber able to twist the halo should
 exceed one Jupiter mass. This threshold is inversely proportional to the
 collisional timescale in the belt and directly proportional to the total dust
 mass.

 \subsection{Dependence on parameters}\label{subsect_ace_variations}
\begin{figure}\centering%
   \includegraphics[width=\linewidth]{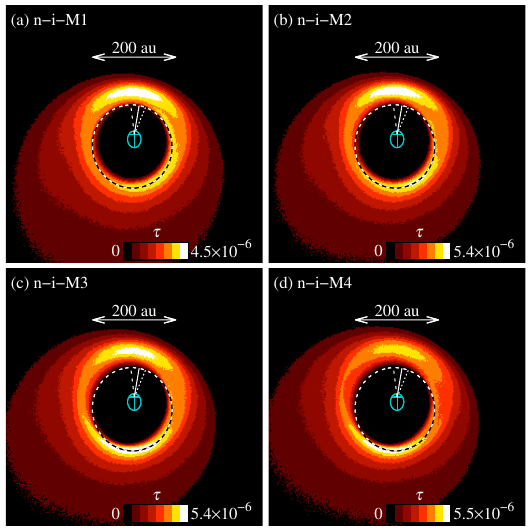}
   \caption{Maps of contributions to face-on optical depth for grains of radius
            $s = \SI{7.5}{\upmu\m}$ in runs: (a) \run{n-i-M1}, (b) \run{n-i-M2},
            (c) \run{n-i-M3}, and (d) \run{n-i-M4}. See Fig.~\ref{fig_ace_sizes} for a
            detailed description.}
   \label{fig_ace_masses_small}
 \end{figure}

 Figure~\ref{fig_ace_masses_small} shows the distributions of small grains for a
 series of runs with different perturber masses. The masses were increased by factors
 of five from run to run, and hence, the perturbation timescales decreased by
 factors of five. As a result, the misalignment of belt and halo increases
 monotonously from run \run{n-i-M1} to run \run{n-i-M4}, as more and more halo
 grains from earlier stages of the precession cycle are still present. However,
 the misalignment is limited because the initial circular belt produced a
 radially symmetric halo that does not contribute to the azimuthal asymmetry.

 The azimuthal distribution of small grains near the parent belt is another
 feature that is modulated by the strength of the perturbations. For a low-mass
 perturber that distribution is dominated by a combination of the overdensity
 near the belt pericenter and a remnant of the left-right asymmetry of the belt.
 For a high-mass perturber, where halos of a wider range of orientations overlap
 with old, symmetric halos, that pericenter overdensity is weakened with respect
 to the belt apocenter.

\begin{figure}\centering%
   \includegraphics[width=\linewidth]{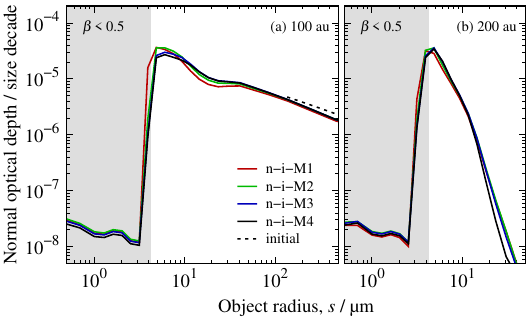}
   \caption{Azimuthally averaged grain size distributions in terms of contribution
   to normal optical thickness at\SI{100}{au} (left) and \SI{200}{au} (right)
   for runs \run{n-i-M1} (red), \run{n-i-M2} (green), \run{n-i-M3} (blue),
   and \run{n-i-M4} (black). The dashed line traces the initial distribution common
   to all runs, with a lower radius limit at $\SI{100}{\upmu\m}$. The gray-shaded
   region is dominated by grains on unbound orbits.}
   \label{fig_ace_size_dist}
 \end{figure}

 The azimuthally averaged size distribution depends only weakly on the degree of
 perturbation. Fig.~\ref{fig_ace_size_dist} shows small differences for grains
 between one and a few blowout radii among runs \run{n-i-M1} through \run{n-i-M4}.
 These differences are caused mostly by the collisional lifetimes of grains
 near the blowout limit (as shown in Fig.~\ref{fig_ace_timescales}) being longer than the
 total time over which the collisional cascade is simulated,
 $t\sbs{settle} + t\sbs{full}$. When translated to spatially unresolved
 spectral energy distributions, the resulting differences are
 small compared to uncertainties that arise from properties
 such as grain composition or dynamical excitation of the disk.

 In a wider parent belt, the angular displacement between inner and outer edge is
 more pronounced, causing a stronger azimuthal asymmetry among the bigger grains
 in the parent belt. Figure~\ref{fig_ace_degen_big}d shows this distribution. In
 contrast, the resulting distribution of small grains
 (Fig.~\ref{fig_ace_degen_small}f), which is always more extended than the parent
 belt, does not differ much from the narrower belt in run \run{n-i-M2}
 (Fig.~\ref{fig_ace_degen_small}a). While a yet wider belt can be expected to
 show differences in the small-grain halo, we refrained from performing additional
 simulation runs because such more diffuse belts would constitute a class of
 objects different from the narrow Fomalhaut-like belts we focus on.
 The results for the high-mass disks in runs \run{m2} and \run{m3} are not shown
 because they exhibit the expected similarity to \run{n-i-M1}, the run with the
 low-mass perturber. As anticipated in Sect.~\ref{subsect_ace_runs}, the higher
 disk masses increase the collision rate, reducing the effect of the secular
 perturbations in the same way that the lower perturber mass does.

\begin{figure}\centering%
   \includegraphics[width=\linewidth]{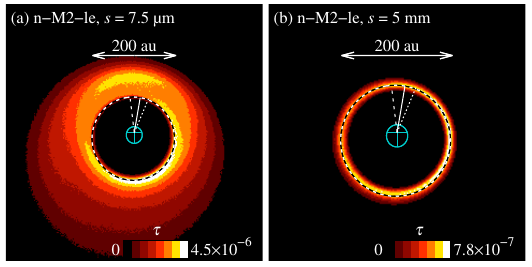}
   \caption{Maps of contributions to face-on optical depth for grains of radii
            $s = \SI{7.5}{\upmu\m}$ (a) and $s = \SI{5}{\milli\metre}$ (b) in
            run \run{n-M2-le}.}
   \label{fig_ace_low_ecc}
 \end{figure}

 In runs \run{le} the orbital eccentricities of the perturbers were halved. The resulting
 belt eccentricities follow suit, reducing the overall asymmetry of the belt and halo.
 The distribution of big and small grains is shown in Fig.~\ref{fig_ace_low_ecc}.
 Compared to the \run{n} runs, the halos in the \run{n-le} runs are more circular: wider near
 the pericenter of the belt and narrower near the belt's apocenter
 \citep[cf.][]{10.1051/0004-6361/201630297}. As a result, the semimajor axes and
 perturbation timescales of grains forming the halo on the apocenter side are
 shorter, closer to that of the belt. The resulting orientation of the halo
 follows that of the belt more closely than for a more eccentric perturber. The
 perturber mass threshold above which the twisted halo becomes notable is higher.

 \subsection{Inner versus outer perturber}\label{subsect_ace_outer}
 The expected degeneracy between runs with different inner perturbers is seen in
 Figs.~\ref{fig_ace_degen_small}a and b. Both runs, \run{n-i-M2} and \run{n-di},
 resulted in practically equal grain distributions because both perturbers exerted
 equal perturbations. Small differences could arise only from higher-order
 corrections to Eqs.~(\ref{eq:Tprec}) and (\ref{eq:ef}). Such corrections were
 included for the involved semimajor axes but not for the eccentricities.
 Hence, the accuracy is limited for small grains on very eccentric orbits
 (see \citet{10.1051/0004-6361/201935199} for a description of how orbits close
 to $e = 1$ are treated in \ACE).

 The \run{o-M3} runs addressed the case of an outer perturber at \SI{500}{au} that causes exactly
 the same perturbation of the belt center as the reference perturber \run{i-M2}, albeit
 at a perturber mass that is five times as high. Figures~\ref{fig_ace_degen_big}c
 and \ref{fig_ace_degen_small}d show the resulting distributions of big and small
 grains, respectively. The outer perturber caused the outer belt edge to lead,
 the inner edge to trail, producing a big-grain asymmetry that is mirrored
 compared to the inner perturber. For actually observed disks and unobserved
 perturbers, these seemingly different cases can appear similar because the sense
 of the orbital motion in the disk, clockwise or counterclockwise, is usually
 unknown.

 The distribution of smaller grains in \run{n-o-M3} (Fig.~\ref{fig_ace_degen_small}d)
 differs from a pure mirror image of \run{n-i-M2} (Fig.~\ref{fig_ace_degen_small}a). Instead
 of two azimuthal maxima, at the belt pericenter and apocenter, the outer
 perturber induced only one maximum, away from the apsidal line. This qualitative
 difference could be used to distinguish between an inner and an outer perturber.
 Another difference is the presence of small grains interior to the belt in
 run~\run{n-o-M3} (Fig.~\ref{fig_ace_degen_small}d). An outer perturber can,
 due to its proximity to the halo, further increase the orbital eccentricity of halo
 grains, shortening their periastron distances compared to the parent belt.
 The brightness maps shown by \citet[their Fig.~10]{10.1051/0004-6361/201219962}
 exhibit a feature that is very similar, including the over-brightness on the periastron side.
 The contribution to this interior asymmetry from grains that are dragged in by the PR
 effect \citep[cf.][]{10.1051/0004-6361/201630297} is small in our case, but might be
 greater for the equilibrated disks discussed by \citet{10.1051/0004-6361/201219962}.
 In our simulations, the distributions had an artificial cut-off at \SI{40}{au}, the open
 inner edge of the pericenter grid described in Sect.~\ref{subsect_ace_params}.
 The observability of these features in and interior to the belt will be explored
 further in Sect.~\ref{subsect_azimuthal_maxima}.

%%%%%%%%%%%%%%%%%%%%%%%%%%%%%%%%%%%%%%%%%%%%%%%%%%%%%%%%%%%%%%%%%%%%%%%%%%%%%%%
%%%%%%%%%%%%%%%%%%%%%%%%%%%%%%%%%%%%%%%%%%%%%%%%%%%%%%%%%%%%%%%%%%%%%%%%%%%%%%%

\section{Observational appearance of sheared debris disks}
\label{sect_obs_appearance}
 Based on the spatial distributions of the dust discussed in
 Sect.~\ref{sect_ace}, we investigated the imprint of planet-debris disk
 interaction for the considered scenarios on selected observable quantities. We
 assumed that the dust distributions are optically thin, that is, the observable
 flux density $S_\nu$ results from the superposition of the individual
 contributions of the thermal emission and scattered stellar light of grains of
 \emph{all} sizes. While -- at a fixed radial distance to the star -- the
 scattered radiation depends on the scattering cross section ($C_\textrm{sca}$)
 and the Müller-matrix element $S_{11}$, the emitted radiation depends on the
 absorption cross section ($C_\textrm{abs}$). For disks seen face-on, hence a
 fixed scattering angle, and assuming compact spherical grains all three
 quantities depend on the grain size, chemical composition, and observing
 wavelength. Therefore, multiwavelength
 observations potentially enable us to separate the contribution of certain
 grain sizes and allow us to conclude on the links between disk structure,
 dynamics, and eventually orbital parameters of an embedded exoplanet.

 We present our method to compute brightness distributions from the spatial dust
 distributions computed with \ACE\ and give a short discussion about the contribution of different
 grain sizes to the total flux in Sect.~\ref{subsect_dms}. Subsequently, in
 Sect.~\ref{subsect_halo_twisting}, we investigate the potential to observe the
 halo twisting discussed in Sects.~\ref{subsect_ace_inner} --
 \ref{subsect_ace_outer}. Any comments about the feasibility of observing our
 simulated distributions of surface brightness using real instruments are based
 on a performance analysis for a system at a stellar distance of
 \SI{7.7}{pc} (e.g., Fomalhaut) that is presented in the appendix (see
 Sect.~\ref{appdx_general observability}). We focus our investigation on the
 systems with the parent belt of the reference parameter set \run{n}.

\subsection{Brightness distributions}\label{subsect_dms}
 \begin{table}
  \caption{Considered observing wavelengths.}
  \label{table_wavelengths}
  \begin{tabular}{l l l}
   \toprule
   $\lambda \left[ \upmu \textrm{m} \right]$  & Dominating radiation & Instruments e.g. \\
   \midrule
   \makecell[cll]{\num{2} \\ $\ $ \\ $\ $} &
        \makecell[cl]{scattered stellar light \\ $\ $ \\ $\ $} &
        \makecell[cll]{VLT/SPHERE$\,$\tablefootmark{a}, \\
                  ELT/MICADO$\,$\tablefootmark{b}, \\
                  JWST$\,$\tablefootmark{c}/NIRCam$\,$\tablefootmark{d}} \\
   \makecell[cl]{\num{10} \\ $\ $} &
        \makecell[cl]{scattered stellar light \\ $\ $} &
        \makecell[cl]{ELT/METIS$\,$\tablefootmark{e}, \\
                      JWST/MIRI$\,$\tablefootmark{f}}\\
   \num{21} & thermal dust emission & JWST/MIRI \\
   \num{70} & thermal dust emission &
        \textit{Herschel}$\,$\tablefootmark{g}/PACS$\,$\tablefootmark{h} \\
   \num{1300} & thermal dust emission & ALMA$\,$\tablefootmark{i} \\
   \bottomrule
  \end{tabular}
  \tablefoot{The denoted dominating radiation is specific for our model setup.}
  \tablebib{
            \tablefoottext{a}{\citet{10.1051/0004-6361/201935251}};
            \tablefoottext{b}{\citet{10.1117/12.2311483, 10.18727/0722-6691/5217}};\\
            \tablefoottext{c}{\citet{10.1007/s11214-006-8315-7}};
            \tablefoottext{d}{\citet{10.1117/12.552281}};
            \tablefoottext{e}{\citet{10.1117/12.2056468, 10.18727/0722-6691/5218}};
            \tablefoottext{f}{\citet{10.1117/12.551717}};
            \mbox{\tablefoottext{g}{\citet{10.1051/0004-6361/201014759}};}
            \tablefoottext{h}{\citet{10.1051/0004-6361/201014535}};
            \tablefoottext{i}{\citet{2002Msngr.107....7K}}.
            }
 \end{table}
 \begin{figure*}
  \centering
  \resizebox{\hsize}{!}{\includegraphics{
                        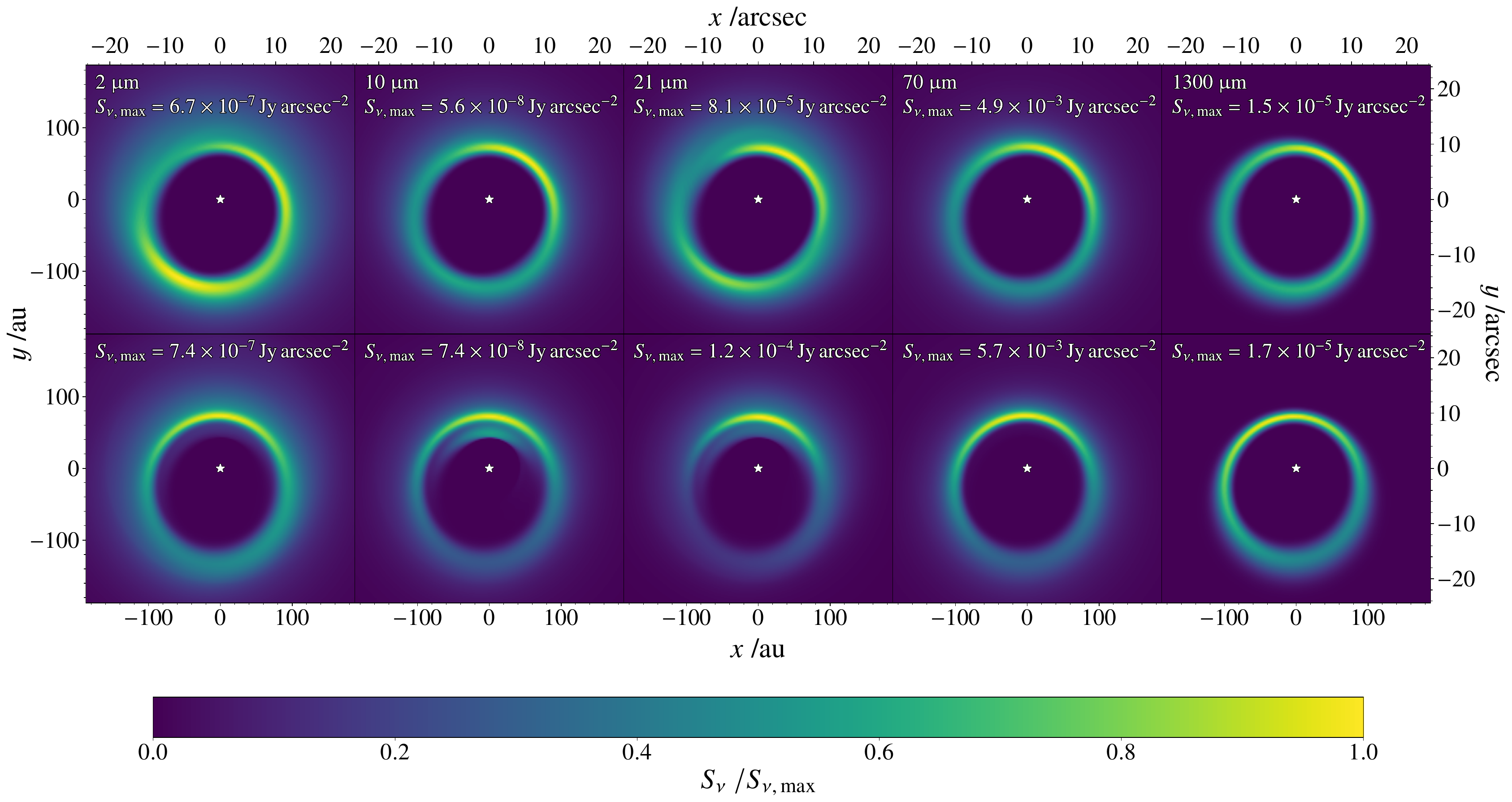}}
  \caption{Surface brightness distributions of the system with parent belt
           \run{n} and perturber \run{i-M2} orbiting inside the belt (\run{n-i-M2};
           \textit{top} row) and the system with belt \run{n} and perturber \run{o-M3}
           orbiting outside the belt (\run{n-o-M3}; \textit{bottom} row) for a
           stellar distance of \SI{7.7}{pc} at five
           wavelengths: \SI{2}{\upmu\m}, \SI{10}{\upmu\m}, \SI{21}{\upmu\m},
           \SI{70}{\upmu\m}, and \SI{1300}{\upmu\m} (from left to right), zoomed in
           on the central $\sim \SI{190}{au}$. Each distribution has been normalized
           by its respective maximum value, $S_{\nu, \, \mathrm{max}}$.
           The white asterisk denotes the position of the central star and
           defines the center of the coordinate system.
           }
  \label{fig_images_all_waves_inner_outer}
 \end{figure*}
 \begin{figure}
  \resizebox{\hsize}{!}{
            \includegraphics{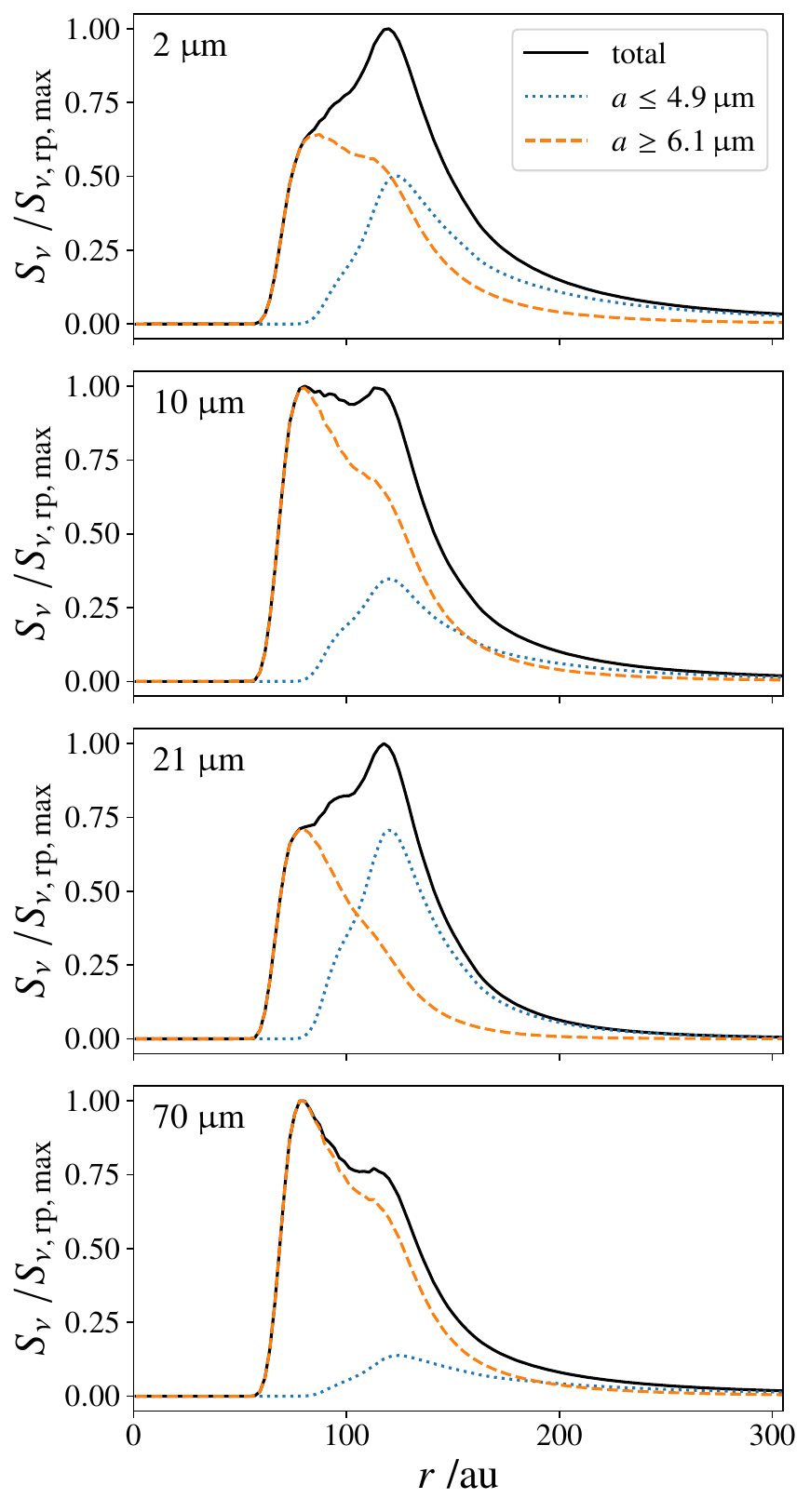}}
  \caption{Radial brightness profiles normalized by their
           respective maximum value, $S_{\nu, \, \mathrm{rp}, \, \mathrm{max}}$,
           at the wavelengths \SI{2}{\upmu\m}, \SI{10}{\upmu\m},
           \SI{21}{\upmu\m}, and \SI{70}{\upmu\m} of the system with the reference
           planetesimal belt \run{n} and the \run{i-M2} inner perturber (\run{n-i-M2}). The dashed
           orange line shows the relative contribution of radiation from
           grains with radii $s \leq \SI{4.9}{\upmu\m}$, the dotted blue line
           of grains with radii $s \geq \SI{6.1}{\upmu\m}$, and the solid black
           line shows the total emission as the sum of the two fractions.
           The radial profiles of each wavelength are normalized to the maximum
           value of the respective total brightness profile.
           }
  \label{fig_radial_profiles}
 \end{figure}
 To compute surface brightness maps, we extended the numerical tool \textbf{D}ebris
 disks around \textbf{M}ain-sequence \textbf{S}tars
 \citep[\DMS;][]{10.1051/0004-6361/201833061} by an interface to process
 the spatial grain distributions computed by \ACE. In \DMS, an
 optically thin dust distribution is assumed to compute maps of thermal
 emission and scattered stellar light. Host star and optical grain properties are modeled
 likewise as described in Sect.~\ref{subsect_ace_params}.

 To establish a vertical structure from the two-dimensional spatial grain
 distributions, we assume a wedge-like structure of the debris disk with a
 half-opening angle of \SI{5}{\degree} and a constant vertical grain number
 density distribution. Furthermore, we considered disks in face-on orientation.
 Therefore, the possible scattering angles
 are all close to \SI{90}{\degree} where the Müller-matrix element $S_{11}$,
 describing the scattering distribution function, shows a smooth behavior.
 Nonetheless, the perfect homogeneous spheres as they are assumed in Mie theory
 are only a coarse approximation, especially for large grains
 \citep[e.g.,][]{10.1051/0004-6361/200913065}.

 With \ACE\ we simulated the evolution of bodies with radii up to
 \SI{481}{\m}. However, grains with sizes much larger than the observing
 wavelength hardly contribute to the total flux. For this reason, we
 neglected large grains in the computation of brightness distributions by applying
 the following criterion: If grains of the next larger size bin would increase
 the total flux of the reference system with parent belt \run{n} and inner
 perturber \run{i-M2} (\run{n-i-M2}) at an observing wavelength of
 \SI{1300}{\upmu\metre} by less than
 \SI{1}{\percent}, neglect these and even larger grains. Using this criterion,
 we included grains with radii of up to \SI{5.65}{\cm}. For the system \run{n-i-M2},
 the dust mass up to that grain size is $\approx \SI{1.1e-8}{M_\odot}$; up to the
 grain size of $\approx \SI{1}{\mm}$ it is $\approx \SI{2.7e-9}{M_\odot}$.
 The latter value fits well into the range of dust masses of cold debris
 disks around A-type stars derived by \citet{10.1088/0004-637X/776/2/111,
 10.3847/0004-637X/831/1/97} who found values between $\approx \num{5.5e-10}$
 and $\SI{4.8e-8}{M_\odot}$. The systems with the reference belt \run{n}
 combined with other perturbers posses slightly different dust masses due to a
 different amount of grain removal and period of simulation $t\sbs{full}$; the
 deviations compared to system \run{n-i-M2} are smaller than $\pm \SI{10}{\percent}$.

 We chose five observing wavelengths motivated by real astronomical instruments
 that allow us to trace grains of sizes ranging from several micrometers to
 several hundreds of micrometers. Neglecting the contribution of the central
 star, in the $K$ band (\SI{2}{\upmu\m}) the flux is entirely dominated by the
 stellar light that has been scattered off small dust grains.
 Likewise, this is the case in the $N$ band (\SI{10}{\upmu\m}) but
 with a small contribution ($\lesssim \SI{10}{\percent}$)
 of thermal emission of small dust grains. In the $Q$ band (\SI{21}{\upmu\m}) and
 at a wavelength of \SI{70}{\upmu\m} we trace the thermal emission of small halo
 grains, while at a wavelength of \SI{1300}{\upmu\m} we trace the thermal
 emission of the largest and coldest grains, which indicate the
 position of the parent belt. At wavelengths of \SI{2}{\upmu\m},
 \SI{10}{\upmu\m}, and \SI{21}{\upmu\m} we will see major improvements regarding
 angular resolving power and sensitivity of imaging instruments due to
 the Multi-Adaptive Optics Imaging Camera for Deep Observations
 \citep[MICADO,][]{10.1117/12.2311483, 10.18727/0722-6691/5217} and
 the Mid-Infrared ELT Imager and Spectrograph
 \citep[METIS,][]{10.1117/12.2056468, 10.18727/0722-6691/5218} at the
 Extremely Large Telescope (ELT)$\,$\footnote{\href{https://elt.eso.org}{https://elt.eso.org}}
 as well as the Near-Infrared Camera \citep[NIRCam,][]{10.1117/12.552281} and
 the Mid-Infrared Instrument \citep[MIRI,][]{10.1117/12.551717} on the
 \textit{James Webb} Space Telescope \citep[JWST,][]{10.1007/s11214-006-8315-7}.
 In Table \ref{table_wavelengths},
 the selected observing wavelengths, the type of radiation dominating the flux
 at those wavelengths, and corresponding exemplary instruments with high
 angular resolving power are listed.

 In Fig.~\ref{fig_images_all_waves_inner_outer}, the central
 $\sim \SI{190}{au}$ of the brightness distributions at all five wavelengths for the
 system \run{n-i-M2} (\textit{top} row) and the system with the reference parent
 belt \run{n} and the \run{o-M3} perturber orbiting outside thereof (\run{n-o-M3};
 \textit{bottom} row) are displayed for illustration.
 These particular systems have been chosen because the perturbation timescales and
 amplitudes of their belt centers are equal (see Sect.~\ref{subsect_ace_outer}).
 Thus, from any possible pair of systems with an inner and an outer perturber, the
 effect on the spatial dust distribution by the parent belt being eccentric is
 most similar. The differences are apparent due to the perturber orbiting inside
 (\run{n-i-M2}) or outside (\run{n-o-M3}) the parent belt.
 For illustrative reason, the
 brightness distributions were normalized to the maximum flux density
 $S_{\nu, \, \mathrm{max}}$ of each map at the respective wavelength. The
 un-normalized distributions are shown in the Appendix (see
 Fig.~\ref{fig_all_waves_unnormalized}).

 All systems with an inner perturber show comparable brightness
 distributions. Taking the system \run{n-i-M2} as an example, we find the
 brightness distributions to be similar at the
 wavelength pairs $\left\{ \SI{2}{\upmu\m}, \SI{21}{\upmu\m} \right\}$ and
 $\left\{ \SI{10}{\upmu\m}, \SI{70}{\upmu\m} \right\}$. At the former two
 wavelengths the ring-like structure is broader and the relative contribution
 from outer regions $\gtrsim \SI{100}{au}$ is stronger than in the
 latter two. The map at \SI{1300}{\upmu\m} differs from the maps at
 \SI{10}{\upmu\m} and \SI{70}{\upmu\m} only by the dimmer emission in the outer
 regions. This findings can be explained by the contribution of grains near the
 blowout limit of $s\sbs{bo} \approx \SI{4}{\upmu\m}$. For illustration, azimuthally
 averaged radial brightness profiles for the system \run{n-i-M2} (see the upper row in
 Fig.~\ref{fig_images_all_waves_inner_outer}) as well as the relative
 contribution of two different grain size groups are shown in
 Fig.~\ref{fig_radial_profiles}. While the distributions of grains near the
 blowout limit can be oriented opposite to the parent
 belt, the distribution of larger grains share the orientation of the belt
 (see Fig.~\ref{fig_ace_sizes}a in contrast to \ref{fig_ace_sizes}b). At
 the wavelength pair $\left\{ \SI{2}{\upmu\m}, \SI{21}{\upmu\m} \right\}$ the
 radiation of the smallest grains makes up a large part of the total emission at
 large separations from the host star $\gtrsim \SI{100}{au}$ while in the
 wavelength pair $\left\{ \SI{10}{\upmu\m}, \SI{70}{\upmu\m} \right\}$ this is not
 the case. The smallest grains scatter the stellar radiation very efficiently at
 \SI{2}{\upmu\m} while their relative contribution drops toward larger wavelengths. At a
 wavelength of \SI{10}{\upmu\m}, scattering by larger grains dominates the net
 flux. At a wavelength of \SI{21}{\upmu\m}, which is dominated by thermal dust
 emission, the smallest grains are the most efficient emitters due to their
 highest temperature. The relative contribution of the smallest grains becomes
 less important at \SI{70}{\upmu\m} and even negligible at \SI{1300}{\upmu\m}.
 At the latter wavelength, only the emission of the largest grains close to the
 parent belt is visible.

 For the systems with an outer perturber and taking the system \run{n-o-M3}
 as an example, we see that the overall appearance at different wavelengths
 is very similar to the one at \SI{1300}{\upmu\m}, while toward shorter
 wavelengths the relative contribution from outer regions
 $\gtrsim \SI{100}{au}$ increases due to the growing contribution of smaller
 grains on extended orbits. This behavior is constant for the other systems with
 different perturber parameters. Furthermore, we see emission from regions
 inward of the belt location, most pronounced at \SI{10}{\upmu\m} and less pronounced
 at \SI{2}{\upmu\m}, \SI{21}{\upmu\m} and \SI{70}{\upmu\m}. This flux stems from
 small grains that are forced by the perturber on orbits leading into that inner
 disk region. The location and intensity of those features in the brightness
 distributions are individual for each system and depend on parameters of the
 perturber (see also Sect.~\ref{subsect_grains_inward_outer}).

%%%%%%%%%%%%%%%%%%%%%%%%%%%%%%%%%%%%%%%%%%%%%%%%%%%%%%%%%%%%%%%%%%%%%%%%%%%%%%%

\subsection{Twisting of the small grain halo}\label{subsect_halo_twisting}
 To analyze the small grain halo, we fitted ellipses to lines of constant
 brightness (isophotes) radially outside the parent belt. First, we produced
 polar grids of the Cartesian flux maps using the Python package
 CartToPolarDetector$\,$\footnote{\href{https://github.com/anton-krieger/CartToPolarDetector}
{https://github.com/anton-krieger/CartToPolarDetector}}
 \citep{2022A&A...662A..99K}: after superposing the Cartesian grid
 with the new polar one, the polar grid cells were computed by summing up the
 values of the intersecting Cartesian cells, each weighted with its relative
 intersecting area with the new polar cell\footnote{We used units of flux
 density per grid cell area, that is, \si{Jy\,arcsec^{-2}}. Thus,
 the quantities are independent of the grid cell area, which varies drastically
 in a polar grid.}.
 The new polar coordinates are the distance to the center, $r$, and the azimuthal
 angle, $\theta$, which is defined such that \SI{0}{\degree} is oriented in the
 direction of the horizontal axis toward the west (i.e., right) side of the map
 and the angle is increasing counterclockwise.

 Second, we determined the polar coordinates of selected brightness levels. To
 trace the halo around the parent belt, we required all isophotes to be
 radially outward of the parent belt. To achieve this, we first defined
 a reference brightness value for each map representing the mean flux level of the
 bright ring: We determined the radial maxima of brightness for each azimuthal angle
 $\theta$ and derived the reference value as the azimuthal average of the radial
 maxima. Taking this reference value, we acquired a reference isophote.
 Lastly, we computed ten isophotes that are radially outward of the reference isophote
 for all azimuthal angles and with a brightness level of at most \SI{80}{\percent} of
 the reference value. As a lower cutoff, we set \SI{1}{\percent} of the reference
 value.

\begin{figure}[h!]
  \begin{subfigure}[b]{\columnwidth}
   \resizebox{\hsize}{!}{
              \includegraphics[scale=0.5]{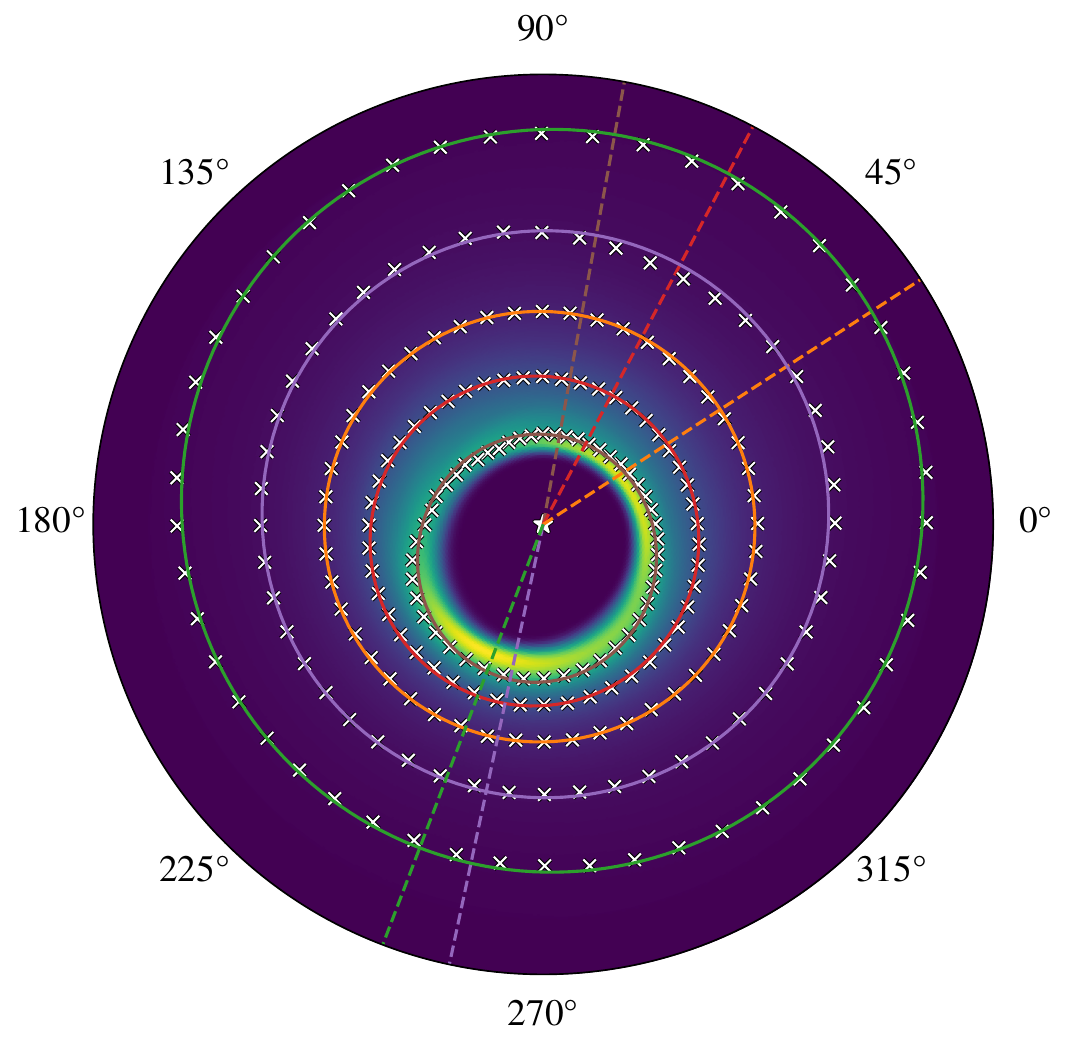}}
   \caption{\run{n-i-M2} at \SI{2}{\upmu\m}.}
   \label{fig_polar_plots_with_isophotes_n-i-M2}
  \end{subfigure}
  \hfill
  \begin{subfigure}[b]{\columnwidth}
   \resizebox{\hsize}{!}{
   \includegraphics[scale=0.5]{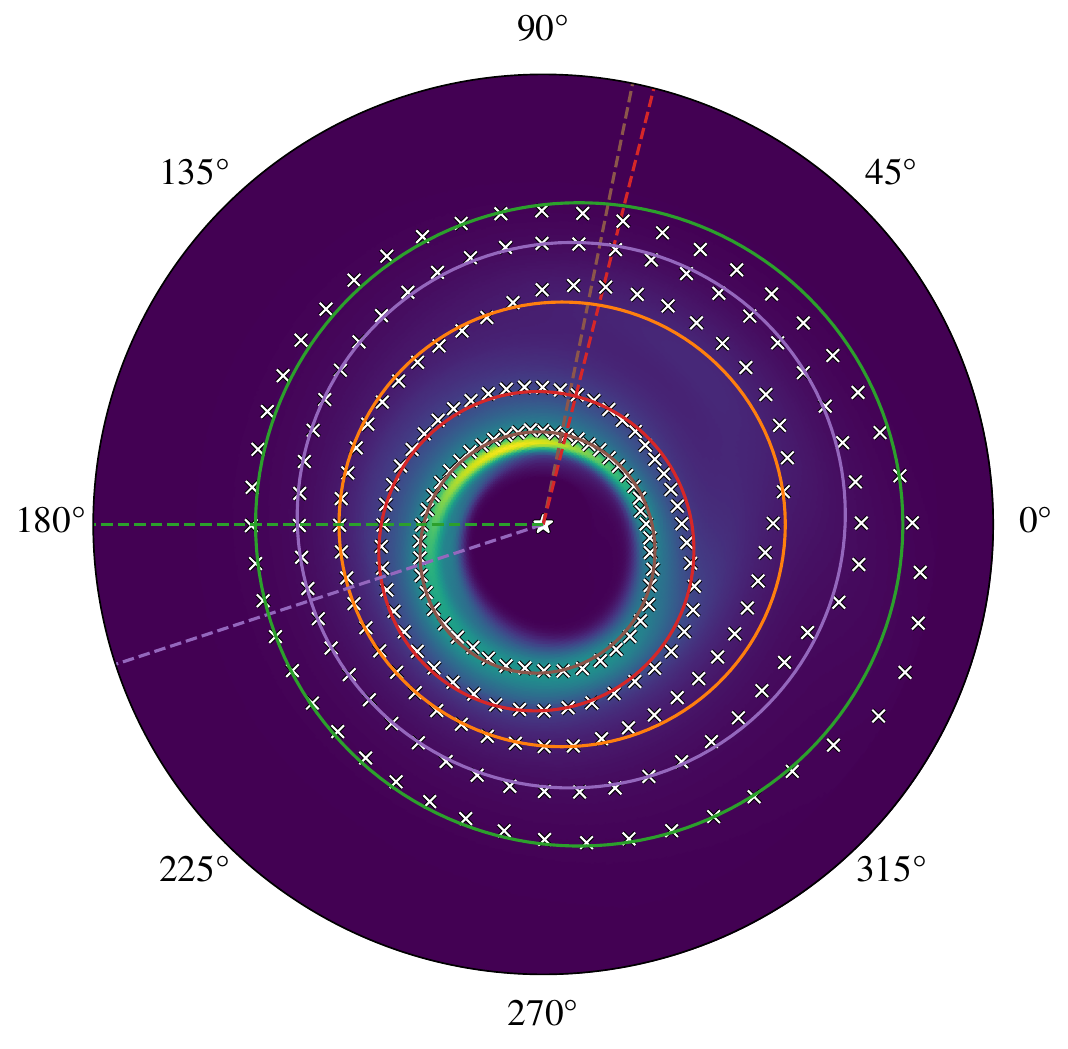}}
   \caption{\run{n-o-M4} at \SI{2}{\upmu\m}.}
   \label{fig_polar_plots_with_isophotes_n-o-M4}
  \end{subfigure}
  \caption{Polar brightness maps of the systems \run{n-i-M2} (same as in
           Fig.~\ref{fig_all_waves_unnormalized_n-i-M2}) and \run{n-o-M4} at a
           wavelength of \SI{2}{\upmu\m} (radius \SI{400}{au}). The white
           asterisk denotes the position of the central star and defines the
           center of the coordinate system. The white crosses trace five lines
           of constant brightness (isophotes) and are drawn every \SI{7.5}{\degree}.
           The colored ellipses have been fitted to the respective isophotes
           using Eq.~\ref{eq:ellipse} with the central star in one of the focal
           points. The dashed lines give the direction from the focal point to
           the periapse, that is, $\phi + \pi$, and are drawn in the same color
           as their corresponding ellipse.}
  \label{fig_polar_plot_with_isophotes}
 \end{figure}

 Assuming the central star to be in one of the focal points and the respective
 focal point to be the coordinate center, we used the ellipse equation
 \begin{equation}
  \label{eq:ellipse}
    r\left(\theta; a, e, \phi  \right) =
    \frac{a \left( 1 - e^2 \right)}
         {1 - e \cos{\left(\theta - \phi \right)}}\, ,
 \end{equation}
 with the azimuthal angle $\theta$ and the parameters semimajor axis $a$,
 eccentricity $e$, and the position angle of the periapse $\phi$, where $\phi$
 corresponds to the longitude of periapse $\varpi$ for the face-on disks
 discussed here$\,$\footnote{We used the routine from
 \href{https://scipython.com/book/chapter-8-scipy/examples/non-linear-fitting-to-an-ellipse/}
 {https://scipython.com/book/chapter-8-scipy/examples/non-linear-fitting-to-an-ellipse/},
 extended by the position angle of the periapse $\phi$.}.
 For illustration, the polar brightness map of the system \run{n-i-M2} with indicated
 brightness levels and fitted isophotes is shown in
 Fig.~\ref{fig_polar_plots_with_isophotes_n-i-M2}. By comparing the orientation of
 isophote ellipses for different semimajor axes, we can analyze whether the
 rotation of the small grain halo with respect to the parent belt, as presented
 in Sect.~\ref{subsect_ace_variations}, can be quantified by analyzing maps of
 surface brightness.

\begin{figure}
  \resizebox{\hsize}{!}{
            \includegraphics[scale=0.6]{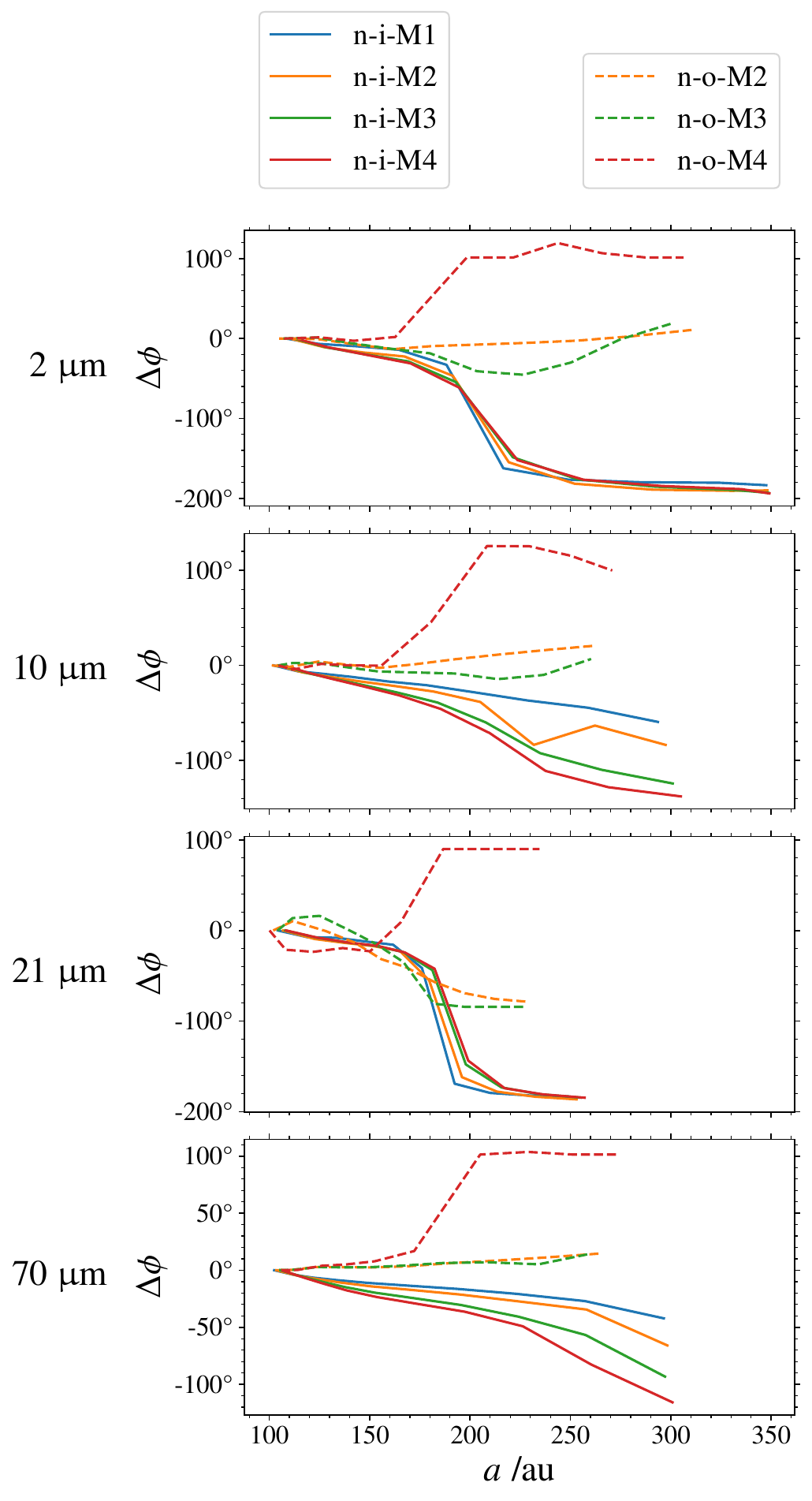}}
  \caption{Orientation, $\Delta \phi$, of the isophote ellipses relative to the
           orientation of the innermost ellipse drawn over semimajor axis $a$.
           A negative value of $\Delta \phi$ means clockwise rotation and a
           positive value counterclockwise rotation.
           Solid lines are used for the systems with inner perturbers and dashed
           lines for those with outer
           perturbers, and rows are used to show four different wavelengths
           tracing small halo grains: \SI{2}{\upmu\m}, \SI{10}{\upmu\m},
           \SI{21}{\upmu\m}, and \SI{70}{\upmu\m}.
           }
  \label{fig_rel_isophote_rotation}
 \end{figure}

 Figure~\ref{fig_rel_isophote_rotation} shows the results of ellipse orientation
 relative to the orientation of the innermost ellipse $\Delta \phi$ for the
 wavelengths of \SI{2}{\upmu\m}, \SI{10}{\upmu\m}, \SI{21}{\upmu\m}, and
 \SI{70}{\upmu\m}. At these wavelengths we trace small grains
 that form an extended halo (see Fig.~\ref{fig_ace_sizes}). Likewise, as
 discussed in Sect.~\ref{subsect_dms}, all results for systems with an inner
 perturber show a similar behavior at a fixed respective wavelengths.
 Furthermore, for a fixed system the maps at wavelength pairs
 $\left\{ \SI{2}{\upmu\m}, \SI{21}{\upmu\m} \right\}$ and
 $\left\{ \SI{10}{\upmu\m}, \SI{70}{\upmu\m} \right\}$ show similar trends,
 respectively.

 At the wavelength pair $\left\{ \SI{2}{\upmu\m}, \SI{21}{\upmu\m} \right\}$,
 we see that isophotes with small semimajor axes $a$ show a small, but monotonous
 change in orientation with
 increasing values of $a$. This is caused by the ongoing differential precession
 \citep[see Sect.~\ref{subsect_ace_variations} and][]{10.1051/0004-6361/201935199}.
 Grains populating regions with increasing distance from the center, marked by
 isophotes with larger semimajor axes, are increasingly older and therefore trail
 the farther precessing parent belt, marked by isophotes with smaller semimajor
 axes, more. Then, this trend is followed by a flip of
 $\Delta \phi \sim \SI{-180}{\degree}$ over a small interval of $a$.
 The flip appears, because at these two wavelengths and large semimajor
 axes the brightness contribution of the smallest, barely bound grains near the
 blowout limit of $s\sbs{bo} \approx \SI{4}{\upmu\m}$ start to dominate (see
 Fig.~\ref{fig_radial_profiles}). These grains form a halo
 inversely oriented to the parent belt (see Fig.~\ref{fig_ace_sizes}a
 and Sect.~\ref{subsect_dms}).
 At \SI{2}{\upmu\m} the flip starts at $a \sim \SI{170}{au}$ and ends
 at $a \sim \SI{250}{au}$. At \SI{21}{\upmu\m} it
 starts at semimajor axes of
 $a \sim \SI{170}{au}$ but is completed sooner at
 $a \sim \SI{210}{au}$ -- $\SI{220}{au}$. The different location of the flip between
 the two observing wavelengths is caused by the varying fraction of the contribution
 of the smallest grains to the total brightness at different separations from the star: at a
 wavelength of \SI{21}{\upmu\m}, the emission of the small grains dominates
 over those of larger grains already at smaller separations from the central
 star than at \SI{2}{\upmu\m} (compare the first and third panel of
 Fig.~\ref{fig_radial_profiles}). Lastly,
 for further increasing semimajor axis $a$ the flip is followed by decreasing
 values of $\Delta \phi$. This is again caused by the ongoing differential
 precession.

 At the wavelength pair $\left\{ \SI{10}{\upmu\m}, \SI{70}{\upmu\m} \right\}$ we do
 not find this behavior. Instead, because the brightness contribution of the smallest
 grains causing
 the flip is much weaker at these wavelengths, we find a mostly monotonous
 decrease in isophote orientation $\Delta \phi$ with increasing semimajor axis
 $a$, explained by differential precession.

 At the wavelength pair $\left\{ \SI{10}{\upmu\m}, \SI{70}{\upmu\m} \right\}$ and
 for separations from the host star $\lesssim \SI{200}{au}$ also at
 \SI{2}{\upmu\m}, variations in the relative isophote orientation
 $\Delta \phi$ are correlated with the perturber mass: the lowest mass
 perturber \run{i-M1} causes the smallest, the highest mass perturber \run{i-M4}
 the largest values of $\Delta \phi$. This is related to the angular velocity of the
 differential precession. The higher the perturber mass, the faster the
 parent belt precesses, causing the small grain halo to lag behind more.

 At the wavelength pair $\left\{ \SI{2}{\upmu\m}, \SI{21}{\upmu\m} \right\}$ we see
 exactly the opposite separation by perturber mass: In the system with the \run{i-M1} perturber
 the isophote orientation flips at the smallest semimajor axes and within the
 smallest axes interval, while in the system with the \run{i-M4} perturber it does so at
 the largest semimajor axes and within the broadest axes interval. This behavior
 is directly related to the semimajor axes distribution of the grain orbits.
 For a certain grain size, the more massive the perturber is, the larger is the
 inner gap of the entire debris disk and the halo of small grains. As the
 location where the flip occurs is
 determined by the radial distance where the brightness contribution of grains near
 the blowout limit starts to
 dominate, a system with a more massive perturber shows the flip
 in isophote orientation $\Delta \phi$ at larger values of the semimajor axis $a$
 than a system with a less massive perturber.

 Contrary to the analysis of systems with an inner perturber, no clear trend is
 found for systems with an outer perturber. For the
 systems \run{n-o-M2} and \run{n-o-M3} we find isophotes strongly different
 than those of \run{n-o-M4}. The first two systems show only small isophote rotations
 over all semimajor axes $a$ in the maps at the wavelengths \SI{2}{\upmu\m},
 \SI{10}{\upmu\m}, and \SI{70}{\upmu\m}. At \SI{21}{\upmu\m} \run{n-o-M2},
 \run{n-o-M3} show a clockwise isophote rotation of up to
 $\Delta \phi \sim \SI{10}{\degree}, \ \SI{20}{\degree}$ for small semimajor
 axes that turns into a counterclockwise rotation of up to
 $\Delta \phi \sim \SI{-80}{\degree}$ for semimajor axes of
 $a \gtrsim \SI{130}{au}, \ \SI{140}{au}$, respectively. The isophote rotation
 of the \run{n-o-M4} perturber maps show a consistent behavior at all
 wavelengths from \SI{2}{\upmu\m} to \SI{70}{\upmu\m}: for small semimajor axes
 we see only small values of
 $\Delta \phi$. At semimajor axes of $a \sim \SI{170}{au}$ -- \SI{190}{au}
 the isophotes rotate clockwise up to $\Delta \phi \sim \SI{100}{\degree}$ --
 \SI{120}{\degree}. The nonuniform behavior of the isophote orientation
 $\Delta \phi$ derived from the maps of systems with an outer perturber results
 from the fact that each outer perturber system shows highly individual spatial
 distributions for small grains. Furthermore, while for the systems with an
 inner perturber the spatial dust distributions change only slowly with grain
 size (see Fig.~\ref{fig_ace_sizes}b and c), for the systems with outer
 perturbers they vary strongly in shape and density with grain size. As a
 consequence, the isophotes deviate more from an elliptical shape and do not
 possess a common orientation. To illustrate this behavior, the polar brightness map
 of the system \run{n-o-M4} with indicated flux levels and fitted isophotes
 thereto is shown in Fig.~\ref{fig_polar_plots_with_isophotes_n-o-M4}.

 The small grain halo is a rather faint part of a debris disk system. Within
 our framework, most promising to detect and constrain it are observations at a
 wavelength of \SI{21}{\upmu\m} using JWST/MIRI. With an exposure time
 of \SI{1}{\hour}, the differential rotation manifested in the monotonous
 change of isophote orientation for small semimajor axes up to
 $a \sim \SI{140}{au}$ -- \SI{160}{au} can be detected. With an exposure time
 of \SI{5}{\hour}, surface brightness levels corresponding to isophotes with
 semimajor axes of $a \sim \SI{190}{au}$ -- \SI{220}{au} can be detected,
 covering almost the entire flip of isophote rotation. Furthermore, observing
 with the Photodetector Array Camera and Spectrometer
 \citep[PACS,][]{10.1051/0004-6361/201014535} on the \textit{Herschel} Space
 Observatory \citep{10.1051/0004-6361/201014759} at the wavelength \SI{70}{\upmu\m} for an
 exposure time of \SI{1}{\hour} would have been sufficient to detect parts of the halo
 corresponding to isophotes with semimajor axes of up to
 $a \sim \SI{220}{au}$ -- \SI{240}{au}. More details on the observational
 analysis are provided in Sects.~\ref{appdx_general observability} and
 \ref{appdx_subsect_halo_twisting}.

%%%%%%%%%%%%%%%%%%%%%%%%%%%%%%%%%%%%%%%%%%%%%%%%%%%%%%%%%%%%%%%%%%%%%%%%%%%%%%%

\section{Feasibility of distinguishing between the existence of an inner or outer
         perturber}
\label{sect_inner_vs_outer}
 Motivated by the notably different distributions of small grains in systems
 with an inner and outer perturber, respectively
 (Sect.~\ref{subsect_ace_outer}), and their corresponding brightness
 distributions (Sect.~\ref{subsect_dms}), we investigated the feasibility of
 distinguishing between those two types observationally. We identify several
 disk properties and features that can be used to accomplish this task
 (Sects.~\ref{subsect_spiral} -- \ref{subsect_azimuthal_maxima}). The
 presentation of each feature is paired with an explanation of its origin by
 discussing the underlying spatial dust distributions and an estimation of
 the feasibility of observing it based on the
 analysis in Sect.~\ref{appdx_general observability}. While restricting the
 detailed discussion on systems with belts simulated with the
 reference parameter set \run{n}, we comment on other systems simulated with the
 other sets \run{w}, \run{m2}, and \run{m3} in Sect.~\ref{subsect_other_belts}.
 In Sect.~\ref{subsect_flowchart} we summarize the analyses to a simple
 flowchart, allowing one to decide whether a debris disk harbors a single inner
 or outer perturbing planet.

%%%%%%%%%%%%%%%%%%%%%%%%%%%%%%%%%%%%%%%%%%%%%%%%%%%%%%%%%%%%%%%%%%%%%%%%%%%%%%%

\subsection{Spiral structure in the $Q$ band}
\label{subsect_spiral}
\begin{figure}
  \resizebox{\hsize}{!}{
            \includegraphics{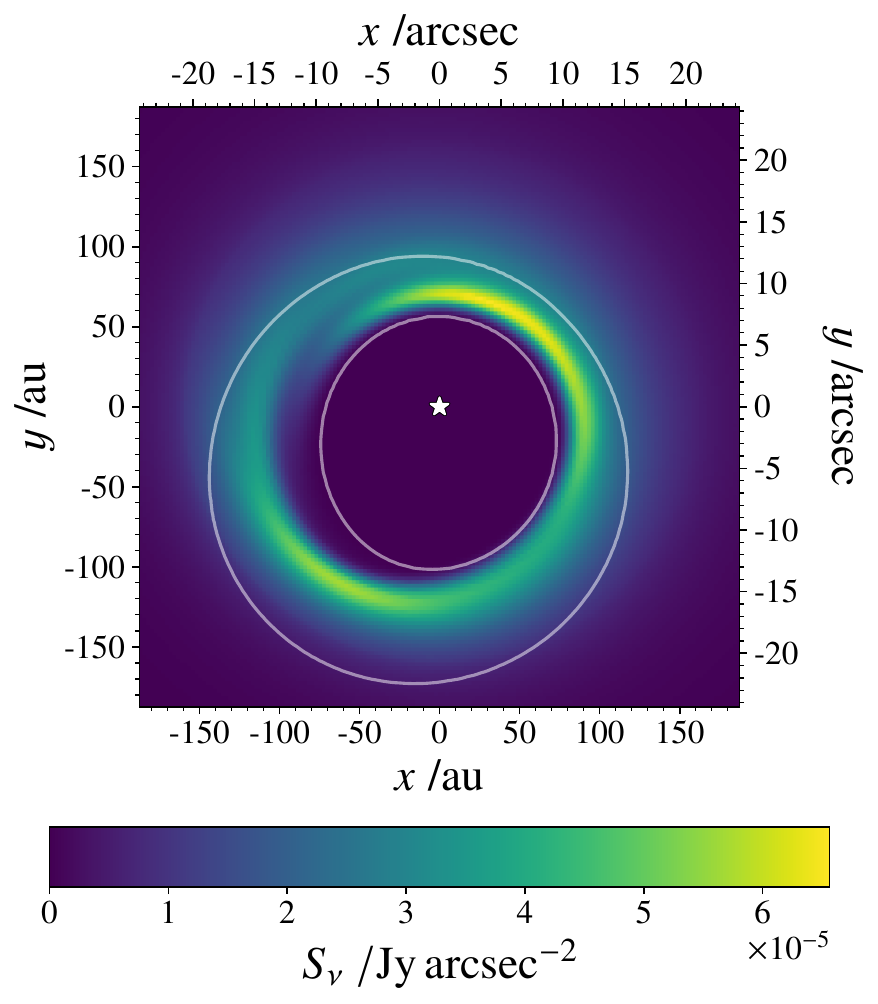}}
  \caption{Surface brightness distribution for a stellar distance of
           \SI{7.7}{pc} at the wavelength \SI{21}{\upmu\m}
           ($Q$ band) of the system with belt
           parameter set \run{n} and the inner perturber \run{i-M4} (\run{n-i-M4}).
           The contour lines denote the isophotes of the corresponding
           \SI{1300}{\upmu\m} map of the same system with a flux density level of
           \SI{1}{\percent} of its maximum value. They enclose the position of the
           parent planetesimal belt. General figure characteristics are the same as in
           Fig.~\ref{fig_images_all_waves_inner_outer}.
           }
  \label{fig_spiral_inner_21micron}
 \end{figure}

 We find that a spiral structure is solely produced by inner perturbers,
 best visible in the $Q$ band at \SI{21}{\upmu\m} (for an illustration, see
 Fig.~\ref{fig_spiral_inner_21micron}). The location of the structure is
 completely consistent with the location of the parent belt (contrary to the
 systems with an outer perturber as discussed in
 Sect.~\ref{subsect_grains_inward_outer}), which we trace by isophotes of
 \SI{1}{\percent} of the maximum brightness in the corresponding \SI{1300}{\upmu\m}
 map.

 The spiral structure originates in the superposed brightness contribution of several grain
 sizes with different spatial distributions. We exemplify this in detail using
 Fig. \ref{fig_spiral_inner_21micron}: At the northeast of the map$\,$\footnote{Using the
 canonical celestial directions for images on the sky and hence north is up and
 west is to the right.} and moving clockwise, the brightness distribution
 forming the inner spiral arm is dominated by the thermal radiation of grains
 of size $s \gtrsim \SI{240}{\upmu\m}$ until the southwest of the map,
 while the brightness distribution forming the remaining spiral arm in
 clockwise direction is dominated by the radiation of grains
 $s \sim \SI{5}{\upmu\m}$ in size (see Fig.~\ref{fig_ace_sizes}a and
 Fig.~\ref{fig_ace_sizes}d for similar distributions of the system \run{n-i-M2}).
 In the brightness distributions of the systems with the lower mass perturbers
 (\run{i-M1} to \run{i-M3}) we find the same spiral structure. However, its
 contrast is lower in these maps but decreases not monotonically
 with the perturber mass. This is because the respective systems show
 similar -- but slightly shifted -- spatial dust distributions
 compared to \run{n-i-M4}. Their distributions overlap more strongly, and
 therefore, the spiral structures in these maps are less rich in contrast. This
 example illustrates that the specific contrast of the spiral structure depends
 on the individual system. The systems with an outer perturber do not show that
 spiral patterns because the grains of the respective sizes have different
 spatial distributions.

 We see strong emission from the
 $s \sim \SI{5}{\upmu\m}$ sized grains at a wavelength of \SI{2}{\upmu\m} (see
 Sect.~\ref{subsect_dms}) as well. However, at that wavelength other similarly
 sized grains with slightly different spatial distributions contribute
 considerably to the net flux, effectively smearing out the spiral structure.
 In the brightness distributions at the wavelengths \SI{10}{\upmu\m} and
 \SI{70}{\upmu\m}, which are not sensitive to emission of $s \sim \SI{5}{\upmu\m}$
 sized grains (see Sect.~\ref{subsect_dms}), the spiral structure is not visible
 at all. While at \SI{2}{\upmu\m} the contrast of the structure is not
 sufficient, and at \SI{10}{\upmu\m} and \SI{70}{\upmu\m} it does not appear, a
 wavelength of \SI{21}{\upmu\m} is suitable to observe such structures with JWST/MIRI:
 With an exposure time of \SI{8}{\hour} it is possible to achieve the required
 contrast to detect the spiral pattern for the system \run{n-i-M4} (see
 Sects.~\ref{appdx_general observability} and \ref{appdx_subsect_spiral}).

\subsection{Grains scattered inward of the parent planetesimal belt}
\label{subsect_grains_inward_outer}

\begin{figure}
  \resizebox{\hsize}{!}
            {\includegraphics{
            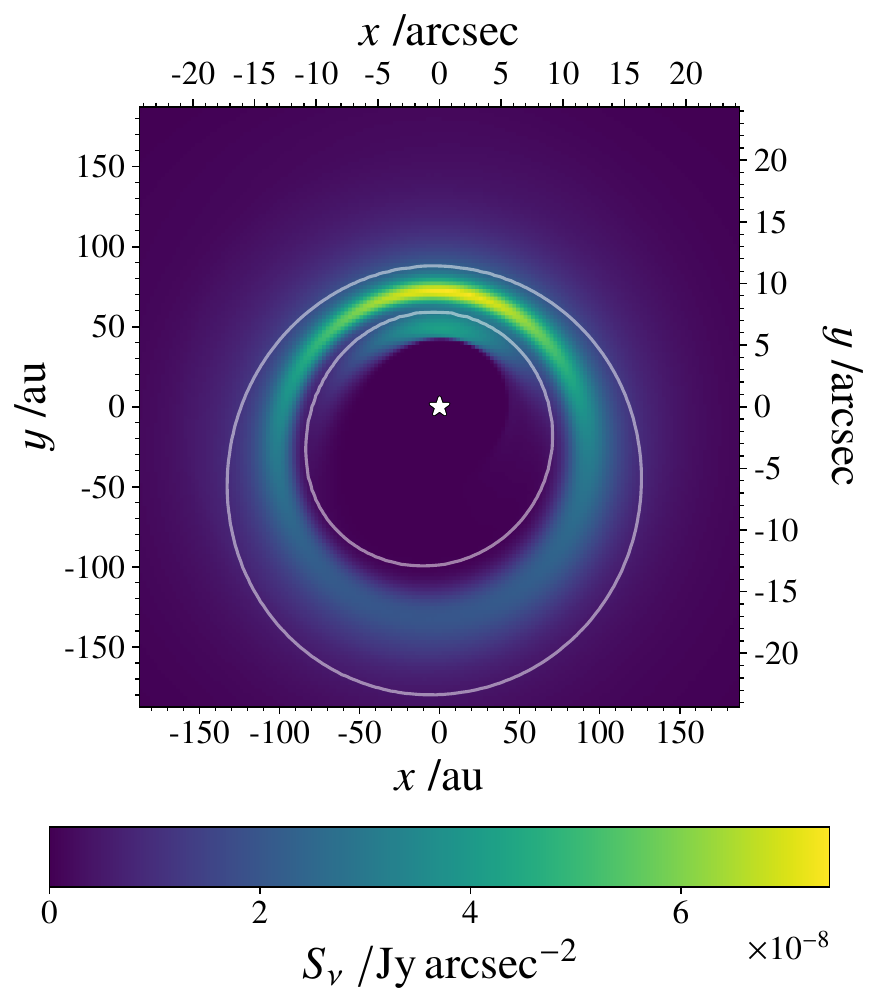}}
  \caption{Surface brightness distribution for a stellar distance of
           \SI{7.7}{pc} at the wavelength \SI{10}{\upmu\m}
           ($N$ band) of the system with
           belt parameter set \run{n} and the outer perturber \run{o-M3} (\run{n-o-M3}).
           General figure characteristics are the same as in
           Fig.~\ref{fig_images_all_waves_inner_outer}, and white contour lines
           have the same meaning as
           in Fig.~\ref{fig_spiral_inner_21micron}.
           }
  \label{fig_grains_inward_outer_10micron}
 \end{figure}

 To analyze the disk regions in the maps inward of the parent belt, we defined this inner
 disk region to be inward of the isophote with a brightness value of \SI{1}{\%} of the
 maximum brightness in the corresponding \SI{1300}{\upmu\m} map. For the systems with an
 inner perturber, we find that the brightness decreases quickly inward of the parent belt
 to levels several orders of magnitude lower than the average, regardless of
 wavelength (see the top row in Fig.~\ref{fig_images_all_waves_inner_outer} and
 Fig.~\ref{fig_spiral_inner_21micron}). However, this is not the case for the
 systems with an outer perturber. At the wavelength of \SI{2}{\upmu\m},
 \SI{10}{\upmu\m}, \SI{21}{\upmu\m}, and weakly at \SI{70}{\upmu\m}, there is an
 considerable amount of light coming from the inner disk region. Depending on
 observing wavelength and perturber parameters it reaches levels of up to several
 tens of the maximum brightness values (for an illustration, see
 Fig.~\ref{fig_grains_inward_outer_10micron}). We note that the sharp cut-off at a
 distance of \SI{40}{au} is an artifact caused by the inner edge of the
 pericenter grid in the \ACE\ simulations. At \SI{10}{\upmu\m}, while the brightness
 from the belt location and outer regions is dominated by scattered stellar
 light, the brightness from those inner regions is dominated by thermal emission.

 This flux originates from grains of the size $s \sim \num{3}$ --
 $\SI{12}{\upmu\m}$. The outer perturber orbits near the small grain halo and
 forces grains on orbits leading inside the parent belt. Although the exact spatial
 grain distributions and thus the characteristics of the brightness distribution
 inside the parent belt strongly depend on the exact perturber parameters, we
 find brightness levels up to several fractions of the maximum brightness at the region
 inward of the parent belt for all investigated systems with an outer perturber but
 for none of those with an inner perturber. We note that the smallest bound grains
 populate orbits that cross or nearly cross the orbit of the outer perturber.
 Hence, the model accuracy is lowest for these grains and the scope of a
 quantitative analysis limited.

 However, the lack of such emission from regions inside the parent belt is not a
 viable criterion for the presence of an inner perturber because in our model
 setup the inner regions are clear from emission regardless of the presence of
 an inner perturber.

 While at the wavelength of \SI{10}{\upmu\m} exposure
 times of $\gtrsim \SI{10}{\hour}$ are required when observing with
 JWST/MIRI, at the wavelength of \SI{21}{\upmu\m} an exposure time of
 \SI{1}{\hour} is sufficient to detect the emission from inside the parent belt
 with the aforementioned caveat of limited accuracy for grains that orbit close
 to the perturber (see Sects.~\ref{appdx_general observability} and
 \ref{appdx_subsect_grains_inward_outer}).

%%%%%%%%%%%%%%%%%%%%%%%%%%%%%%%%%%%%%%%%%%%%%%%%%%%%%%%%%%%%%%%%%%%%%%%%%%%%%%%

\subsection{Number of azimuthal radial flux maxima}
\label{subsect_azimuthal_maxima}

\begin{figure}
  \resizebox{\hsize}{!}{
            \includegraphics{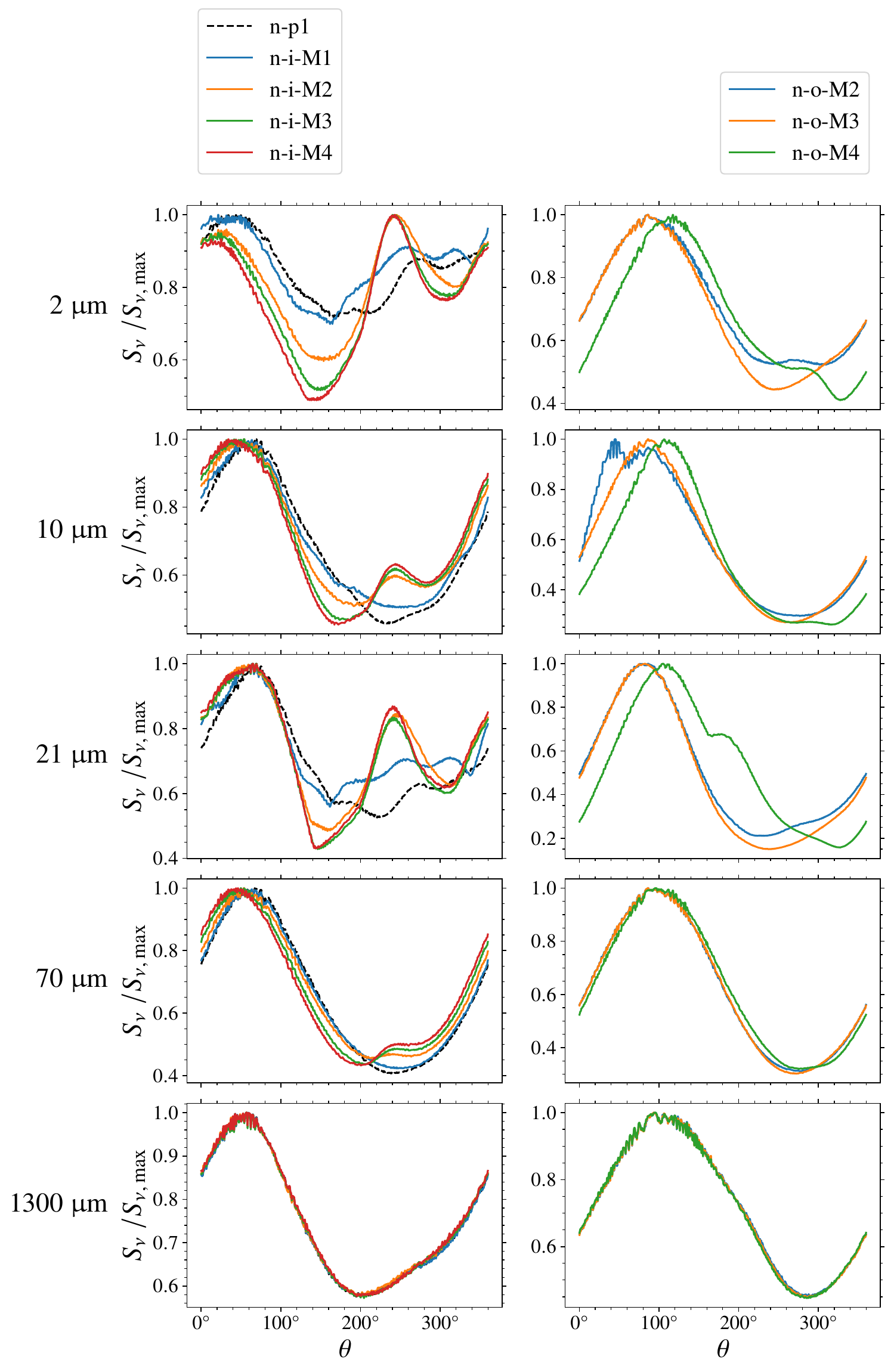}}
  \caption{Azimuthal distribution of radial maxima of brightness normalized by their
           respective maximum value, $S_{\nu, \, \mathrm{max}}$, in the five
           investigated wavelengths: \SI{2}{\upmu\m}, \SI{10}{\upmu\m}, \SI{21}{\upmu\m},
           \SI{70}{\upmu\m}, and \SI{1300}{\upmu\m}. \textit{Left}: Systems with
           an inner perturber. System \run{n-p1} is displayed as a reference
           that shows the case without ongoing precession. \textit{Right}:
           Systems with an outer perturber.
           }
  \label{fig_azimuthal_maxima}
 \end{figure}

 As mentioned in Sect.~\ref{subsect_ace_outer}, the system \run{n-o-M3}
 with an outer perturber shows one azimuthal maximum in the distribution of
 small halo grains while the system \run{n-i-M2} with an inner perturber shows
 two (for an illustration, see Fig.~\ref{fig_ace_degen_small}d and
 \ref{fig_ace_degen_small}a). Likewise, we find this to be the case for the
 systems with perturbers of different masses as well. In the following we
 investigate whether this feature appears in the brightness distributions
 by analyzing the bright ellipsoidal structure.
 In Fig.~\ref{fig_azimuthal_maxima}, the radial maxima, as computed in
 Sect.~\ref{subsect_halo_twisting}, over the azimuthal angle $\theta$
 are displayed, separated for systems with an inner and with an outer perturber.

 First, we discuss the systems with an inner perturber: At a wavelength of
 \SI{1300}{\upmu\m}, we find one global maximum and one minimum at angles of
 $\theta \sim \SI{60}{\degree}$ and \SI{200}{\degree}, respectively.
 At the wavelengths
 \SI{2}{\upmu\m}, \SI{10}{\upmu\m}, and \SI{21}{\upmu\m}, the first maximum at around
 $\theta \sim \SI{20}{\degree}$ -- \SI{60}{\degree} is congruent with the
 maximum at \SI{1300}{\upmu\m}, the second is located at $\theta \sim \SI{240}{\degree}$.
 While the second maximum is very pronounced at \SI{2}{\upmu\m} and \SI{21}{\upmu\m},
 at \SI{10}{\upmu\m} it appears less pronounced. At \SI{70}{\upmu\m}, it is only visible
 as a small shoulder on the increasing flank of the global maximum for all systems
 except \run{n-i-M1}. The system \run{n-i-M1} generally represents an
 exception: At the wavelength of \SI{2}{\upmu\m} the second maximum is misplaced compared to
 the other systems and located at $\theta \sim \SI{260}{\degree}$ and
 it shows another smaller local maximum at $\theta \sim \SI{320}{\degree}$; at
 \SI{21}{\upmu\m}, there are several small local maxima located around
 $\theta \sim \SI{190}{\degree}$ -- \SI{320}{\degree}; at \SI{10}{\upmu\m} there
 is no second maximum. The system \run{n-i-M1}, whose perturber has the lowest mass,
 represents an intermediate step between the stronger perturbations in the systems
 with perturber \run{i-M2} -- \run{i-M4} and the ceased perturbations in system \run{n-p1}.
 To illustrate this, the system \run{n-p1} is added to the inner perturbers as a reference
 in Fig.~\ref{fig_azimuthal_maxima}.

 The first maximum around $\theta \sim \SI{20}{\degree}$ -- \SI{60}{\degree} is
 produced by the superposed contributions of various grain sizes
 $s \gtrsim \SI{15}{\upmu\m}$. Therefore, that maximum appears over a large
 wavelength range. As the grain distributions vary slightly between systems with
 different perturbers and the relative contribution of the individual grain sizes
 on the net flux varies with wavelength, the azimuthal position of the
 radial maximum changes slightly with perturber parameters and wavelength. The second
 maximum at $\theta \sim \SI{240}{\degree}$ originates from $\sim \SI{5}{\upmu\m}$
 grains. As shown in \ref{subsect_dms}, these
 grains contribute a large fraction of the net flux at the wavelengths of
 \SI{2}{\upmu\m} and \SI{21}{\upmu\m}, but not at \SI{10}{\upmu\m} and \SI{70}{\upmu\m}.
 This is the reason for the smaller height of the second maximum at \SI{10}{\upmu\m}
 and its absence at \SI{70}{\upmu\m}.

 For the systems with an outer perturber we find no second maximum and only one
 global maximum and minimum, respectively, regardless of wavelength. In contrast
 to systems with an inner perturber, the systems
 with an outer perturber do not show a prominent spatial overdensity of
 $\sim \SI{5}{\upmu\m}$ grains able to produce a pronounced second maximum. The only
 exception is the system \run{n-o-M2}, where a less pronounced second maximum is located at
 $\theta \sim \SI{280}{\degree}$. Furthermore, run \run{n-o-M2} shows a minor
 deviation from the general trend at \SI{10}{\upmu\m}, a small additional local
 maximum is located at $\theta \sim \SI{50}{\degree}$. The global maximum is
 located at around $\theta \sim \SI{80}{\degree}$ -- \SI{110}{\degree} for the systems with
 perturbers \run{n-o-M1} to \run{n-o-M3}. The system with the highest mass
 perturber \run{n-o-M4} shows a slightly deviating position.

 Concluding, by using the azimuthal location of the maximum of radial flux
 density, we identified two ways of distinguishing between systems with an inner
 versus those with an outer perturber. First, if we find two azimuthal maxima at the wavelengths
 \SI{10}{\upmu\m} or \SI{21}{\upmu\m}, an inner perturber is
 indicated. As inner perturbers with low masses may produce obscure second
 maxima (such as the system \run{n-i-M1}) or the maxima might be undetectable
 because of observational limitations,
 the absence of such a second maximum is a necessary, but not sufficient
 criterion for an outer perturber system. At \SI{2}{\upmu\m}, a clear
 differentiation is not possible, as a second maximum, much smaller than the
 global, can also be produced by systems with an outer perturber. Nonetheless,
 if that second maximum at \SI{2}{\upmu\m} is accompanied by second maxima at
 larger wavelengths, an inner perturber is indicated.

 A second way to distinguish between the two types of systems is to measure relative
 location of a maximum and minimum of the contrast $\approx 1.7$ ($2.2$) for inner
 (outer) perturber systems at the wavelength of \SI{1300}{\upmu\m}. For the
 systems with an inner perturber, the angular distance between these extrema is
 $\Delta \phi_{\textrm{min-max}} \approx \SI{140}{\degree}$ while the systems with
 an outer perturber appear symmetric with
 $\Delta \phi_{\textrm{min-max}} \approx \SI{180}{\degree}$. Within our
 investigated parameter space, this behavior is independent of perturber mass.
 The relative location of the maxima at \SI{1300}{\upmu\m} depends on which parts of
 the parent belt lead and trail during precision. For the systems with an
 inner perturber the inner edge of the belt leads while the outer trails and
 vice versa for systems with an outer perturber (see Sect. \ref{subsect_ace_outer}).

 We estimated the exposure time required to achieve a sufficient contrast
 for distinguishing the second local maximum at $\theta \sim \SI{240}{\degree}$
 from its consecutive minimum for the system \run{n-i-M2} when observing with
 JWST/MIRI at \SI{21}{\upmu\m} and found \SI{1}{\hour} to be sufficient (see
 Sects.~\ref{appdx_general observability} and
 \ref{appdx_subsect_azimuthal_maxima}). Furthermore, we compiled the flux
 contrasts between the different azimuthal maxima and minima in
 Table~\ref{table_contrast}. For disks with an outer perturber, the contrast
 is notably higher, in particular at mid- and far-infrared wavelengths.
 This is a direct consequence of the lack of a second local brightness maximum.
 \begin{table*}
  \centering
  \caption{Flux contrast of azimuthal flux maximum to minimum.}
  \label{table_contrast}
  \begin{tabular}{l | l l l l l l l} % centered columns
   \toprule
   $\lambda \left[ \upmu \textrm{m} \right]$ &
   \run{n-i-M1} & \run{n-i-M2} & \run{n-i-M3} & \run{n-i-M4} &
   \run{n-o-M2} & \run{n-o-M3} & \run{n-o-M4}  \\
   \midrule
   \num{2} & 1.4/1.1 & 1.7/1.2 & 1.9/1.2 & 2.0/1.3 & 1.9/1.03 & 2.3 & 2.4 \\
   \num{10} & 2.0 & 2.0/1.1 & 2.1/1.1 & 2.2/1.1 & 3.4 & 3.7 & 3.8 \\
   \num{21} & 1.8/1.1 & 2.1/1.4 & 2.3/1.4 & 2.3/1.4 & 4.7 & 6.7 & 6.3 \\
   \num{70} & 2.4 & 2.2 & 2.3 & 2.3 & 3.2 & 3.3 & 3.1 \\
   \num{1300} & 1.7 & 1.7 & 1.7 & 1.7 & 2.2 & 2.2 & 2.2 \\
   \bottomrule
  \end{tabular}
  \tablefoot{Ratio of the flux maxima around
             $\theta \sim \SI{20}{\degree}$ -- \SI{60}{\degree} to their
             consecutive flux minima in Fig.~\ref{fig_azimuthal_maxima}. An optional
             second value denotes the ratio of the maximum around
             $\theta \sim \SI{240}{\degree}$ to its consecutive minimum.}
 \end{table*}

%%%%%%%%%%%%%%%%%%%%%%%%%%%%%%%%%%%%%%%%%%%%%%%%%%%%%%%%%%%%%%%%%%%%%%%%%%%%%%%

\subsection{Other parent planetesimal belts}\label{subsect_other_belts}

 To investigate the impact of the parent belt parameters on the observational appearance of
 the debris disk system, we analyzed brightness distributions of the
 systems with the parameter sets of the parent belt \run{w}, \run{m2}, and
 \run{m3} combined with the parameter set of the inner perturber \run{i-M2} and
 the outer perturber \run{o-M3} and compared them to the results of the system
 simulated with the reference belt parameter set \run{n}, that is, \run{n-i-M2}
 and \run{n-o-M3}.

 In the brightness distributions of the inner perturber systems with the different
 belt parameter sets, we can find the same halo twisting as for the
 system \run{n-i-M2}. Furthermore, we see the expected difference in the level of
 twisting: the system with parameter set \run{w-i-M2}, which possesses the longest
 collision timescale, shows the most trailing halo while the system with parameter
 set \run{m3-i-M2}, which possesses the shortest collision timescale, shows the least
 trailing halo (see Sect.~\ref{subsect_ace_params}).

 Regarding the disk characteristics used to differentiate between systems
 with an inner or an outer perturber, the systems with parent belt parameter
 sets \run{w}, \run{m2}, and \run{m3} show no major differences compared to
 those with set \run{n}. The system with
 parameter set \run{w-i-M2} shows an increased contrast of the spiral structure in the
 $Q$ band compared to \run{n-i-M2} due to the spatially more extended emission.
 All systems with an outer perturber show nonzero brightness in the region inward
 of the parent belt. However, a variation in the belt parameters causes its shape
 and position to vary, similar to a variation of the perturber parameters.

 Concerning the second azimuthal maximum at around $\theta \sim \SI{240}{\degree}$
 (see Sect.~\ref{subsect_azimuthal_maxima}) for systems with an inner perturber:
 As predicted in Sect.~\ref{subsect_ace_runs}, the system with belt parameter set
 \run{w} shows an increased contrast of that second maximum. Furthermore, we
 find an increased contrast between maximum and minimum at a wavelength of
 \SI{1300}{\upmu\m}.
 The broader parent belt of the system causes the dust distribution at the belt
 apocenter to be more diluted, increasing the contrast of the extrema.

 To the contrary, the belts with parameter sets \run{m2} and \run{m3} show a smaller
 second maximum. The higher belt
 mass decreases the collisional timescale, which causes small grains to be
 replenished more quickly, in turn decreasing the relative contribution of older small
 grains on more extended orbits to the net flux. Furthermore, we see
 the predicted degeneracy between a system with a low mass perturber and a
 system with a high mass parent belt: the brightness distributions of the systems with
 parameter sets \run{n-i-M1} and \run{m3-i-M2} appear similarly.
 For systems with an outer perturber, a variation of the parent belt
 parameters causes almost no change in the azimuthal trend of radial maxima.
 However, we identify two minor effects: the parent belt with parameter set
 \run{w} causes an higher contrast between maximum and minimum at a wavelength
 of \SI{1300}{\upmu\m}, similar to the system with an inner perturber
 (\run{w-i-M2}). Furthermore, the system with parameter set \run{m3-o-M3} shows a
 small secondary maximum at a wavelength of \SI{2}{\upmu\m}, similar to that of
 the system with set \run{n-o-M2}, which is again an expression of the
 degeneracy of perturber and parent belt mass.
 At a wavelength of \SI{1300}{\upmu\m} we find no difference in the azimuthal
 trend of radial maxima for varying parent belt mass; the systems with the
 belt parameter sets \run{n}, \run{m2}, and \run{m3} appear the same.

%%%%%%%%%%%%%%%%%%%%%%%%%%%%%%%%%%%%%%%%%%%%%%%%%%%%%%%%%%%%%%%%%%%%%%%%%%%%%%%

\subsection{Flowchart for distinguishing between inner and outer perturbers}
\label{subsect_flowchart}

 We compile the results of the analysis presented in Sects. \ref{subsect_spiral}
 -- \ref{subsect_azimuthal_maxima} in a decision flowchart, displayed
 in Fig.~\ref{fig_flowchart}. For this flowchart we assumed that exactly
 one (dominant) inner or outer perturber orbits the system.
 \tikzstyle{start} = [rectangle, rounded corners, minimum width=3cm,
                      minimum height=1cm,text centered, draw=black,
                      fill=orange!30]
 \tikzstyle{criterion} = [rectangle, rounded corners, minimum width=3cm,
                          minimum height=1cm,text centered, text width=5cm,
                          draw=black, fill=blue!30]
 \tikzstyle{perturber} = [ellipse, minimum width=3cm, minimum height=1cm,
                          text centered, draw=black, fill=green!30]
 \tikzstyle{indifferent} = [ellipse, minimum width=3cm, minimum height=1cm,
                            text centered, draw=black, fill=red!30]
 \tikzstyle{arrow} = [thick,->,>=stealth]
 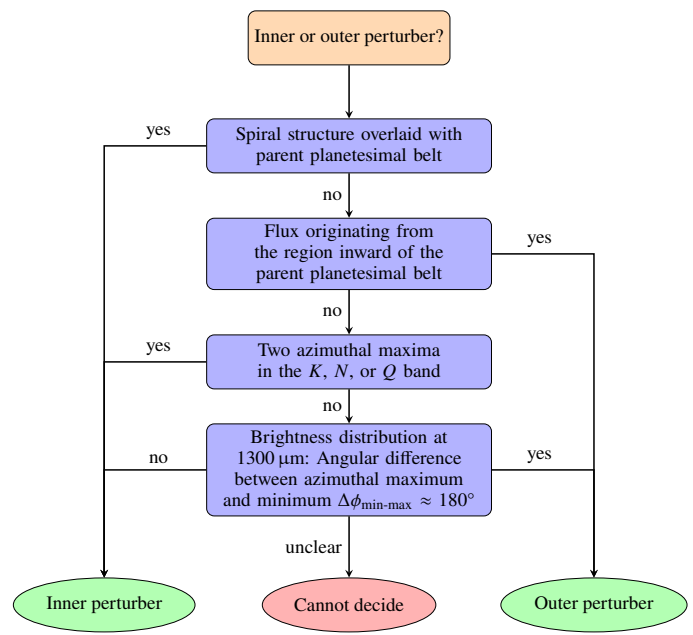
\begin{figure}
  \centering
  \resizebox{\hsize}{!}{
   \begin{tikzpicture}[node distance=2cm]
   \node (start) [start] {Inner or outer perturber?};
   \node (spiral) [criterion, below of=start]
         {Spiral structure overlaid with parent planetesimal belt};
   \node (inner_radiation) [criterion, below of=spiral]
         {Flux originating from the region inward of the parent planetesimal belt};
   \node (two_max) [criterion, below of=inner_radiation]
         {\mbox{Two azimuthal maxima} in the $K$, $N$, or $Q$ band};
   \node (1300micron_minmax_180deg) [criterion, below of=two_max]
         {Brightness distribution at \SI{1300}{\upmu\m}: Angular difference
          between azimuthal maximum and minimum $\Delta \phi_{\textrm{min-max}}
          \approx \SI{180}{\degree}$};
   \node (indifferent) [indifferent, below of=1300micron_minmax_180deg,
                        yshift=-0.5cm]
         {Cannot decide};
   \node (inner_perturber) [perturber, left of=indifferent, xshift=-2.5cm]
         {Inner perturber};
   \node (outer_perturber) [perturber, right of=indifferent, xshift=2.5cm]
         {Outer perturber};

   % Arrows
   \draw [arrow] (start) -- (spiral);
   \draw [arrow] (spiral) -| node[anchor=south, xshift=1cm] {yes}
         (inner_perturber);
   \draw [arrow] (spiral) -- node[anchor=east] {no} (inner_radiation);
   \draw [arrow] (inner_radiation) -| node[anchor=south, xshift=-1cm] {yes}
         (outer_perturber);
   \draw [arrow] (inner_radiation) -- node[anchor=east] {no} (two_max);
   \draw [arrow] (two_max) -| node[anchor=south, xshift=1cm] {yes}
         (inner_perturber);
   \draw [arrow] (two_max) -- node[anchor=east] {no} (1300micron_minmax_180deg);
   \draw [arrow] (1300micron_minmax_180deg) -| node[anchor=south, xshift=1cm] {no}
         (inner_perturber);
   \draw [arrow] (1300micron_minmax_180deg) -| node[anchor=south, xshift=-1cm] {yes}
         (outer_perturber);
   \draw [arrow] (1300micron_minmax_180deg) -- node[anchor=east] {unclear} (indifferent);
  \end{tikzpicture}
  }
 \caption{Flowchart for deciding whether an inner or outer perturber orbits in the
          system based on distinct observable features of a debris disk system.}
 \label{fig_flowchart}
 \end{figure}

%%%%%%%%%%%%%%%%%%%%%%%%%%%%%%%%%%%%%%%%%%%%%%%%%%%%%%%%%%%%%%%%%%%%%%%%%%%%%%%
%%%%%%%%%%%%%%%%%%%%%%%%%%%%%%%%%%%%%%%%%%%%%%%%%%%%%%%%%%%%%%%%%%%%%%%%%%%%%%%

\section{Discussion}\label{sect_discussion}

\subsection{Numerical dispersion}\label{subsect_dispersion}
 We could show that a grid-based model of a collisional cascade can be combined
 with a TVD advection scheme for PR drag and secular dynamical
 perturbations. The upwind scheme used in previous \ACE\ versions is suitable
 for scenarios were a stream of small particles springs from a continuous supply
 in a static belt of bigger objects. The population of small grains decays
 collisionally, for example during PR-induced drift toward the star, resulting in
 smooth distribution. The distribution of bigger source objects is confined more
 sharply, but does not migrate. In the case of the secular perturbations discussed
 here, that static picture no longer applies. When dynamical perturbations
 move the source region itself through the phase space, the belt's sharp
 features will be dispersed over time, artificially broadening narrow disks or
 exciting dynamically cold disks. After some time, the simulation results do no
 longer reflect the intended physical scenario
 \citep{10.1051/0004-6361/201935199}. We addressed this problem with the introduction
 of a second-order TVD scheme to the ACE code, reducing the dispersion.
 While dispersion is still significant, integration is now possible over a
 longer period of time (see Appendix~\ref{app_ace_dispersion} for more details).

 Further improvements could include a return to a semimajor axis grid instead
 of the current pericenter grid. That step would eliminate the dispersion in
 one dimension because secular
 perturbations do not affect the semimajor axes. The pericenters are affected
 because they depend on eccentricity: $q = a(1 - e)$. On the other hand, a
 pericenter-based grid can represent the orbits of halo grains on highly
 eccentric orbits more accurately, preserving their origin in the parent belt
 and their collision rates and velocities there. For the problem at hand, where
 both the halo of small grains and the parent belt are important, none of the
 two options provides a clear advantage.

 The angular dispersion in $\varpi$ could be reduced further by a
 ``co-precessing'' grid that follows the parent belt's orientation. A mean
 precession $\langle\dot\varpi\rangle$, weighted by the masses in individual
 bins, could be calculated for each time step and then subtracted. The advection
 through the grid would be reduced to differential precession relative to the
 mean belt. We will implement and test this modification in future work.

 A consequent iteration on a grid that follows the mean precession would be a
 grid where the individual bins completely follow the advection stream. Instead
 of moving material from bin to bin, the bins themselves would move. No
 dispersal could occur. However, the current rigid discretization with separate
 grids for $m$, $q$, $e$, and $\varpi$ allows for optimizations that make use
 of an azimuthal symmetry in the collision physics, even if the material
 distribution itself is not azimuthally symmetric. Variable, asymmetric
 discretization would increase computational costs of the collisional cascade
 model drastically, necessitating a reduction in grid resolution. The non-static
 bins in such an approach would be similar to the so-called tracers used in the
 LIPAD code \citep{10.1088/0004-6256/144/4/119}. The superparticles adopted in
 LIDT-DD \citep{10.1051/0004-6361/201321398} and SMACK
 \citep{10.1088/0004-637X/777/2/144} differ more strongly because the particle
 ensembles are not orbit-averaged there. The same applies to the destruction-only
 approach called collisional grooming, which is used by
 \citet{10.1088/0004-637X/707/1/543} and in the DyCoSS code
 \citep{10.1051/0004-6361/201117899,10.1051/0004-6361/201219962}.

%%%%%%%%%%%%%%%%%%%%%%%%%%%%%%%%%%%%%%%%%%%%%%%%%%%%%%%%%%%%%%%%%%%%%%%%%%%%%%%

\subsection{Scope of the perturbation scenario}\label{subsect_scope}
 We modeled debris disks that are secularly perturbed by a single companion
 on an eccentric orbit distant from the planetesimal parent belt. The induced
 differential precession leads to
 asymmetries in belt and halo that are reflected in scattered light and thermal
 emission maps. The considered perturber masses are high, ranging from half a
 Jupiter mass to roughly 0.06 solar masses. The models are not applicable to
 Neptune or Earth-mass planets. Firstly, perturber masses should match or exceed disk
 masses in order to perturb with a significant amplitude. The dust-rich debris
 disks that are most easily detected are expected to be collisionally
 replenished from massive planetesimal belts, possibly totaling hundreds of
 Earth masses \citep{10.1016/j.icarus.2006.01.022,10.1093/mnras/stx2932},
 assuming biggest planetesimals with diameters on the order of $\sim 100$~km.
 Even minimum total disk masses, assuming at most km-sized objects, reach tens
 of Earth masses \citep{10.1093/mnras/staa2385}.

 Secondly, the described effects of differential precession require the
 perturber to overcome the self-gravity of the disk. The
 precession induced in the halo by the planetesimal belt will drag the halo
 along with the belt, reducing the differential precession and the trailing of
 the halo caused by an inner perturber. Both timescale and amplitude of the
 perturbations favor the planetesimal belt over the planetary perturber because
 the belt is closer to the outer halo than an inner planet. When belt and
 perturber are of similar mass, the belt will dominate the halo's dynamical
 evolution because of its proximity. The further away from the belt the inner
 perturber is, the greater must be its mass in excess of the belt mass.
 In the case of a distant outer perturber, the halo is located between belt and perturber,
 weakening the relative strength of the perturbations exerted by a massive belt.
 However, as illustrated by Eq.~\ref{eq:Tprec}, an outer perturber at a given separation
 from or distance ratio with the belt would always need to be more
 massive than an inner perturber that exerts the same perturbations.

 Thirdly, only high-mass planets induce perturbations on timescales short enough
 to compete with the collisional replenishment of the small-grain halos of bright disks.
 The secular perturbation rate is proportional to perturber mass, while the
 collision rate is proportional to disk mass squared. A ten-fold reduction
 in disk mass would require a hundred-fold reduction in planet mass, from half
 a Jupiter mass to roughly one Earth mass, to result in similar disk asymmetries.
 For a given perturber mass, less massive disks are thus more likely to exhibit
 the discussed features.

 Aside from perturber mass, the distance between disk and perturber has a lower
 bound in our approach. Both in the collision model and the description of the
 secular perturbations we assumed orbit averaging, and thus, could not cover resonant
 interactions or close encounters. As a result of this intrinsic property, the
 model is constrained to perturbers more distant from the disk.

 In the described scenario the belt eccentricity was caused and modified by a
 single perturber on an eccentric orbit. The problem of the original cause of
 the asymmetry is not solved. It is only transferred from the disk to the
 perturber. The eccentric orbit of the perturber could itself be the result of
 a short-term event, such as a close encounter with another body
 \citep[e.g.,][]{10.1086/590227}, or long-term
 perturbations. In the case of a close encounter, the fates of the perturber and
 the disk depend on whether the other body is still present or has been ejected
 in the process. If it remained present, further encounters can occur, limiting
 the time available for steady secular perturbation of the disk. If long-term
 perturbations drive the orbital eccentricity of the primary perturber of the
 disk, the combined action of both perturbers mutually and on the disk would
 need to be taken into account, including possible secular resonances
 \citep{10.1093/mnras/sty1678}. \citet{2022MNRAS.tmp.2411R} explored the
 effects of interaction with a former protoplanetary disk and of repeated
 planet--planet scattering, both of which could drive perturber eccentricity
 while keeping the planetesimal belt narrow.

 There are three main causes for asymmetry in our modeled belts and halos:
 (i) the global belt eccentricity, which translates into an asymmetric halo
 \citep{10.3847/0004-637X/827/2/125,10.1051/0004-6361/201630297},
 (ii) the differential precession within the belt, with density variations along
 the belt and resulting spatial variations in the production rates of smaller
 grains, and (iii) the differential precession of the halo relative to the belt.
 The third was first discussed in \citep{10.1051/0004-6361/201935199} and is also
 the main focus of our present work.
 In our model, both (i) the belt eccentricity and (ii) the differential precession
 of the belt are caused by a perturber on an eccentric orbit and still evolving.
 However, the effects of (iii) the twisting of the halo are independent from this
 particular scenario. If the belt eccentricity originated from earlier phases
 \citep[e.g.,][]{10.1098/rsos.200063}, a perturber on a circular orbit would
 suffice to drive secular precession in both belt and halo. Or, if the belt itself
 were already in dynamical equilibrium, its complex eccentricity centered on
 the eccentricity forced by the perturber
 \citep[all but the top-right panels in Fig.~2 of][]{10.1098/rsos.200063},
 the small-grain halo would still be twisted with respected to the belt.

%%%%%%%%%%%%%%%%%%%%%%%%%%%%%%%%%%%%%%%%%%%%%%%%%%%%%%%%%%%%%%%%%%%%%%%%%%%%%%%%

\subsection{Observational appearance of secularly perturbed debris disks}
 We set up a debris disk model with an A3$\,$V central star and dust grains
 composed of a water ice-silicate mixture. Then, we derived spatial dust
 distributions for various combinations of parent belt (see
 Table~\ref{table_belt}) and perturber parameters (see
 Table~\ref{table_perturber}) using \ACE.
 Based on the simulated spatial dust distributions we computed maps of
 surface brightness for disks seen face-on using \DMS\ in five wavelengths
 important for debris disk
 observations (see Table~\ref{table_wavelengths}). Based on that, we
 performed an observational analysis and investigated whether the surface
 brightness distributions and some features thereof for an assumed distance to
 the stellar system of \SI{7.7}{pc} can be detected with the observing
 facilities and instruments MICADO and METIS at the ELT, NIRCam
 and MIRI on the JWST, and PACS on \textit{Herschel} (see
 Sect.~\ref{appdx_general observability}).
 We find that multiwavelength observations combining
 JWST/MIRI to resolve the disk morphology and trace the small grain halo
 and ALMA millimeter observations to trace the position of the parent planetesimal
 belt are suitable for inferring the existence of unseen perturbers in the system.
 The debris disks in our study have intermediate dust masses up to a grain size
 of \SI{1}{\mm} of $\SI{2.7e-9}{M_\odot} \pm \SI{10}{\percent}$ thereof. A
 specific debris disk can have dust masses and hence surface brightness
 values by a magnitude higher than our systems
 \citep{10.1088/0004-637X/776/2/111, 10.3847/0004-637X/831/1/97}, resulting
 in a significant decrease of the required exposure times derived here.
 When applying this observational feasibility study to another system, both (i) the
 impact of the limited angular resolution on the achievable spatial resolution
 and (ii) the distance and mass-dependent brightness together with the
 sensitivity of the observing instrument have to be taken into account
 accordingly.

 The key characteristics of the brightness distributions of the considered debris
 disk systems are strongly wavelength dependent. In the case of a perturber orbiting
 radially inward of the parent planetesimal belt, the smallest grains near the blowout limit
 of $s\sbs{bo} \approx \SI{4}{\upmu\m}$ have very extended spatial distributions
 that differ strongly from those of larger grains. Within the considered
 parameter space, we find that observations in
 the $K$ and $Q$ bands are best suited for tracing those smallest grains,
 while the $N$ band is only suited for tracing larger grains.
 At longer wavelengths, the apocenter glow effect described by
 \citet{10.3847/0004-637X/832/1/81} is not visible in our data because the belt
 width is resolved at both apocenter and pericenter. An instrumental
 point spread function (PSF) would blur preferentially the narrower pericenter,
 reducing the peak surface brightness there.

 In the case of a system with a perturber orbiting radially outward of the parent
 planetesimal belt, we found the smallest grains to be of minor importance for the
 overall appearance of the system. Therefore, for those systems the general
 shape of the brightness distributions varies little with observing wavelength.

 We showed that for systems with an inner perturber it is possible to investigate the
 difference in precession between the small grain halo and the parent belt using
 their brightness distributions. To do so we characterize the shape of the halo by lines
 of constant surface brightness (isophotes) and fitted ellipses to them.
 A higher mass perturber causes the parent belt to precess with a larger angular
 velocity than a lower mass perturber does. Therefore, the small grain halo lags behind
 the parent belt more when being secularly perturbed by a more massive perturber.
 In an observational analysis, using this information alone does not permit the
 perturber mass to be constrained, because the dependence of the halo precession
 on the perturber mass is degenerate with the distance of the perturber to
 the parent belt. Nonetheless, this information can add a useful puzzle piece
 to the larger picture of a stellar system.
 The effect of differential precession is best pronounced at
 \SI{10}{\upmu\m} and \SI{70}{\upmu\m} where the smallest grains
 are of minor importance. However, at \SI{10}{\upmu\m} the
 surface brightness of the halo is too low to be detected by JWST/MIRI assuming
 a reasonable exposure time. At \SI{70}{\upmu\m}, \textit{Herschel}/PACS provided
 enough sensitivity to detect the halo. Therefore, for close debris disk systems
 that are sufficiently resolved with that observatory, we suggest a review of
 archive data. At \SI{2}{\upmu\m} and \SI{21}{\upmu\m} where the contribution of
 the smallest grains is of importance, the orientation of the ellipses we used
 to describe the halo
 shape undergoes a flip of $\Delta \phi \sim \SI{-180}{\degree}$ for increasing
 distances from the central star. This flip originates in a change of the grain
 size dominating the surface brightness: closer to the star larger grains
 $> \SI{5}{\upmu\m}$ dominate, while in outer regions the smallest grains
 $\lessapprox \SI{5}{\upmu\m}$ near the blowout limit dominate. Observing at
 \SI{21}{\upmu\m} the regions of the halo undergoing that flip can be detected
 entirely using JWST/MIRI with an exposure time of \SI{5}{\hour}.

 The difference in the spatial distribution of different grain size intervals
 causes distinct features in the brightness distributions.
 We compared the observational appearance of systems with an inner to
 those with an outer perturber and identified multiple
 features that can be used to distinguish between those two types of systems and
 compiled the results in a flowchart presented in Fig.~\ref{fig_flowchart}.
 The differences are the following:

 First, for systems with an
 inner perturber, a spiral structure appears in the $Q$ band (see
 Sect.~\ref{subsect_spiral}). Using JWST/MIRI, an exposure time of
 $\gtrsim \SI{8}{\hour}$ is required to achieve the required contrast for the
 system with a substellar mass companion \run{n-i-M4}. However, the contrast
 of the spiral structure can be higher for different sets of parameters,
 requiring less exposure time. Regarding angular resolution, resolving the
 spiral structure in close-by systems such as Fomalhaut is possible with
 JWST/MIRI.

 Second, systems with an outer perturber show brightness levels of up to a
 factor of several tens of the maximum brightness from
 regions radially inward of the parent belt (see Sect.~\ref{subsect_grains_inward_outer}).
 However, for the corresponding grains orbiting close to the perturber, the
 accuracy of our model is limited. With that caveat, observing at the wavelength of
 \SI{21}{\upmu\m} with JWST/MIRI is most promising to detect this emission:
 an exposure time of \SI{1}{\hour} is sufficient.

 Third, the number of azimuthal maxima of surface brightness at various
 wavelengths differs between systems with an
 inner and with an outer perturber (see Sect.~\ref{subsect_azimuthal_maxima}).
 Observing at the wavelength of \SI{21}{\upmu\m} with JWST/MIRI, an exposure
 time of \SI{1}{\hour} is sufficient to distinguish the pair of azimuthal
 maximum-minimum with the lowest contrast.
 To provide a basis for further observational investigations we compiled
 the contrast values of these extrema in Table~\ref{table_contrast}.

 Lastly, the relative location of the azimuthal maximum and minimum in the surface
 brightness distribution at the wavelength of \SI{1300}{\upmu\m} differs
 between systems with an inner and an outer perturber. While for the former
 systems the angular difference between the azimuthal angles of maximum and
 minimum is $\approx \SI{140}{\degree}$, the latter appear axisymmetric with
 an angular difference of $\approx \SI{180}{\degree}$.

 We investigated synthetic brightness asymmetries solely of disks seen face-on to
 ensure that the effects discussed in this study are clearly distinguished from
 effects caused by observing an inclined system.
 At short wavelengths, where the intensity is dominated by scattered
 stellar light, inclined disks can - without any asymmetry in orbital dust
 distribution - show significant brightness asymmetries. This is due to
 an asymmetric scattering phase function causing the strong forward scattering
 nature of small grains, but also the backscattering nature of larger grains
 \citep[e.g.,][]{10.1051/0004-6361/200911854, 10.1051/0004-6361/200913065}.
 An extended study of how
 disk morphology in scattered stellar light depends on the viewing angle is
 presented in \citet{10.3847/0004-637X/827/2/125}. To
 interpret spatially resolved observations of such systems using our present
 study, first the scattering phase function has to be determined from the
 respective observation \citep[e.g.,][]{10.1051/0004-6361/201527838,
 10.1051/0004-6361/202038237} or adapted from others
 \citep[e.g.,][]{10.1088/0004-637X/811/1/67}; second the surface brightness maps
 have to be deprojected to the face-on case
 \citep[as done for example in][]{10.1038/nature03601,
 10.1051/0004-6361/201527838}. Nonetheless, by opting for a wavelength of
 \SI{21}{\upmu\m} (where thermal dust emission dominates),
 which is promising to detect most of the features of surface brightness
 distributions discussed in this study, or even longer wavelengths this
 problem can be circumvented.

 Besides secular perturbation by a substellar companion as discussed in this
 study and effects of observing an inclined disk discussed above, there are
 other sources of disk asymmetry and non trivial structures. Giant impacts
 \citep[see, e.g.,][]{10.1007/s11214-016-0248-1} can happen in the outer reaches
 of a stellar system. The aftermath of those impacts was investigated
 numerically for example by \citet{10.1051/0004-6361/201321398},
 \citet{10.1093/mnras/stu476}, and \citet{10.1051/0004-6361/201424309}:
 While a spiral structure emerging in the aftermath of the impact is rather short
 lived and is dispersed on a \SI{}{kyr} timescale, a narrow, eccentric belt
 with a clear brightness asymmetry is a long lasting \citep[$\sim$ \SI{1}{Myr} at
 \SI{50}{au} distance to the central star, ][]{10.1093/mnras/stu476} product
 of a giant impact and could mimic the presented effects of secular
 perturbations.

 Resonant clumps are a further source of potential confusion with the azimuthal
 brightness variations discussed here. Particles caught in individual
 mean-motion resonances cluster at regular intervals, potentially creating
 azimuthal clumps in the belts \citep{10.1086/379064,10.1051/0004-6361:20077934}.
 With bigger grains being more easily trapped in these resonances
 \citep{10.1051/0004-6361:20065584}, the clumps are expected to be more
 pronounced at longer wavelengths, that is, in the millimeter  wavelength range.
 Long-lasting spiral arms of unbound grains that emanate from the clumps
 \citep{10.1086/499487} would show up at shorter wavelengths. Resonant clumps
 co-orbit with their respective captor, which provides a possible way to
 identify them with multi-epoch observations. However, this approach is limited
 by large orbital periods. On a different note,
 \citet{10.1093/mnras/stab760} showed that even the narrow, eccentric, and
 seemingly smooth dust ring around Fomalhaut is consistent with being produced
 by populations of left-over planetesimals in different mean-motion resonances
 with a perturber on a belt-crossing, eccentric orbit. That scenario does not
 require ongoing secular perturbation, but assumes a very high initial disk
 mass and strong subsequent depletion thereof.

 Stellar flybys as a source of confusion are unlikely for our systems, as
 violent, short-term causes would produce belts broader than the ones considered
 in this study \citep[e.g.,][]{10.1046/j.1365-8711.2001.04212.x,
 10.1006/icar.2001.6700}.

%%%%%%%%%%%%%%%%%%%%%%%%%%%%%%%%%%%%%%%%%%%%%%%%%%%%%%%%%%%%%%%%%%%%%%%%%%%%%%%%
%%%%%%%%%%%%%%%%%%%%%%%%%%%%%%%%%%%%%%%%%%%%%%%%%%%%%%%%%%%%%%%%%%%%%%%%%%%%%%%%

\section{Summary}\label{sect_summary}

 We have implemented a second-order TVD\ scheme in \ACE,
 effectively reducing numerical dispersion and allowing the simulation of a
 longer period of time than before. With this improved numerical code, we
 simulated the collisional evolution of debris disks around an \mbox{A3$\,$V}
 star that is being secularly perturbed by a companion far from the planetesimal
 belt and analyzed the resulting spatial grain distributions. We find that
 such systems exhibit differential precession, which leads to asymmetries in the
 grain distributions of the belt and grain halo.

 Subsequently, we simulated brightness distributions of the resulting grain
 distributions in the $K$, $N$, and $Q$ bands and at wavelengths of
 \SI{70}{\upmu\m} and \SI{1300}{\upmu\m} using \DMS. The simulation
 results show that the appearance of a secularly perturbed debris disk strongly
 depends on the combination of the observing wavelength and the grain blowout limit. We
 were able to recover the difference in precession between the small grain halo
 and the parent planetesimal belt by analyzing lines of constant brightness in
 the spatially resolved surface brightness maps. Furthermore, we identified
 several features in the brightness distributions characteristic of a system
 with a perturber orbiting either within or outside the parent belt and compiled
 them into a decision flowchart (see Fig.~\ref{fig_flowchart}). Based on our
 computed observables, we investigated the feasibility of observing the small
 grain halo and the aforementioned features: observations with JWST/MIRI in the
 $Q$ band together with \SI{1300}{\upmu\m} wavelength observations that trace the
 position of the parent planetesimal belt are best suited for distinguishing between
 systems with either an inner or outer perturber; observations with JWST/MIRI
 and \textit{Herschel}/PACS are well suited for investigating the small grain halo.

 The model is applicable only to systems where the perturber is more massive
 than the disk, the perturber is on an eccentric orbit $e \gtrsim \num{0.3}$
 that is distant from the disk, the disk is massive enough to be dominated by
 collisions, and the star is sufficiently luminous to remove grains below a
 size threshold via radiation pressure. Observability constraints and our
 specific application to cold debris disks at around \SI{100}{au} from the
 host star further limit the scope to disks around nearby stars, preferably
 seen face-on.

%%%%%%%%%%%%%%%%%%%%%%%%%%%%%%%%%%%%%%%%%%%%%%%%%%%%%%%%%%%%%%%%%%%%%%%%%%%%%%%%
%%%%%%%%%%%%%%%%%%%%%%%%%%%%%%%%%%%%%%%%%%%%%%%%%%%%%%%%%%%%%%%%%%%%%%%%%%%%%%%%

\section*{ORCID iDs}
T. A. Stuber \orcidlink{0000-0003-2185-0525}
\href{https://orcid.org/0000-0003-2185-0525}
     {https://orcid.org/0000-0003-2185-0525}\\
T. Löhne \orcidlink{0000-0003-2584-5280}
\href{https://orcid.org/0000-0003-2584-5280}
     {https://orcid.org/0000-0003-2584-5280}\\
S. Wolf \orcidlink{0000-0001-7841-3452}
\href{https://orcid.org/0000-0001-7841-3452}
     {https://orcid.org/0000-0001-7841-3452}

%%%%%%%%%%%%%%%%%%%%%%%%%%%%%%%%%%%%%%%%%%%%%%%%%%%%%%%%%%%%%%%%%%%%%%%%%%%%%%%%
%%%%%%%%%%%%%%%%%%%%%%%%%%%%%%%%%%%%%%%%%%%%%%%%%%%%%%%%%%%%%%%%%%%%%%%%%%%%%%%%

\begin{acknowledgements}
 The authors thank the anonymous referee for their comments and suggestions that
 helped to improve the article,
 M. Booth for helpful comments regarding the sensitivity of
 \textit{Herschel}/PACS, J. Sende regarding TVD advection schemes,
 D. C. Hines, B. Hilbert, and J. H. Girard from
 the JWST Help Desk for crucial support in computing surface brightness
 sensitivities for MIRI and NIRCam, and all members of the Astrophysics
 Department Kiel for discussions and comments about this work in general.
 This research has made use of NASA's Astrophysics Data System Bibliographic
 Services, \texttt{adstex} (\href{https://github.com/yymao/adstex}
 {https://github.com/yymao/adstex}), a modified A\&A bibliography style
 file (\href{https://github.com/yangcht/AA-bibstyle-with-hyperlink}
 {https://github.com/yangcht/AA-bibstyle-with-hyperlink}),
 Ipython \citep{10.1109/MCSE.2007.53}, Jupyter notebooks \citep{jupyter},
 \texttt{Astropy} (\href{https://www.astropy.org}
 {https://www.astropy.org}), a community-developed core Python package for
 Astronomy \citep{10.1051/0004-6361/201322068, 10.3847/1538-3881/aabc4f},
 Matplotlib \citep{10.1109/MCSE.2007.55}, Numpy
 \citep{10.1038/s41586-020-2649-2}, Scipy \citep{10.1038/s41592-019-0686-2},
 and scikit-image \citep{10.7717/peerj.453}.

 This work was supported by the Research Unit FOR 2285 “Debris Disks in
 Planetary Systems” of the Deutsche Forschungsgemeinschaft (DFG). The authors
 acknowledge the DFG for financial support, TS and SW under grant WO 857/15-2,
 TL under grant WO 1715/2-2.
\end{acknowledgements}

%%%%%%%%%%%%%%%%%%%%%%%%%%%%%%%%%%%%%%%%%%%%%%%%%%%%%%%%%%%%%%%%%%%%%%%%%%%%%%%%
%%%%%%%%%%%%%%%%%%%%%%%%%%%%%%%%%%%%%%%%%%%%%%%%%%%%%%%%%%%%%%%%%%%%%%%%%%%%%%%%

\begin{appendix}

\section{Advection scheme with reduced dispersion}\label{app_ace_dispersion}
\begin{figure}
  \centering
  \includegraphics{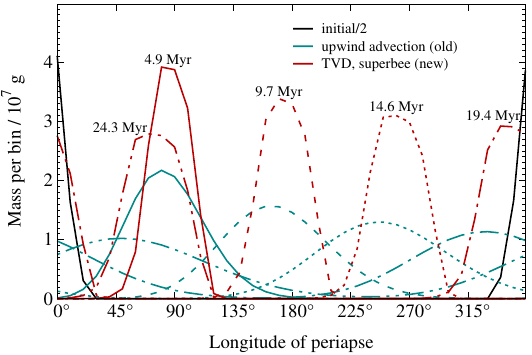}
  \caption{Azimuthal dispersion due to numerical limitations over time for the
           upwind advection scheme (blue-green) and the TVD scheme
           (red) with the \texttt{superbee} flux limiter. The labeled snapshots are
           isochronously separated by one-quarter of a full precession period.
           Equal styles of dashed lines correspond to equal times. The solid
           black line shows
           the initial distribution, scaled down by a factor of 1/2. Results are
           shown for a case with purely azimuthal precession:
           $M\sbs{p} = \SI{2.5}{M_{\textrm{Jup}}}$, $a\sbs{p} = \SI{20}{au}$,
           $e\sbs{p}=0$, $a\sbs{b} = 102$, $e\sbs{b} = 0.31$.}
  \label{fig_ace_dispersion}
 \end{figure}

 Figure~\ref{fig_ace_dispersion} shows that the numerical dispersion is strongly
 reduced if the TVD scheme is used, but is still significant.

\section{Observability of companion-induced shear}
\label{appdx_general observability}
\begin{figure*}
  \centering
  \begin{subfigure}[b]{\textwidth}
   \resizebox{\hsize}{!}{
             \includegraphics[width=1\linewidth]{
             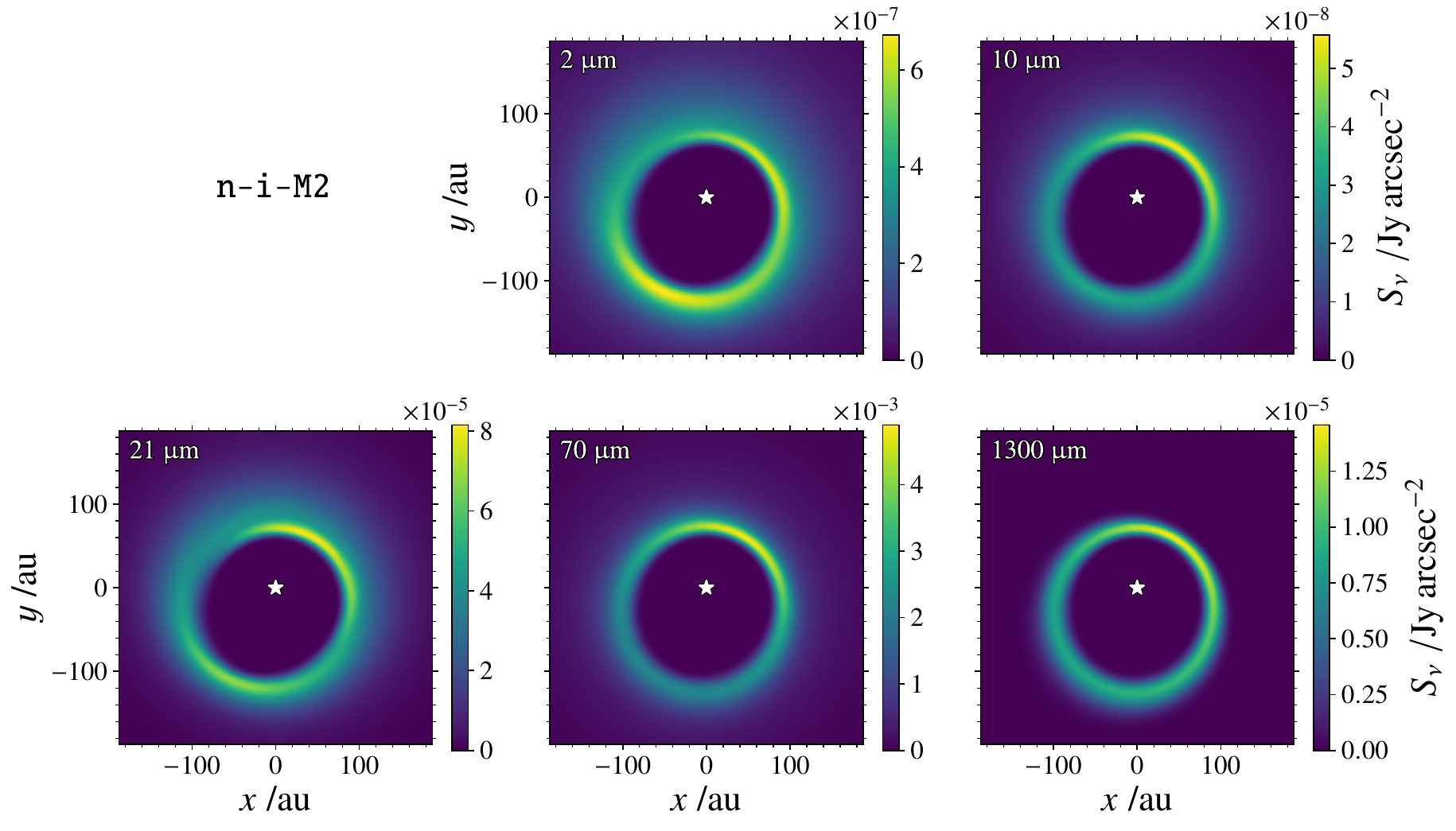}}
   \caption{}
   \label{fig_all_waves_unnormalized_n-i-M2}
  \end{subfigure}
  \hfill
  \begin{subfigure}[b]{\textwidth}
   \resizebox{\hsize}{!}{
             \includegraphics[width=1\linewidth]{
             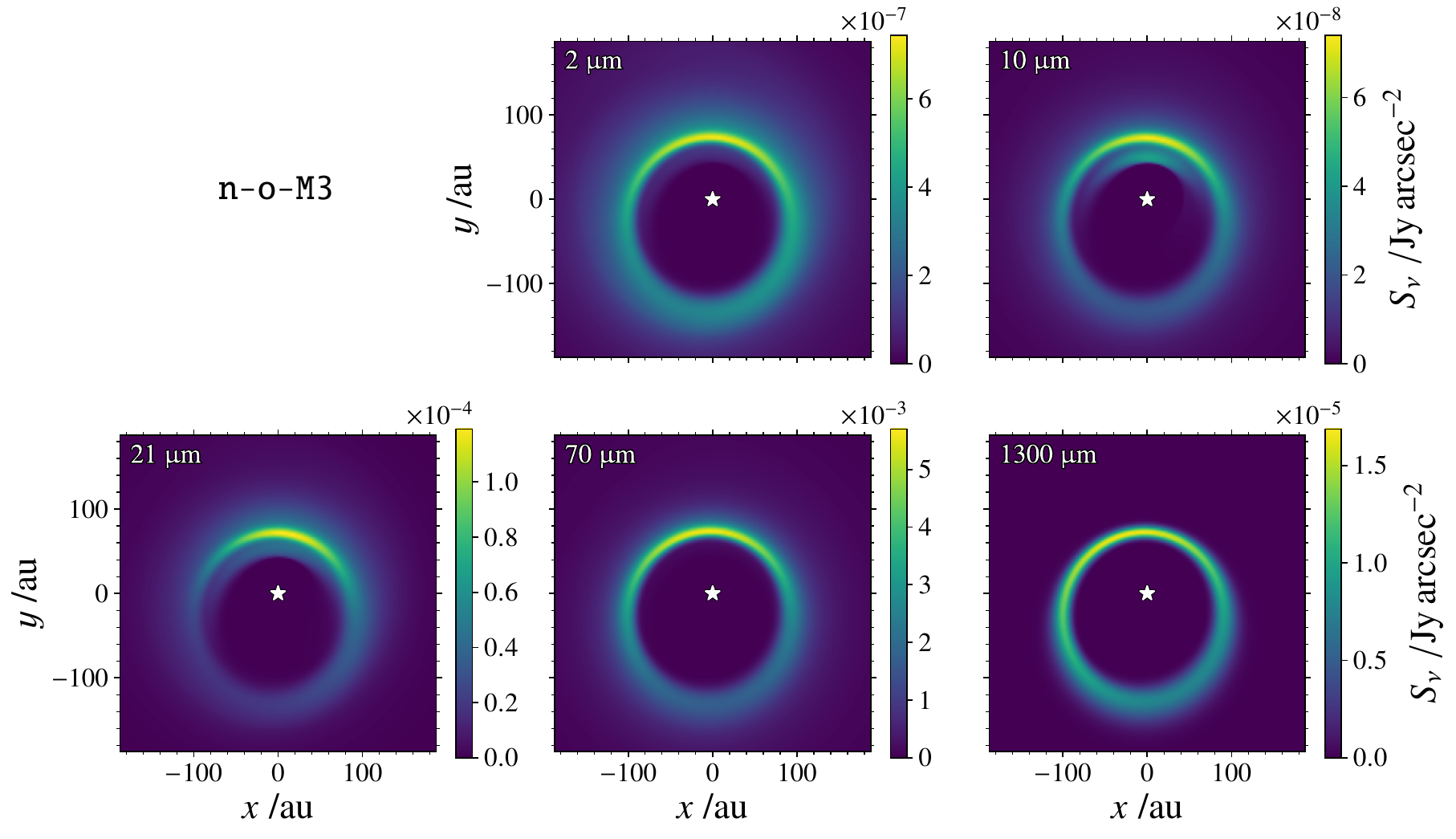}}
    \caption{}
    \label{fig_all_waves_unnormalized_n-o-M3}
  \end{subfigure}
  \caption{Surface brightness distributions of the systems \run{n-i-M2} (a) and
           \run{n-o-M3} (b) at the five
           wavelengths, \SI{2}{\upmu\m}, \SI{10}{\upmu\m}, \SI{21}{\upmu\m},
           \SI{70}{\upmu\m}, and \SI{1300}{\upmu\m}, zoomed in on the central
           $\sim \SI{190}{au}$. The white asterisk denotes the position of the
           central star and defines the center of the coordinate system. Maps
           are the same as in Fig.~\ref{fig_images_all_waves_inner_outer},  but now
           with the absolute scale of surface brightness for a stellar distance
           of \SI{7.7}{pc}.}
  \label{fig_all_waves_unnormalized}
 \end{figure*}

 In this section we explore the feasibility of observing the features of our
 modeled planetary sheared debris disks as presented in
 Sects.~\ref{sect_obs_appearance} and \ref{sect_inner_vs_outer} for an assumed
 distance of \SI{7.7}{pc} (e.g., Fomalhaut) to the stellar system. In
 Fig.~\ref{fig_all_waves_unnormalized}, the same maps as in
 Fig.~\ref{fig_images_all_waves_inner_outer} are displayed, but with an absolute
 scale of surface brightness. Information about
 the corresponding dust masses is given in Sect.~\ref{subsect_dms}.

 We investigated the potential of observing our simulated systems using
 JWST/NIRCam and ELT/MICADO
 at a wavelength of \SI{2}{\upmu\m}, JWST/MIRI and ELT/METIS at \SI{10}{\upmu\m},
 JWST/MIRI at \SI{21}{\upmu\m}, and \textit{Herschel}/PACS at \SI{70}{\upmu\m}.
 We estimated surface brightness sensitivities based on the literature and
 scaled them to the desired signal-to-noise ratio (\SNR) and exposure time
 ($t\sbs{exp}$) with the canonical relation, that the \SNR\ is proportional to the
 square root of the exposure time and to the square root of the number of
 photons received.

 \noindent\textbf{JWST/NIRCam} at \SI{2}{\upmu\m}:
 To obtain an estimation of the surface brightness sensitivity we used the JWST exposure
 time calculator \citep{10.1117/12.2231768} in version 1.7. We employed a uniform brightness
 distribution, the preset readout pattern DEEP2, and a low background at the position
 of Fomalhaut. For the value of surface brightness we chose
 \SI{5e-7}{Jy\,arcsec^{-2}}, a value intermediate for our systems
 (see the first panel of Fig.~\ref{fig_all_waves_unnormalized_n-i-M2} and
 \ref{fig_all_waves_unnormalized_n-o-M3}, respectively). Using one exposure with six
 integrations of ten groups each we found an exposure time of
 $t\sbs{exp} \approx \SI{11780}{\s} \approx \SI{3.3}{\hour}$
 to be sufficient to reach a \SNR\ of \num{3.1}. We neglected any influence by light not
 blocked or scattered of a possible coronagraph. The angular distance of disk emission to
 the position of the central star of several arcseconds is sufficiently large compared to the inner
 working angle of the suitable coronagraphic mask MASK210R, \SI{0.40}{arcsec}
 (JWST Help Desk, priv. communication).

 \noindent\textbf{ELT/MICADO} at \SI{2}{\upmu\m}:
 \citet{10.1117/12.2311483} expect the point-source sensitivity of ELT/MICADO to be comparable
 to the JWST, but with an angular size of one instrument pixel of \SI{4e-3}{arcsec} for the
 low resolution imager.
 Therefore, as for ELT/MICADO the angular area on the sky of one
 pixel is smaller than that of
 JWST/NIRCam$\,$\footnote{\href{https://jwst-docs.stsci.edu/jwst-near-infrared-camera}
{https://jwst-docs.stsci.edu/jwst-near-infrared-camera}}
 by a factor of
 \begin{equation}
  \left( \frac{d_{\mathrm{pix, NIRCam}}}{d_{\mathrm{pix, MICADO}}} \right)^2 =
  \left( \frac{\SI{3.1e-2}{arcsec}}{\SI{4e-3}{arcsec}} \right)^2 \approx \num{60}\, ,
 \end{equation}
 to detect a surface brightness of \SI{5e-7}{Jy\,arcsec^{-2}} with a \SNR\
 of \num{3}, an exposure time of $t\sbs{exp} \approx \SI{186}{\hour}$ is required.
 Accordingly, to increase the \SNR, a binning of several pixels will be required. To estimate
 the instrument performance in that case requires detailed modeling of the instrument noise.
 This is out of the scope of this estimation and we do not consider ELT/MICADO in
 the further analysis.

 \noindent\textbf{ELT/METIS} at \SI{10}{\upmu\m}:
 \citet{10.18727/0722-6691/5218} denote for an
 observation with the N2 filter a required surface brightness sensitivity of
 \SI{7.2}{Jy\,arcsec^{-2}} to achieve a \SNR\ of \num{5} with an exposure time
 of \SI{1}{\hour}.
 To detect our peak surface brightness of
 $\sim \SI{1e-7}{Jy\,arcsec^{-2}}$ (see the second panel of
 Fig.~\ref{fig_all_waves_unnormalized_n-i-M2} and
 \ref{fig_all_waves_unnormalized_n-o-M3}, respectively) with a \SNR\ of \num{3} using ELT/METIS,
 an unfeasibly long exposure time would be required ($\sim$ years). As for
 ELT/MICADO a binning of pixels will be required and we do not consider
 ELT/METIS in the further analysis.

 \noindent\textbf{JWST/MIRI} at \SI{10}{\upmu\m} and \SI{21}{\upmu\m}:
 We follow the suggestions
 of the JWST Help Desk (priv. communication) and use the
 values of minimum detectable flux density for a \SNR\ of 10 for an exposure time
 $t\sbs{exp}$ of \SI{10000}{\s}, listed at
 the JWST User Documentation
 Webpage$\,$\footnote{\href{https://jwst-docs.stsci.edu/jwst-mid-infrared-instrument/miri-performance/miri-sensitivity}
{https://jwst-docs.stsci.edu/jwst-mid-infrared-instrument/miri-performance/miri-sensitivity}}.
 These values are updated versions of the ones presented in Table~3 of
 \citet{10.1086/682259} and will be updated during commissioning of the JWST. We
 selected the filters F1000W, F2100W for the wavelengths \SI{10}{\upmu\m},
 \SI{21}{\upmu\m} and retrieved values of \SI{0.52e-6}{Jy}, \SI{5.14e-6}{Jy},
 respectively. These values of sensitivity are valid for an unresolved point source.
 Using Eq.~12 from \citet{10.1086/682259}, which approximates the fraction of the total number
 of photons received by the central pixel, together with the field of view of a MIRI pixel
 of $\approx \SI{0.012}{arcsec^{2}}$ we transformed these values to
 surface brightness sensitivities of
 $\approx \SI{4.2e-5}{Jy\,arcsec^{-2}}$,
 \SI{4.2e-4}{Jy\,arcsec^{-2}}, respectively.

 Like for the discussion of JWST/NIRCam, for this general and qualitative discussion
 of JWST/MIRI performance we neglected any influence of a possible coronagraph. The
 Lyot coronagraph of MIRI, especially suited for the observation of circumstellar
 disks, has an angular radius on the sky of \SI{2.16}{arcsec}
 \citep{10.1086/682256}. That is smaller by at least a factor of two than the
 angular distance of disk emission to central star in the investigated systems.

 To detect our peak surface brightness of
 $\sim \SI{1e-7}{Jy\,arcsec^{-2}}$ (see the second panel of
 Fig.~\ref{fig_all_waves_unnormalized_n-i-M2} and
 \ref{fig_all_waves_unnormalized_n-o-M3}, respectively) with a \SNR\ of \num{3}
 at a wavelength of \SI{10}{\upmu\m} using JWST/MIRI an exposure time of
 $t\sbs{exp} \approx \SI{9}{\hour}$ is required.

 At the wavelength of \SI{21}{\upmu\m}, the situation is more favorable for JWST/MIRI
 than at \SI{10}{\upmu\m}: For example, to detect intermediate surface brightness
 values of $\SI{5e-5}{Jy\,arcsec^{-2}}$ (see the third panel of
 Fig.~\ref{fig_all_waves_unnormalized_n-i-M2} and
 \ref{fig_all_waves_unnormalized_n-o-M3}, respectively) with a \SNR\ of \num{3}, an
 exposure time of $\approx \SI{10}{\min}$ is required.

 \noindent\textbf{\textit{Herschel}/PACS} at \SI{70}{\upmu\m}:
 We adopt values from Table~1 of
 \citet{10.1051/0004-6361/201118581}, who observed the Fomalhaut debris disk:
 a surface brightness of $\approx \SI{16.5e-3}{Jy\,arcsec^{-2}}$
 with a \SNR\ of \num{77} and a total exposure time of $t\sbs{exp} = \SI{10956}{\s}$.
 Based on these values and assuming a similar observing mode as it was used
 by \citet{10.1051/0004-6361/201118581}, our surface brightnesses at
 \SI{70}{\upmu\m} of $\sim \SI{5e-3}{Jy\,arcsec^{-2}}$ are
 well detectable.

%%%%%%%%%%%%%%%%%%%%%%%%%%%%%%%%%%%%%%%%%%%%%%%%%%%%%%%%%%%%%%%%%%%%%%%%%%%%%%%%

\subsection{Twisting of the small grain halo}\label{appdx_subsect_halo_twisting}
\begin{figure}
  \resizebox{\hsize}{!}{
            \includegraphics[scale=0.6]{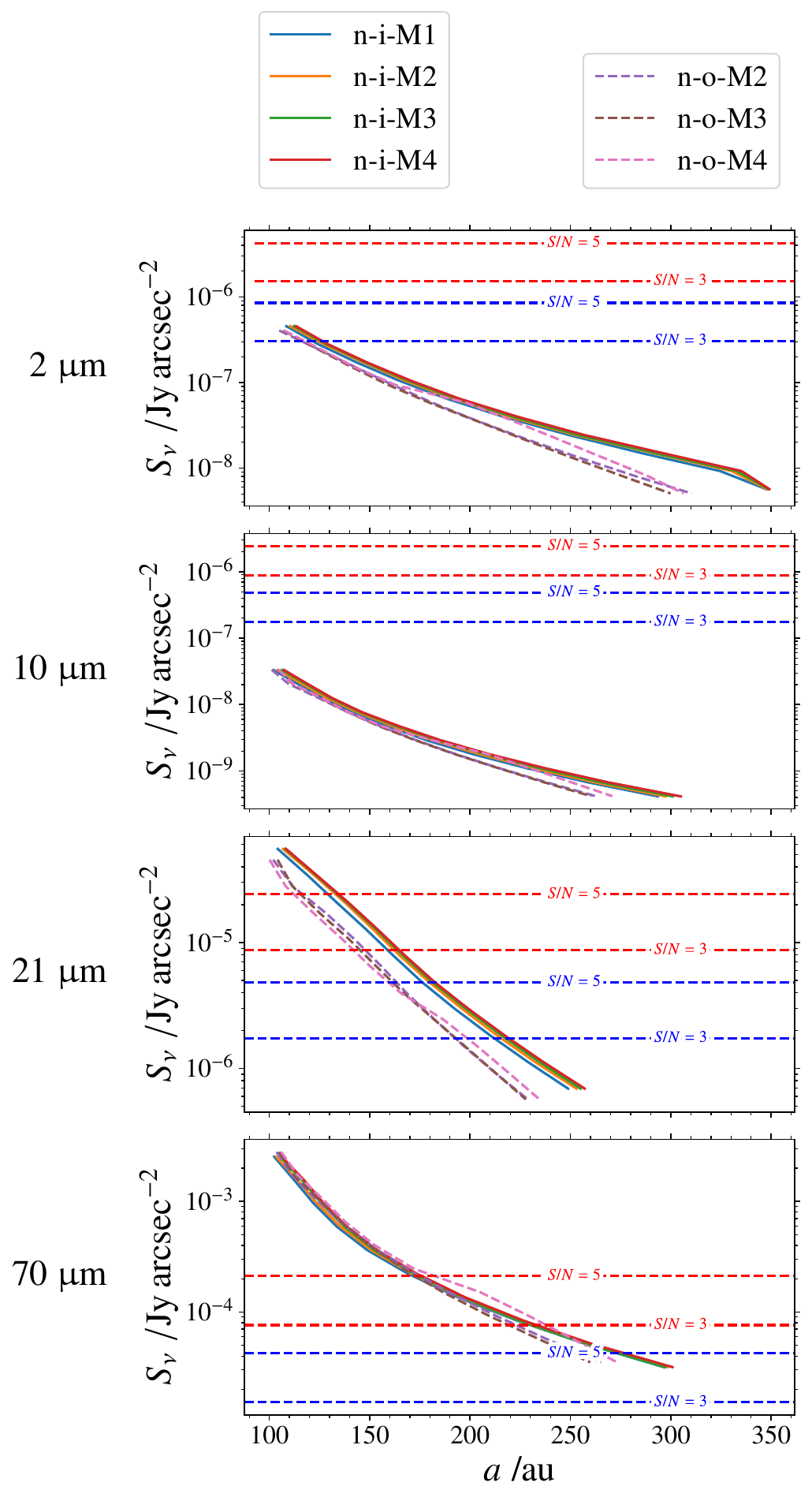}}
  \caption{Brightness level of the isophotes used to trace the small grain halo (see
           Sect.~\ref{subsect_halo_twisting}) drawn over semimajor axis $a$.
           Solid lines are used for the systems with inner perturbers and dashed lines for
           those with outer perturbers, and rows are used to show four different
           wavelengths that trace small halo grains: \SI{2}{\upmu\m},
           \SI{10}{\upmu\m}, \SI{21}{\upmu\m}, and \SI{70}{\upmu\m}. The
           horizontal dashed lines denote the minimum surface brightness
           required to achieve a certain detection significance (\SNR), denoted
           by a multiple of the noise, $\sigma$. Red lines are for an exposure
           time of $t\sbs{exp} = \SI{1}{\hour}$, blue for
           $t\sbs{exp} = \SI{5}{\hour}$.
           For the wavelength of \SI{2}{\upmu\m} the corresponding instrument
           is JWST/NIRCam, for \SI{10}{\upmu\m} and \SI{21}{\upmu\m} it is
           JWST/MIRI, and for \SI{70}{\upmu\m} it is \textit{Herschel}/PACS.
           }
  \label{fig_isophotes_flux_levels}
 \end{figure}

 To illustrate the feasibility of characterizing the properties of the small
 grain halo, the values of surface brightness corresponding to the isophotes
 that are presented in Sect.~\ref{subsect_halo_twisting} are displayed in
 Fig.~\ref{fig_isophotes_flux_levels}. Here we neglect the instrument
 PSF to keep this discussion independent
 from the distance to the system. We note that especially in the case of
 \textit{Herschel}/PACS, convolving with the instruments PSF can significantly
 decrease the surface brightness of the systems. Additionally, horizontal lines
 indicate the surface brightnesses required to achieve a \SNR\ of \num{3} and
 \num{5} for exposure times of $t\sbs{exp} = \SI{1}{\hour}$ and \SI{5}{\hour}.
 At the wavelength \SI{2}{\upmu\m} this is done for JWST/NIRCam, at \SI{10}{\upmu\m}
 and \SI{21}{\upmu\m} for JWST/MIRI, and at \SI{70}{\upmu\m} for
 \textit{Herschel}/PACS.

 When observing at \SI{2}{\upmu\m} with JWST/NIRCam, only the innermost parts of the
 halo can be detected with a \SNR\ of at least \num{3} and an exposure time of
 $t\sbs{exp} = \SI{5}{\hour}$, while at \SI{10}{\upmu\m}, the halo is too faint
 to be detected within that time.
 At the wavelength of \SI{21}{\upmu\m}, substantial parts of the halo can
 be detected with JWST/MIRI. An observation with $t\sbs{exp} = \SI{1}{\hour}$
 is sufficient to detect emission with values of surface brightness
 corresponding to isophotes with semimajor axes of up to
 $a \sim \SI{140}{au}$ -- \SI{160}{au} with a \SNR\ of \num{3}. That allows
 halo regions undergoing differential rotation to be detected, but not those
 with flipping isophote orientation (see Fig.~\ref{fig_rel_isophote_rotation}).
 To do so, five hours of observation time are required to detect
 emission with values of surface brightness corresponding to isophotes with
 semimajor axes of up to $a \sim \SI{190}{au}$ -- \SI{220}{au} with a \SNR\
 of \num{3}, enough to measure almost the entire halo region undergoing that
 flip.

 By observing at the wavelength of \SI{70}{\upmu\m} with \textit{Herschel}/PACS,
 major parts of the halo can be detected. An observation with
 $t\sbs{exp} = \SI{1}{\hour}$ would have been sufficient to detect emission with
 values of surface brightness corresponding to isophotes with semimajor axes of up to
 $a \sim \SI{220}{au}$ -- $\SI{240}{au}$ with a \SNR\ of \num{3}.

%%%%%%%%%%%%%%%%%%%%%%%%%%%%%%%%%%%%%%%%%%%%%%%%%%%%%%%%%%%%%%%%%%%%%%%%%%%%%%%%

\subsection{Spiral structure in the $Q$ band}\label{appdx_subsect_spiral}
\begin{figure}
  \resizebox{\hsize}{!}{
            \includegraphics{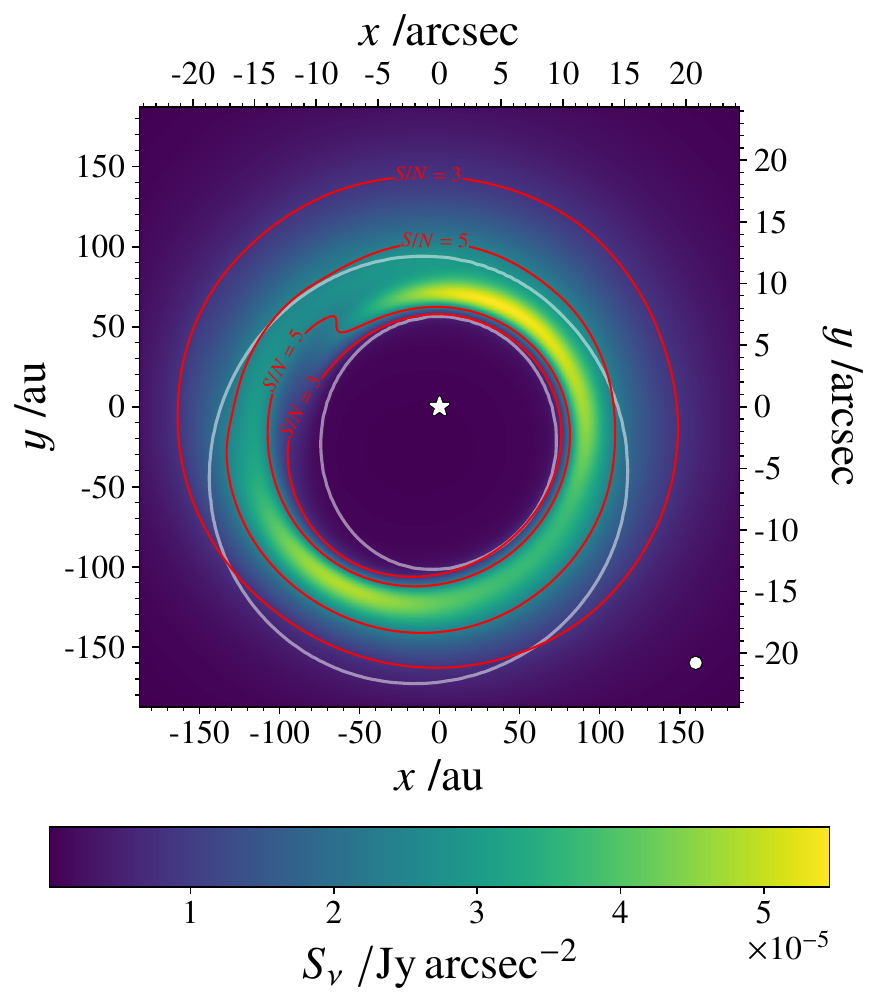}}
   \caption{Surface brightness distribution for a stellar distance of
            \SI{7.7}{pc} at a wavelength of \SI{21}{\upmu\m}
            of the system \run{n-i-M4}, convolved with a PSF suitable
            for JWST/MIRI with the filter F2100W that has been obtained with the
            WebbPSF tool using a flat input source. To illustrate the PSF,
            a white circle is depicted in the lower-right edge with a diameter
            of \SI{0.8}{\arcsec} (suitable for a wavelength of \SI{21}{\upmu\m}
            and a telescope diameter of \SI{6.5}{\m}), similar to the used, more
            complex PSF. The red contour lines denote
            the regions succeeding the minimum surface brightness required to
            achieve a certain \SNR\ when observing with JWST/MIRI for an
            exposure time of $t\sbs{exp} = \SI{1}{\hour}$.
            General figure characteristics are the same as in
            Fig.~\ref{fig_images_all_waves_inner_outer}, and white contour lines have same meaning as in
            Fig.~\ref{fig_spiral_inner_21micron}.}
   \label{fig_n-i-M4_21micron_MIRI_sensitivity}
 \end{figure}

 To illustrate the observability of the spiral structure appearing in the
 $Q$ band (see Sect.~\ref{subsect_spiral}), the surface brightness distribution
 of the system \run{n-i-M4} such as
 in Fig.~\ref{fig_spiral_inner_21micron} is displayed in
 Fig.~\ref{fig_n-i-M4_21micron_MIRI_sensitivity}, now convolved with a PSF
 suitable for JWST/MIRI. The PSF was calculated with version 1.0.0 of the tool
 WebbPSF$\,$\footnote{\href{https://www.stsci.edu/jwst/science-planning/proposal-planning-toolbox/psf-simulation-tool}
{https://www.stsci.edu/jwst/science-planning/proposal-planning-toolbox/psf-simulation-tool}}
 \citep{10.1117/12.925230, 10.1117/12.2056689}
 using a flat input source for the filter F2100W. Beforehand, the surface
 brightness map  was remapped to the JWST/MIRI pixel scale used by WebbPSF, which is
 \SI{0.1108}{arcsec}. Overlaid on the surface brightness map are contour
 lines denoting significance regions for an exposure time of
 $t\sbs{exp} = \SI{1}{\hour}$. The whole area of the spiral structure can be
 measured with a \SNR\ of more than \num{3} and most parts of it with a \SNR\ of
 more than \num{5}.

 However, the contrast of the spiral structure would not be sufficient to
 detect the structure with such
 an observational setup. We assumed the structure to be detectable when the
 radial minimum and maxima producing the spiral structure have non overlapping
 noise intervals (computed by dividing the surface brightness by its \SNR).
 Based on that assumption, an exposure time of
 $t\sbs{exp} \gtrsim \SI{8}{\hour}$ with JWST/MIRI would be required to
 significantly detect the spiral pattern for the system \run{n-i-M4}.

%%%%%%%%%%%%%%%%%%%%%%%%%%%%%%%%%%%%%%%%%%%%%%%%%%%%%%%%%%%%%%%%%%%%%%%%%%%%%%%%

\subsection{Grains scattered inward of the parent planetesimal belt}
\label{appdx_subsect_grains_inward_outer}
\begin{figure}
  \resizebox{\hsize}{!}{
            \includegraphics{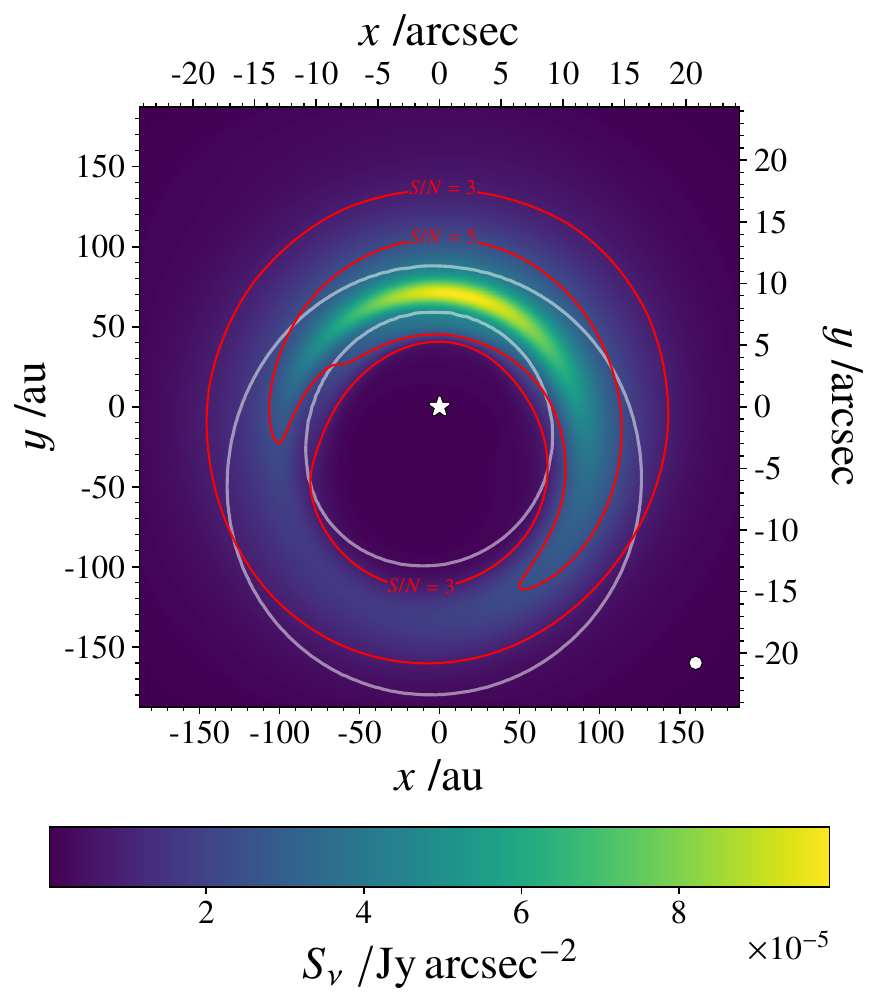}}
   \caption{Same as Fig.~\ref{fig_n-i-M4_21micron_MIRI_sensitivity}, but for the
            system \run{n-o-M3}.}
   \label{fig_n-o-M3_21micron_MIRI_sensitivity}
 \end{figure}

 The overall surface brightness of our systems is much higher at a wavelength
 of \SI{21}{\upmu\m} than at \SI{10}{\upmu\m} (see
 Fig.~\ref{fig_all_waves_unnormalized}, and Sect.~\ref{subsect_dms} and
 Fig.~\ref{fig_radial_profiles} for the different contribution to surface
 brightness of different grain sizes). Therefore, we opt for
 \SI{21}{\upmu\m} to illustrate the observability of emission from the region
 inward of the parent planetesimal belt (see
 Sect.~\ref{subsect_grains_inward_outer}) with JWST/MIRI. In
 Fig.~\ref{fig_n-o-M3_21micron_MIRI_sensitivity}, the surface brightness map
 at \SI{21}{\upmu\m} of the system \run{n-o-M3} is displayed. The map has been
 treated the same way as Fig.~\ref{fig_n-i-M4_21micron_MIRI_sensitivity} as
 described in Sect.~\ref{appdx_subsect_spiral}. Evidently, with a exposure time
 of $t\sbs{exp} = \SI{1}{\hour}$ using JWST/MIRI, the emission inside the belt
 can be detected with a \SNR\ of at least \num{3} and mostly of at least \num{5}.

\subsection{Number of azimuthal radial flux maxima}
\label{appdx_subsect_azimuthal_maxima}
 We estimated the exposure time required to demarcate the second azimuthal maximum
 at $\theta \sim \SI{240}{\degree}$ from its consecutive minimum of the system
 \run{n-i-M2} at the wavelength of \SI{21}{\upmu\m} with JWST/MIRI. As in
 Sect.~\ref{appdx_subsect_spiral} we assumed the structure to be detectable when
 the extrema values of surface brightnesses have non overlapping noise
 intervals. We found that an exposure time of $t\sbs{exp} \approx \SI{1}{\hour}$
 would be sufficient to achieve that contrast.
\end{appendix}

%%%%%%%%%%%%%%%%%%%%%%%%%%%%%%%%%%%%%%%%%%%%%%%%%%%%%%%%%%%%%%%%%%%%%%%%%%%%%%%%
%%%%%%%%%%%%%%%%%%%%%%%%%%%%%%%%%%%%%%%%%%%%%%%%%%%%%%%%%%%%%%%%%%%%%%%%%%%%%%%%

\title{Using debris disk observations to infer substellar companions orbiting
       within or outside a parent planetesimal belt\\ \textit{(Corrigendum)}}
\titlerunning{Infer substellar companions from debris disk observations}
\author{T. A. Stuber\inst{\ref{inst_tucson}, \ref{inst_kiel_2}} \orcidlink{0000-0003-2185-0525}
        \and T. Löhne\inst{\ref{inst_jena_2}} \orcidlink{0000-0003-2584-5280}
        \and S. Wolf\inst{\ref{inst_kiel_2}} \orcidlink{0000-0001-7841-3452}
}
\authorrunning{T. A. Stuber et al.}
\institute{}
\institute{
	   Department of Astronomy and Steward Observatory,
	   The University of Arizona, 933 North Cherry Ave, Tucson, AZ 85721, USA\\
	   \email{tstuber@arizona.edu}
	   \label{inst_tucson}
	   \and Institut für Theoretische Physik und Astrophysik,
           Christian-Albrechts-Universität zu Kiel,
           Leibnizstr. 15, 24118 Kiel, Germany
           \label{inst_kiel_2}
           \and Astrophysikalisches Institut und Universitätssternwarte,
           Friedrich-Schiller-Universität Jena, Schillergässchen 2–3,
           07745 Jena, Germany\label{inst_jena_2}
          }

\date{A\&A 669, A3 (2023) \href{https://doi.org/10.1051/0004-6361/202243240}{https://doi.org/10.1051/0004-6361/202243240}}


\fancyhead{} % avoid running title
\maketitle

%%%%%%%%%%%%%%%%%%%%%%%%%%%%%%%%%%%%%%%%%%%%%%%%%%%%%%%%%%%%%%%%%%%%%%%%%%%%%%%%
%%%%%%%%%%%%%%%%%%%%%%%%%%%%%%%%%%%%%%%%%%%%%%%%%%%%%%%%%%%%%%%%%%%%%%%%%%%%%%%%

\noindent In Appendix B of the original article we investigated the sensitivity of JWST/MIRI for the filters F1000W and F2100W. We transformed sensitivities
valid for an unresolved point source into surface brightness sensitivities using Eq.~12 from \citet{10.1086/682259}. For the filter F1000W, the resulting detection
limit is given in the original article as $\approx \SI{4.2e-5}{Jy\,arcsec^{-2}}$. This limit should have been read $\approx \SI{3.5e-6}{Jy\,arcsec^{-2}}$. This is only a
typographical error, and the correct value was used for the analysis. For the filter F2100W, the resulting detection limit is given in the original article as
$\approx \SI{4.2e-4}{Jy\,arcsec^{-2}}$. However, in our analysis, a limit of $\approx \SI{3.5e-5}{Jy\,arcsec^{-2}}$ was used. This value was erroneously
calculated; the correct limit is $\approx \SI{7.9e-6}{Jy\,arcsec^{-2}}$. This detection limit is lower than the used limit, and hence the performance of JWST/MIRI at
a wavelength of \SI{21}{\upmu\m} was underestimated.

Consequently, the exposure times required to detect structures in the surface brightness distributions using JWST/MIRI at \SI{21}{\upmu\m} are shorter than
what is given in the original article. To detect intermediate surface brightness values of \SI{5e-5}{Jy\,arcsec^{-2}} with a \SNR\ of \num{3} an exposure time of
$t\sbs{exp} \approx \SI{2.5}{\min}$ is required (instead of the $t\sbs{exp} \approx \SI{10}{\min}$ given in the original article).

\begin{figure}
  \resizebox{\hsize}{!}{
            \includegraphics[scale=0.6]{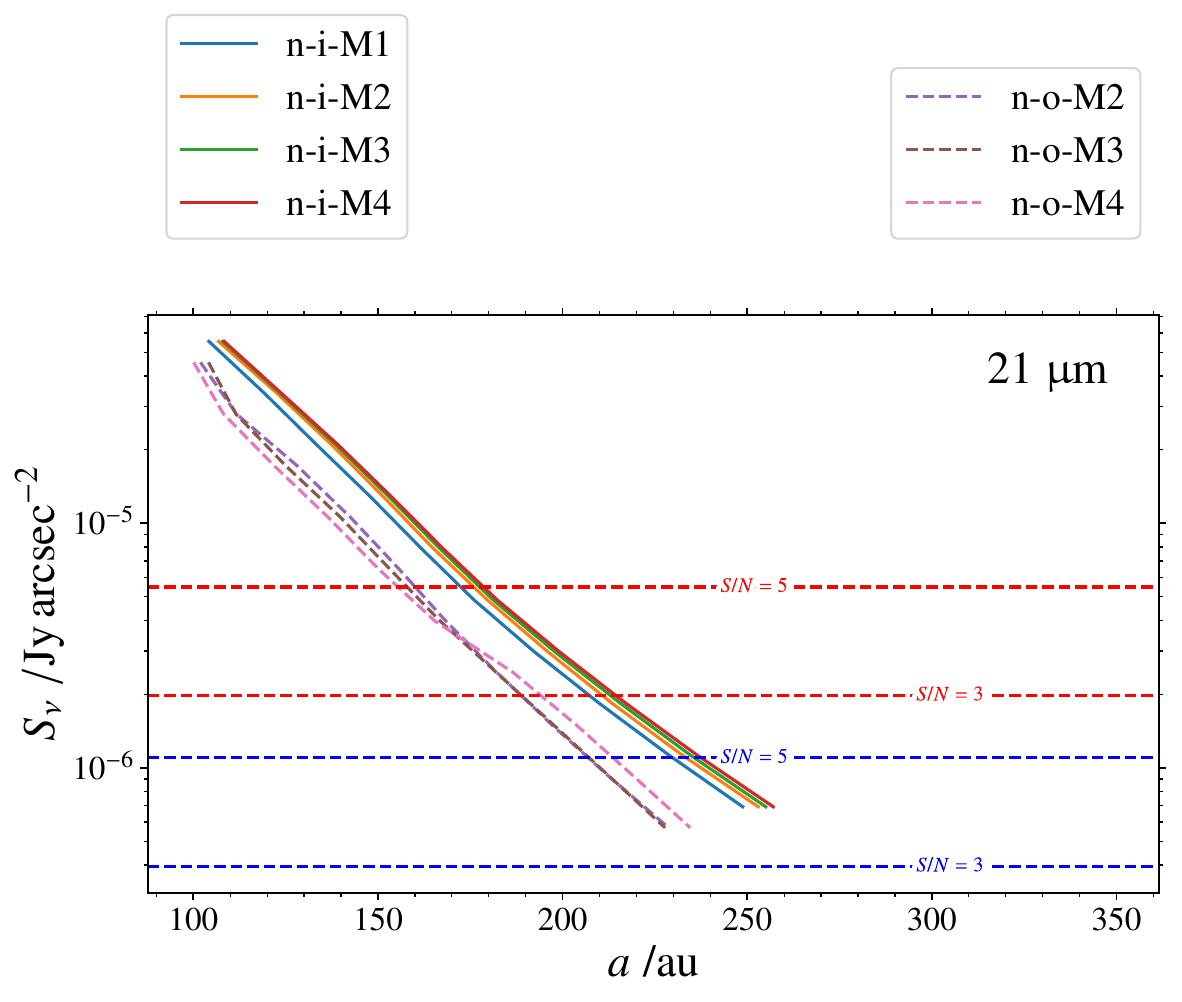}}
  \caption{
      Same as the third panel of Fig. B.2 in the original article, but with the correct sensitivity limits used to compute the surface brightness values of the horizontal
      dashed lines. These horizontal lines denote the minimum surface brightness required to achieve a certain detection significance (\SNR) using JWST/MIRI
      at a wavelength of  \SI{21}{\upmu\m}.
      Red lines are for an exposure time of $t\sbs{exp} = \SI{1}{\hour}$, blue lines for $t\sbs{exp} = \SI{5}{\hour}$.
           }
  \label{figure_corrigendum}
 \end{figure}

The third panel of Fig. B.2 shows horizontal lines with too high sensitivity limits. This is corrected here in Fig.~\ref{figure_corrigendum}. Using the correct
sensitivity limits, an exposure time of $t\sbs{exp} = \SI{1}{\hour}$ is sufficient to detect emission with surface brightness values corresponding to isophotes with
semimajor axes of up to $a \sim \SI{190}{au} - \SI{220}{au}$ with a \SNR\ of 3. With $t\sbs{exp} = \SI{5}{\hour}$, all halo regions investigated can be detected
with a \SNR\ of at least \num{3}.

The red contour lines in Figs. B.3 and B.4 are valid for an exposure time of $t\sbs{exp} \approx \SI{14}{\min}$ (instead of the $t\sbs{exp} = \SI{1}{\hour}$ given
in the original article). For the system \run{n-i-M4}, an exposure time of  $t\sbs{exp} \gtrsim \SI{2}{\hour}$ would be required to significantly detect the spiral
pattern (instead of the $t\sbs{exp} \gtrsim \SI{8}{\hour}$ given in the original article).\\

\noindent The error only affects the conclusions about the feasibility of detecting certain disk features with JWST/MIRI at a wavelength of \SI{21}{\upmu\m}:
a shorter exposure time than what is stated in the original article is required. The main conclusions of the original article remain unchanged.

%%%%%%%%%%%%%%%%%%%%%%%%%%%%%%%%%%%%%%%%%%%%%%%%%%%%%%%%%%%%%%%%%%%%%%%%%%%%%%%%
%%%%%%%%%%%%%%%%%%%%%%%%%%%%%%%%%%%%%%%%%%%%%%%%%%%%%%%%%%%%%%%%%%%%%%%%%%%%%%%%

\newpage
\bibliographystyle{aa_url}
\bibliography{bibliography}

\end{document}